\documentclass[journal]{IEEEtran}
\usepackage{cite}

  \usepackage{graphicx,epstopdf,epsfig}
  \graphicspath{{figures/}}
  \DeclareGraphicsExtensions{.eps}

\usepackage{amsmath}
\usepackage{array}

\usepackage{fixltx2e}

\usepackage{url}

\usepackage{booktabs}       
\usepackage{float}
\usepackage[caption=false,font=footnotesize,labelfont=sf,textfont=sf]{subfig}

\begin{document}
%

\title{Collision Selective Visual Neural Network Inspired by LGMD2 Neurons in Juvenile Locusts}

\author{Qinbing~Fu,
        Cheng~Hu,
        and~Shigang~Yue,~\IEEEmembership{Member,~IEEE}
\thanks{Q. Fu is with the School of Computer Science, University of Lincoln, Lincoln, LN6 7TS, UK. (first author,Email: qifu@lincoln.ac.uk)}
\thanks{C. Hu is with the School of Computer Science, University of Lincoln, Lincoln, LN6 7TS, UK. (Email: chu@lincoln.ac.uk)}%
\thanks{S. Yue is with the School of Computer Science, University of Lincoln, Lincoln, LN6 7TS, UK. (corresponding author, Email: syue@lincoln.ac.uk)}}%


\maketitle

\begin{abstract}
For autonomous robots in dynamic environments mixed with human, it is vital to detect impending collision quickly and robustly. The biological visual systems evolved over millions of years may provide us efficient solutions for collision detection in complex environments. In the cockpit of locusts, two Lobula Giant Movement Detectors, i.e. LGMD1 and LGMD2, have been identified which respond to looming objects rigorously with high firing rates. Compared to LGMD1, LGMD2 matures early in the juvenile locusts with specific selectivity to dark moving objects against bright background in depth while not responding to light objects embedded in dark background - a similar situation which ground vehicles and robots are facing with. However, little work has been done on modeling LGMD2, let alone its potential in robotics and other vision-based applications. In this article, we propose a novel way of modeling LGMD2 neuron, with biased ON and OFF pathways splitting visual streams into parallel channels encoding brightness increments and decrements separately to fulfill its selectivity. Moreover, we apply a biophysical mechanism of spike frequency adaptation to shape the looming selectivity in such a collision-detecting neuron model. The proposed visual neural network has been tested with systematic experiments, challenged against synthetic and real physical stimuli, as well as image streams from the sensor of a miniature robot. The results demonstrated this framework is able to detect looming dark objects embedded in bright backgrounds selectively, which make it ideal for ground mobile platforms. The robotic experiments also showed its robustness in collision detection - it performed well for near range navigation in an arena with many obstacles. Its enhanced collision selectivity to dark approaching objects versus receding and translating ones has also been verified via systematic experiments.
\end{abstract}


%
\IEEEpeerreviewmaketitle

\section{Introduction}
\label{InSec}
\IEEEPARstart{C}{ollision} detection is crucial to many animals' survival in searching for food and/or escaping from predators. For future intelligent robots, ability to detect collision timely and efficiently is also critical to navigate in a dynamic environment mixed with human hosts. There are now a few state-of-the-art collision detectors based on either infra-red, ultrasound, laser, radar, vision sensor or combination of these sensors \cite{LGMD1-Yue2006}. However, those solutions are restricted heavily from wider application due to their size, efficiency, reliability and/or energy consumption. Amongst different sensing modalities, vision plays an irreplaceable role in responding to a dynamic environment for many animals. On the other hand, the traditional segmentation and registration based computer vision methods can not cope with the degree of complexity in real physical world for real time collision detection tasks \cite{NeuroVisionSensor-2000,LGMD1-robot2010}.

As the result of hundreds of millions of years evolution, biological vision systems have provided abundant source of inspirations for constructing artificial visual systems for collision detection. Especially the insects' vision systems, which have demonstrated amazing ability in interacting with the dynamic world yet with very limited number of neurons compared to the vertebrates' brains, could be ideal models to design collision free artificial vision systems. In locusts, for example, much progress has been made in understanding the cellular mechanisms underlying motion detection \cite{LGMD-anatomy,DCMD-1974,LGMD-ONOFF,LGMD1-1996,LGMD2-1997,LGMD1-synaptic2015,LGMD2-cockpit}. A group of Lobula Giant Movement Detectors (LGMDs) in the third stack of neuropiles in the locusts’ optic lobe have been discovered \cite{LGMD1-1996}, - two of them, which are identified as LGMD1 and LGMD2, respond selectively to looming objects in depth \cite{LGMD-anatomy,LGMD1-1996,LGMD2-cockpit,LGMD2-1997} with high frequency spikes.

\begin{figure}[!t]
	\centering
	\includegraphics[width=0.4\textwidth]{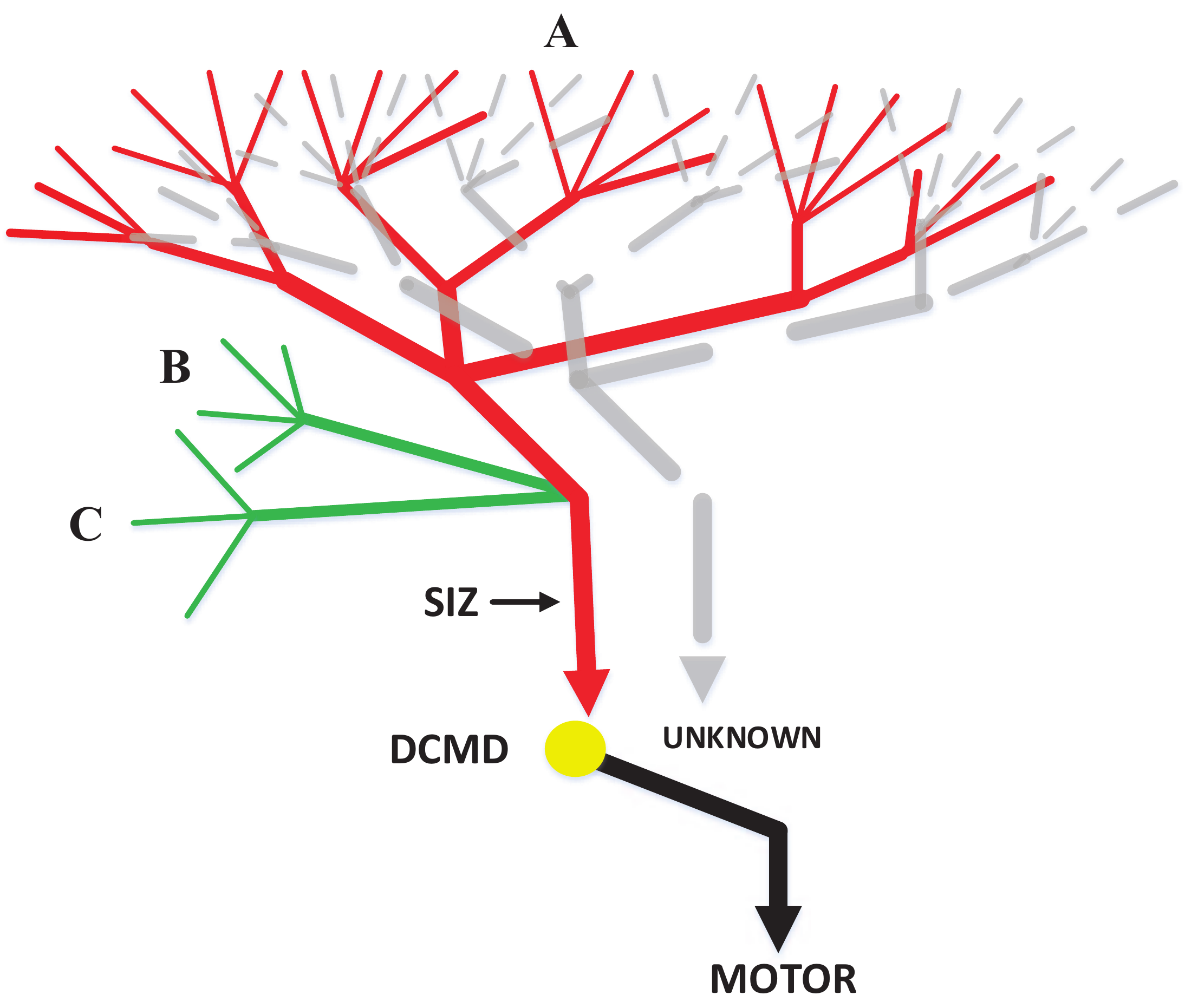}
	\caption{The schematic illustration of LGMDs neural circuitry. The red area indicates the LGMD1's dendritic tree whilst the gray-dashed one denotes the LGMD2's. DCMD (yellow neuron) as a one-to-one connected post-synaptic partner to LGMD1 passes signals further to the motion motor; the target to post-synaptic area of LGMD2 remains unknown. SIZ indicates the spiking initiation zones of both LGMDs. The LGMD1's dendritic tree consists of additional two ventral subfields B and C, which are absent from LGMD2.}
	\label{circuitry}
\end{figure}

In morphology, both LGMDs have a characteristic, extensive fan-shaped arbor \cite{LGMD2-1997}, as shown in Fig.\ref{circuitry}. The lobula arbor of LGMD2 is beneath that of LGMD1; there are also two dendritic subfields (B and C in Fig.\ref{circuitry}) located more ventrally in the lobula area of LGMD1, which are lacking from LGMD2. The Descending Contralateral Motion Detector (DCMD) is a one-to-one post-synaptic target neuron to LGMD1, which is directly excited through a chemical synapse between them and conveys information to further motion control system \cite{DCMD-1974,DCMD-1992one,DCMD-1992two,DCMD-2013}. However, the post-synaptic partner to LGMD2 is still unknown \cite{LGMD2-1997}. Compared to LGMD1, recent research revealed LGMD2 matures early in juvenile, even the newly hatching locusts \cite{LGMD2-cockpit}. The juveniles are lacking of wings so that living mainly on the ground, whereas they are already capable of reacting to imminent dangers, especially swooping predators from the sky. Recent investigation suggests that the LGMD2 neuron could be a dominant collision-detector for juvenile's hiding behaviors \cite{LGMD2-cockpit}.

Computationally modeling the fascinating collision-detecting neurons such as LGMD1 and LGMD2 will not only deepen our understanding of the visual pathways in locusts, but also shed lights to vision systems for future robots. In the past decades, LGMD1 neuron has been modeled and tested in vehicles and robots for collision detection \cite{LGMD1-car2006,LGMD1-car2007,LGMD1-robot2005,LGMD1-robot2010,LGMD1-robot2012,LGMD1-frombiologytorobot,LGMD1-Silva1,LGMD1-Silva2}. On the other hand, for LGMD2 in juvenile locusts, although it shows unique selectivity on dark looming objects against bright background, yet very little LGMD2 modeling work has been done in the past\cite{LGMD2-Fu,LGMD2-BMVC}. In this paper, we attempt to fill this gap by modeling LGMD2 and testing the model with systematic experiments to demonstrate its characteristic and potential.

Unlike the previous LGMD1 models, e.g. \cite{LGMD1-Glayer,LGMD1-1996,LGMD1-nonlinear}, we proposed a new structure to separate the ON and OFF channels from the photoreceptor layer. For the first time, the LGMD2 model works well in responding selectively to only dark looming objects against bright background but not to white/bright objects embedded in dark background or to other translating moving objects, demonstrating the unique characteristics of LGMD2 neurons in juvenile locusts. We have also implemented spike frequency adaptation (SFA) mechanism \cite{SFA-Gabbiani,SFA-2009,SFA-2009Role}, to further enhance the selectivity of LGMD2 model on dark looming objects amongst other visual stimuli such as receding or translating movements, although there are other mechanisms in shaping LGMDs' looming selectivity \cite{SFA-2009}.

In the following sections, we will first review the related works in Section \ref{ReSec}. We then illustrate the detailed LGMD2 model with parameters setting in Section \ref{NNSec}. The systematic experiments with results, analysis and further discussion are followed up in Section \ref{EASec}. Finally, we give a conclusion.

\section{Related Work}
\label{ReSec}
In this section, we introduce the relevant works to the proposed LGMD2 model in the areas of - neural properties of LGMDs and related modeling works, ON and OFF visual pathways, neural signal competition, and biophysical mechanism of spike frequency adaptation.

\subsection{LGMDs Neural Characteristics and Models}
In the lobula area in locusts, LGMD1 was first identified as a movement detector \cite{LGMD-anatomy,LGMD-ONOFF} and gradually recognized as a looming objects detector, e.g. \cite{LGMD1-1996}. In the same place, LGMD2 was also identified but with unique characteristics that are different to the LGMD1 \cite{LGMD2-1997,DCMD-2013,LGMD2-cockpit}. Both LGMD1 and LGMD2 respond selectively to looming stimuli, with increasing firing rates, peaked before the objects reach a particular angular size in the retina \cite{LGMD2-1997,LGMD1-1996,LGMD1-Escapes2010,LGMD2-cockpit}. They are both inhibited during either the whole-field luminance change or grating movements \cite{LGMD1-1996,LGMD2-1997,DCMD-2013}. However, the LGMD2 neuron matures very early in juvenile locusts \cite{LGMD2-cockpit}, and one of its unique features is that it only responds to light-to-dark luminance change, which may be representing swooping predators from the sky. This special selectivity, i.e, it is able to detect moving dark objects embedded in the bright background in depth selectively while not responding to light objects approaching against the dark background, makes it outside of normal expectation and an unique neuron to model. On the other aspect, early researches have demonstrated LGMD1 neurons respond to both situations of illumination and darkening \cite{LGMD1-1996,LGMD1-seeing1999,DCMD-2013}. In addition, when challenged against translating stimuli, both LGMDs are excited for a short while then inhibited soon even early before the end of movements \cite{LGMD2-1997,LGMD1-1996}.

To realize the neural characteristics of LGMDs, a few computational models have been proposed for LGMD1 \cite{LGMD1-Glayer,LGMD1-Meng,LGMD1-Silva1,LGMD1-Silva2,LGMD1-Yue2006} and successfully utilized in vision-based platforms such as vehicles \cite{LGMD1-car2007,LGMD1-car2006} and robots \cite{LGMD1-frombiologytorobot,LGMD1-robot2005,LGMD1-Yue2009,LGMD1-robot2010,LGMD1-robot2012,LGMD1-nonlinear} for collision detection. Nevertheless, very little modeling works have been conducted for LGMD2. In this article, we will propose a visual neural network to fulfill the specific properties of LGMD2 and explore its potential with systematic experiments. The preliminary results of this research has been partially published in \cite{LGMD2-BMVC,LGMD2-Fu}.

\subsection{ON and OFF Visual Pathways}
In recent years, the ON and OFF visual pathways have been found in motion detection circuitry of not only insects like the drosophila \cite{Joesch_2010,Joesch_2013}, but also vertebrates like the rabbit \cite{Circuit-motion}, which reveal the fundamental principle of processing visual information - signals are separated into parallel ON and OFF pathways encoding brightness increments (onset events) in ON channels and decrements (offset events) in OFF channels respectively \cite{Circuit-genetic,Circuit-2Q,Circuit-correlation2013,Clark_2011}. As a matter of fact, such a structure has been asserted to play an irreplaceable role underlying separated pathways in motion detection circuit \cite{Circuit-motion}. Although there is little evidence that such pathways exist in locusts, as early in 1970s, LGMDs were proposed to be fed by a homogeneous population of ON and OFF cells in their pre-synaptic areas \cite{LGMD-ONOFF}. As depicted in Fig. \ref{circuitry}, the signals conveyed in subfield A of LGMDs dendritic tree were put forward to be mediated by such polarity cells \cite{LGMD1-multiplication}.

In the LGMD1 state-of-the-art frameworks, e.g. \cite{LGMD1-Glayer,LGMD1-Yue2006,LGMD1-Yue2009}, the visual signals are only processed in a single pathway. However, very recently, a LGMD1 modeling work  \cite{LGMD1-nonlinear} with relevant mechanism of ON and OFF cells was demonstrated, which shed lights on that such a dual-channel structure could also be conducted for LGMD neurons. Moreover, when modeling practical collision-detecting vision systems, the selective response to objects approach rather than recession should be enhanced. Some methods have been proposed to help discriminate approach from recession, such as comparing the angular speeds/sizes \cite{LGMD2-1997}, or monitoring the membrane potential change gradient \cite{LGMD1-Meng}, etc. However, such avenues either require costly computational power or have no biological plausibility. On this aspect, the separated ON and OFF visual pathways could be the secret mechanism underlying unique collision selectivity, especially for LGMD2 neuron, i.e., its unique feature of only selectively responding to dark looming objects embedded in bright background can be fulfilled via such a biological structure.

\subsection{Neural Signal Competition}
Within such collision-detecting neurons, like the LGMD1 \cite{LGMD1-1996,LGMD1-seeing1999,LGMD1-multiplicative,LGMD1-multiplication} and also the LGMD2 \cite{LGMD2-1997,DCMD-2013,LGMD2-cockpit}, there are two kinds of signal flows - excitation and inhibition, interacting and competing with each other. If the excitatory flow wins, the neuron will immediately spike; otherwise, it remains quiet. In addition, two kinds of inhibitions coexist to compete with excitation. First is the pre-synaptic lateral inhibition originates in the second neuropile - the medulla \cite{LGMD1-seeing1999}, and conveyed in the subfield A of dendritic trees in both LGMD1 and LGMD2 (Fig. \ref{circuitry}), which aims to cut down the excitations of motion-dependent pathway when an object growing on the retina \cite{LGMD1-multiplication,LGMD1-1996,LGMD2-1997}. There are also two extra ventrally located areas in LGMD1 (B and C in Fig. \ref{circuitry}), receiving the object-size corresponded inhibitory flows - feed forward inhibitions (FFI) \cite{LGMD1-multiplication}, which could directly suppress the neuron. Although such FFI circuit has not yet been anatomically explored for LGMD2 \cite{LGMD2-1997}, the vigorous inhibition recorded from the intracellular recordings of LGMD2 neuron, either after the end of looming or the start of recession, was demonstrated be shaped by the similar feed forward pathway \cite{LGMD2-1997}, activated when luminance changes rapidly over a large field of the retina. Therefore, we also construct the FFI pathway in LGMD2 neuron model to achieve the related biological findings \cite{LGMD2-1997}.

In the LGMD1 state-of-the-art models, e.g. \cite{LGMD1-Glayer,LGMD1-Yue2006,LGMD1-Yue2009,LGMD1-DSN-competing}, the lateral inhibition is always time-delayed relative to the excitation. However, the excitation has also been assumed to be time-delayed relative to the inhibition \cite{LGMD1-manyways,LGMD1-nonlinear,LGMD1-multiplication}. With the similar idea in the modeling of collision-detecting neuron \cite{LGMD1-nonlinear}, we demonstrate the ON and OFF mechanism leads effects on different relatively delayed information in the LGMD2 neuron model: luminance increments will activate ON cells to elicit onset responses - the excitation is time advanced relative to the inhibition in ON pathway; otherwise, the excitation is time delayed relative to the inhibition in OFF pathway, i.e, OFF cells generate offset responses by brightness decrements.

\subsection{Spike Frequency Adaptation}
LGMDs are well-known to respond better to looming than translating stimuli which are independent of velocity, size and direction of motion \cite{SFA-2009,SFA-2003}. In such collision-sensitive neurons, there are many mechanisms cooperating to shape their specific selectivities \cite{SFA-Gabbiani}. For LGMDs, two intrinsic processes have been well-studied which decline firing rate elicited during rapid change or sustained stimuli. The first is FFI as mentioned above, which mainly deals with transient luminance change over a large field. On the aspect of sustained stimuli, the biophysical mechanism - spike frequency adaptation also takes part in mediating looming selectivity in the spiking initiation zone (SIZ in Fig. \ref{circuitry}). When stimulated with constant stimuli, the neuron initially responds with a high spike rate then decays down to a lower steady state frequency \cite{SFA-2014}. Actually, the neural circuitry of LGMD2 shaping collision avoidance must be tuned to approaching rather than translating visual stimuli, as only the former type should reliably evoke collision-avoidance behaviors.

To be more specific, when challenged by constant translations, a fixed number of photoreceptors in the retina are activated which makes the neuron susceptible to adaptation. However, when facing approach, an increasing number of photoreceptors will be activated which likely leads the neuron to overcome adaptation. The SFA suppresses neural response to translating stimuli but has little effect on approaching ones underlying its indispensable role of shaping the selectivity of LGMDs for looming over translating stimuli \cite{SFA-2009Role,SFA-Gabbiani,SFA-2009}.

In addition, the receding stimuli will bring about a decreasing number of stimulated photoreceptors which also causes adaptation \cite{SFA-2009,SFA-2009Role}. As a result of that, the SFA is a reduction of neurons' firing rate to a stimulus of constant intensity, which makes it ideal in shaping LGMD2' looming selectivity for approach over recession and translation. There are some computational roles of SFA, one of which is acting as a high-pass filter on largely stationary inputs \cite{SFA-2009,SFA-2014}. More broadly, such a selective mechanism has also been proved useful in the neural processing of not only visual, but also auditory and electro-sensory systems \cite{SFA-2009}.

\section{The Visual Neural Network}
\label{NNSec}
In this section, we present the proposed LGMD2 neural network (or model) in detail. The key of the proposed LGMD2 model is an architecture of ON and OFF pathways splitting visual signals into parallel channels, each of which involves multiple layers. The brightness increments flow into ON channels whilst the decrements flow into OFF channels. Signals in separated pathways are spatially and temporally filtered, then pooled to form the membrane potential which is later mapped to invoke the spikes. Finally, a few continuous spikes elicited in a short time window corresponds to a potential collision recognition.

\begin{figure*}[!t]
	\centering
	\includegraphics[width=0.9\linewidth]{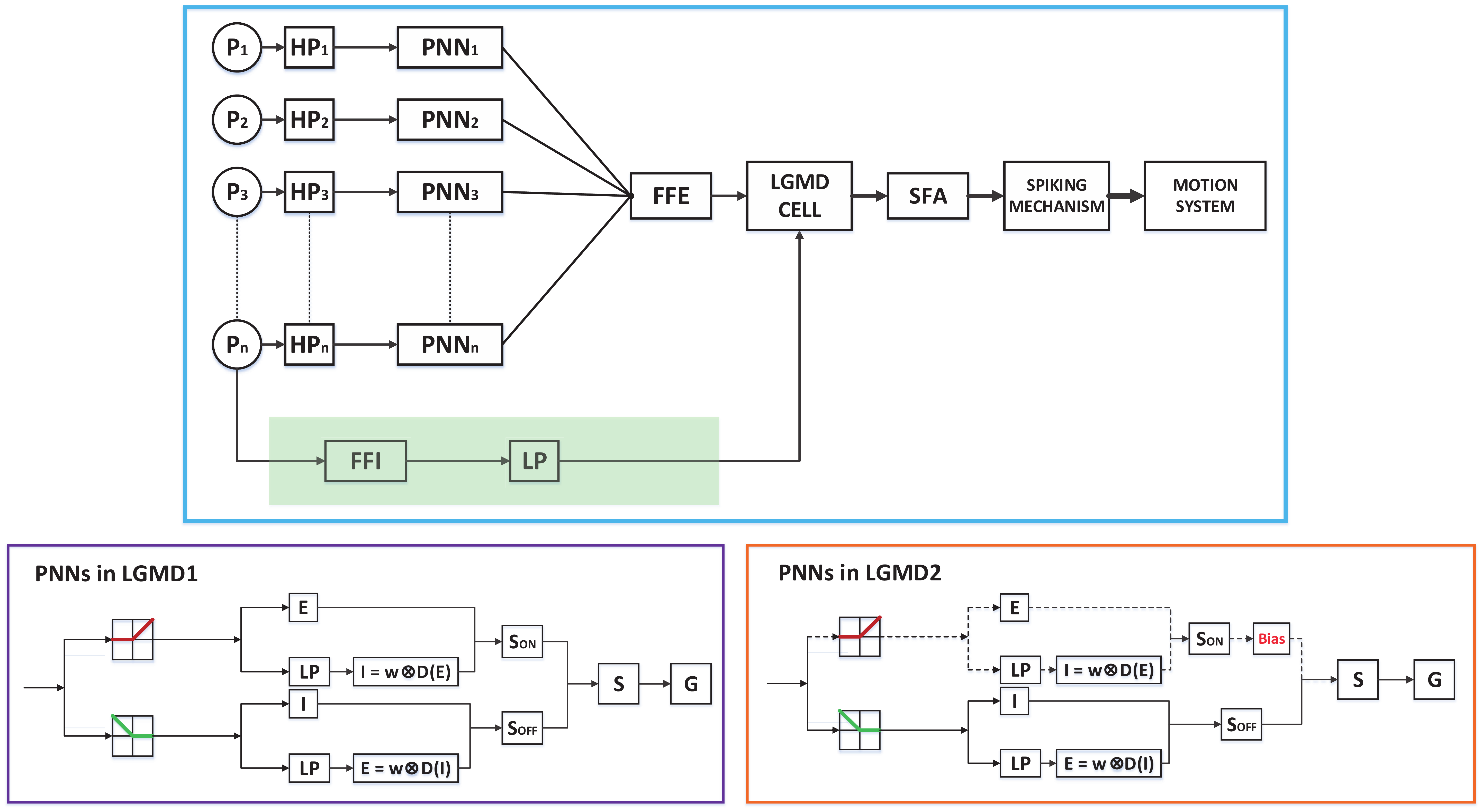}
	\caption{The LGMDs visual neural network with a general circuit in the upper box: $P_1$ to $P_n$ indicate total $n$ numbers of photoreceptors in the first neuropile layer. $HP_s$ denote the high-pass filters. $PNN_s$ indicate the partial neural networks varying between LGMD1 (bottom-left box) and LGMD2 (bottom-right box) corresponding to each local pixel. ON and OFF half-wave rectifiers split signals into two parallel pathways encoding brightness increments and decrements respectively. $LP_s$ denote the low-pass filters. In ON channels, the inhibitions ($I$) are convoluted by the surrounding time-delayed excitations ($D(E)$). In OFF channels, the excitations ($E$) are convoluted by the periphery time-delayed inhibitions ($D(I)$). $S_{ON}$ and $S_{OFF}$ denote the local summation cells. $S$ and $G$ indicate the Summation and Grouping layers integrate local dual-channel excitations. The bias is put forth in ON channels of LGMD2 circuit (dashed lines). $FFE$ indicates the feed forward excitation pooled from the intact pre-synaptic area. $FFI$ denotes the feed forward inhibition which is conveyed in another pathway(light green). The LGMD cell elicits sigmoid membrane potential which then goes through the SFA and spiking mechanism for generating spikes towards the motion system.}
	\label{model}
\end{figure*}

The LGMD2 neuron model is shown in Fig.\ref{model}. It is worth emphasizing the internal partial neural networks (PNNs) vary between LGMD1 and LGMD2. Specifically, for modeling LGMD2 neuron, the ON pathway is rigorously sieved to achieve LGMD2's unique collision selectivity to dark objects embedded in bright background. It is also necessary to note that compared to other vision-based collision detectors, the proposed framework only involves low-level image processing methods, detects potential collision by reacting to the expanding edges of an object. Those computationally expensive methodologies, such as target classification, scene analysis will not be applied in this study at all. Therefore, this neuron model has the potential of hardware realization for robotic applications.

\subsection{Photoreceptors}
The first layer of the visual neural network consists of photoreceptors arranged in a two-dimensional matrix form. The number of photoreceptors equals to the amount of pixels in the receptive field ($P_1$ to $P_n$ in Fig. \ref{model}, where subscript $n$ indicates the total quantity of receptive cells). Photoreceptors capture the discrete gray-scale visual inputs frame-by-frame. We apply a first-order high-pass filter (HP in Fig. \ref{model}) to retrieve the luminance change between every two continuous frames:
\begin{equation}
P_{x,y}(t) = (L_{x,y}(t) - L_{x,y}(t-1)) + \sum_{i}a_i \cdot P_{x,y}(t-i)
\label{high-pass}
\end{equation}
where $P_{x,y}(t)$ is the change of luminance corresponds each pixel at frame $t$, subscripts $x$ and $y$ are the 2-D coordinates. $L(t)$ and $L(t-1)$ are the original brightness of two successive frames with $t$, $t-1$ denoting the current and previous frames. The persistence of luminance change could last for a while: $i$ indicates the number of frames constitute the luminance duration; the coefficient $a_i$ is defined by $a_i = (1 + e^{u \cdot i})^{-1}$ and $u \in (-\infty,+\infty)$. We just take one pixel to represent the procedures before further pooling, which will not be restated later. If there is no difference between continuous frames, the photoreceptor will not be activated.

\subsection{ON and OFF Half-wave Rectifiers}
As described in Section \ref{ReSec}, the pre-synaptic areas of both LGMDs are reconciled by ON and OFF transient cells \cite{LGMD-ONOFF}. There are sufficient and identical densities of both polarity afferent units arranging to cover the intact retina, eliciting onset and offset events respectively depending on brightness change at each local cell. As depicted in PNNs of Fig. \ref{model}, the signals after high-pass filtering are fed into two parallel half-wave rectifiers, one forming the input to ON pathway, another to OFF pathway. There is a cutoff in both rectifiers which is set at $0$ in our case. Such mechanisms filter out or invert negative input, along with brightness increments flowing into ON channels, decrements of reverse-sign into OFF:
\begin{equation}
\begin{aligned}
&P^{ON}_{x,y}(t) = (P_{x,y}(t) + |P_{x,y}(t)|) / 2,\\
&P^{OFF}_{x,y}(t) = |(P_{x,y}(t) - |P_{x,y}(t)|)| / 2
\end{aligned}
\label{half-wave}
\end{equation}
where $P^{ON}_{x,y}$ denotes the ON cell value at $(x,y)$ and similarity for the OFF cell value $P^{OFF}_{x,y}$. Interestingly, despite building collision-sensitive neurons, the ON and OFF cells have also been proposed the potential to set up directional selective neurons \cite{LGMD1-nonlinear,Circuit-motion,Circuit-genetic,Circuit-2Q,Clark_2011}. One could manually decide the arrangement of both polarity cells, like neighboring placing a pair-wise of combination along the axis where a first onset sensitive cell and a second offset cell, etc. Alternately placing ON and OFF cells in the same layer could also encode for sensory neurons to the directions of translating stimulus. In the LGMD2 neuron model, we only concern its looming sensitivity in depth; therefore, each local pixel connects with a pair-wised ON and OFF cells respectively (Fig. \ref{model}).

In addition, we also allow a small fraction of original signals in parallel from the ON and OFF half-wave rectifiers to pass through, which mimics the absolute brightness in the motion detection system \cite{Circuit-2Q}. $\sigma_{1}$ indicates a fraction number to get residual information of previous frame.
\begin{equation}
\begin{aligned}
&P^{ON}_{x,y}(t) = P^{ON}_{x,y}(t) + \sigma_{1} \cdot P^{ON}_{x,y}(t-1),\\
&P^{OFF}_{x,y}(t) = P^{OFF}_{x,y}(t) + \sigma_{1} \cdot P^{OFF}_{x,y}(t-1)
\end{aligned}
\label{DC}
\end{equation}

\subsection{Multi-Layers in ON and OFF Pathways}
In LGMD1 state-of-the-art works, all the visual information are processed in a single pathway, e.g. \cite{LGMD1-Glayer,LGMD1-Yue2009}. The output of photoreceptors forms the input to two separate cell types in the next layers. One type is the excitatory cell, through the excitation is passed directly to the retinotopical counterpart of the following layer. Another type is the one-frame-delayed lateral inhibitory cell, relaying inhibition to its retinotopical counterpart’s neighboring cells in the next layer.

With respect to the mechanism of ON and OFF cells, in the proposed visual neural network, signals are split downstream into two separated pathways, each of which comprises a cascade of layers respectively. First, in ON channels, the signals conveyed by ON cells form the input into two separated flows in next I (inhibition) and E (excitation) Layers as shown in Fig. \ref{model}. ON cells elicit onset response by brightness increments, so that the excitatory flow passes directly to E-Layer and the counterpart's cell in following Summation (S) Layer (Eq. \ref{L1_EI}); meanwhile, it is fed into a first-order low-pass filtering which gives feedback on a time-delayed information. Let $P^{ON}_{x,y}$ be $Y$ and the delayed signal $D^{ON}_{x,y}$ be $X$, we could deduce the following equation:

\begin{figure}[!t]
	\centering
	\includegraphics[width=0.45\textwidth]{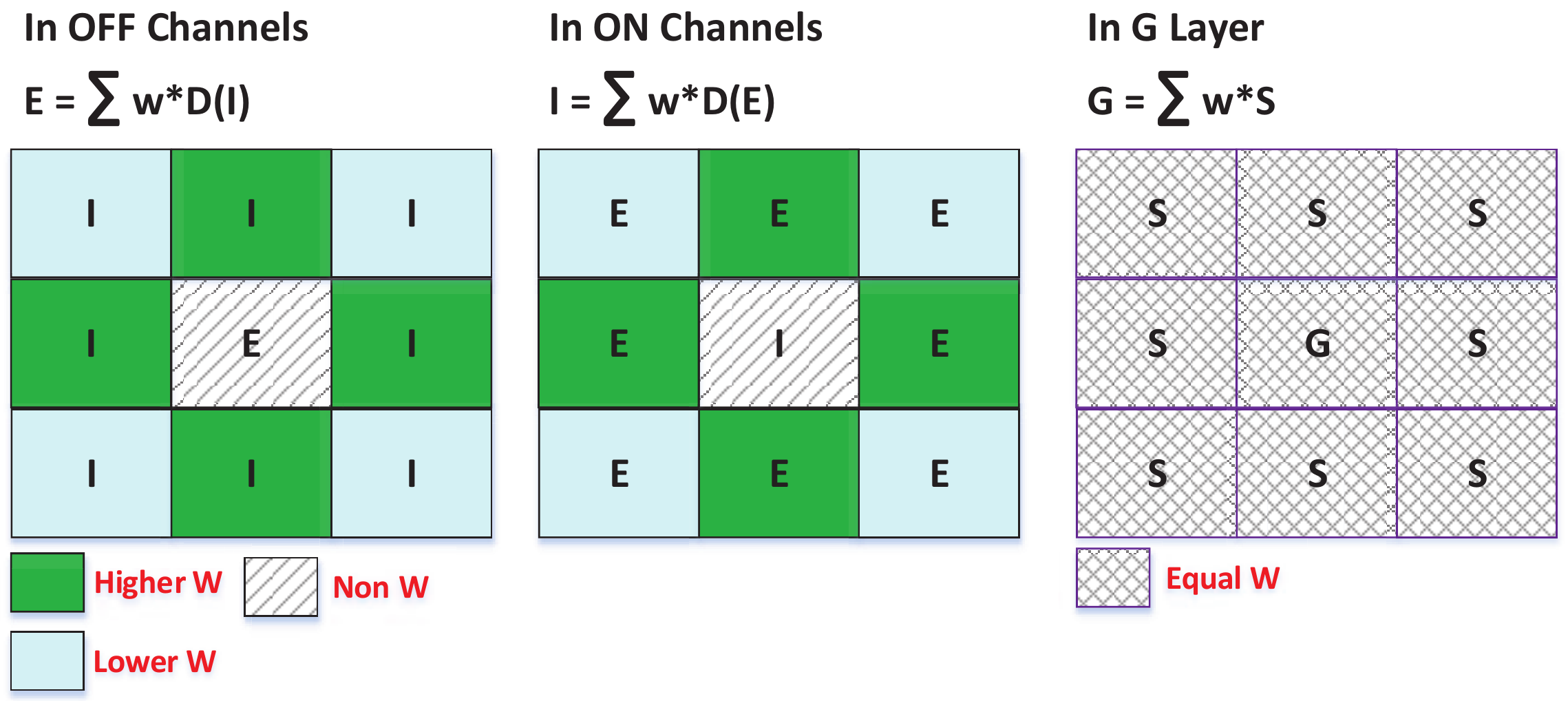}
	\caption{The illustration of spatial-temporal convolution processes: in OFF channel, the excitation ($E$) is convoluted by periphery time-delayed inhibitions ($D(I)$); while in ON channel, the inhibition ($I$) is convoluted by surrounding time-delayed excitations ($D(E)$). The weights in four nearest neighbor cells are higher than those in four diagonal locations, and zero in the center of matrix. The right-side kernel denotes the G cell in Grouping Layer convolving its direct counterpart and surrounding non-delayed Summation cells, each pixel of which shares an equal weight.}
	\label{convolve}
\end{figure}
\begin{equation}
\frac{d X(t)}{d t} = \frac{1}{\tau_1}(Y(t) - X(t))
\label{L12_lowpass}
\end{equation}
where $\tau_1$ is a time constant in milliseconds. After that, the inhibitory flow is convoluted by periphery time-delayed excitations in I-Layer:
\begin{equation}
\begin{aligned}
E^{ON}_{x,y}(t) = &P^{ON}_{x,y}(t),\\
I^{ON}_{x,y}(t) = &\sum_{i=-r}^{r}\sum_{j=-r}^{r}D^{ON}_{x+i,y+j}(t) \cdot W(i,j),(i \ne j, if\ i = 0)
\end{aligned}
\label{L1_EI}
\end{equation}
where $r$ denotes the radius of convolution kernel (size of inhibited area) which is normally set at $1$. One could increase it which nevertheless will require much more computational power as the convolving procedure goes through each local cell within the dual-channel. $W$ indicates the convolution matrix as illustrated in Fig. \ref{convolve}. In addition, it is clear to notice that the delayed information only spread out to their neighboring cells rather than to their direct counterparts, wherein the index $i$ and $j$ are not agreed to be equal at $0$ simultaneously.

Similarity for OFF channels, the signals conveyed by OFF cells form the input downstream to two flows in E and I Layers. While compared to the delayed information in ON channels, the excitatory flow in OFF is time-delayed relative to the inhibitory flow by offset response of brightness decrements. The inhibitions are directly fed into the I-Layer and counterpart's cell in S-Layer, meanwhile undergoing the first-order low-pass filtering. And on the other hand, the excitations are convoluted in E-Layer by corresponded surrounding delayed signals before flowing into S-Layer. Let $P^{OFF}_{x,y}$ be $Y$, the time-delayed information $D^{OFF}_{x,y}$ be $X$, the low-pass filtering procedure pertains to Eq. \ref{L12_lowpass}. In addition, the following Eq. \ref{L2_IE} illustrates the signal processing in E and I Layers of OFF pathway. The other notations conform to those in Eq. \ref{L1_EI}.
\begin{equation}
\begin{aligned}
I^{OFF}_{x,y}(t) = &P^{OFF}_{x,y}(t),\\
E^{OFF}_{x,y}(t) = &\sum_{i=-r}^{r}\sum_{j=-r}^{r}D^{OFF}_{x+i,y+j}(t) \cdot W(i,j),(i \ne j, if\ i = 0)
\end{aligned}
\label{L2_IE}
\end{equation}
Next, there are local summation cells for each single channel. The excitation and inhibition depict a purely linear summation:
\begin{equation}
\begin{aligned}
S^{ON}_{x,y}(t) = &E^{ON}_{x,y}(t) - w_{bias} \cdot I^{ON}_{x,y}(t),\\
S^{OFF}_{x,y}(t) = &E^{OFF}_{x,y}(t) - w_{bias} \cdot I^{OFF}_{x,y}(t)
\end{aligned}
\label{localS}
\end{equation}
where $w_{bias}$ denotes a local bias suppressing inhibitions in either polarity channel.

\subsection{Summation and Grouping Layers}
There are interactions between ON and OFF channels at each local pixel in the S-Layer and G-Layer of a PNN, as shown in Fig. \ref{model}. According to a biological research in insects' vision system \cite{Circuit-correlation2013}, we depict a supralinear computation between polarity channels in S-Layer wherein excitations interact both linearly and multiplicatively:
\begin{equation}
S_{x,y} = \theta_1 \cdot S^{ON}_{x,y} + \theta_2 \cdot S^{OFF}_{x,y} + \theta_3 \cdot S^{ON}_{x,y} \cdot S^{OFF}_{x,y}
\label{supralinearS}
\end{equation}
where $\theta_1$, $\theta_2$ and $\theta_3$ denote the combinations of term coefficients allow us to represent different 'balances' of interactions, which could realize either pure linear computation by setting $\theta_3$ at zero, or nonlinear relationship between polarity channels. Such a supralinear interaction of excitations plays a crucial role of achieving the specific collision selectivity of LGMD2 neuron by allowing the 'bias' in ON channels as depicted in Fig. \ref{model}. Otherwise, balancing $\theta_1$ and $\theta_2$ can realize the general functions of LGMD1 neuron. Furthermore, there is a local threshold gate in S-Layer:
\begin{equation}
S_{x,y}^{'}(t) = \left\{
\begin{aligned}
S_{x,y}(t), \quad &if \quad S_{x,y}(t) \ge T_s \\
0, \quad &else \\
\end{aligned}
\right.
\label{threS}
\end{equation}
where $T_s$ denotes the threshold to allow the excitation of each local pixel to reach the summation cell. In this neural network, the expanded edges represented by clustered excitations are enhanced to extract colliding objects against complex backgrounds through a Grouping (G) Layer following S-Layer (Fig. \ref{model}). It is essentially another convolving course with an equal-weighted kernel $W_g$ as shown in Fig. \ref{convolve}:
\begin{equation}
G_{x,y}(t) = \sum_{i=-r}^{r}\sum_{j=-r}^{r}S_{x+i,y+j}^{'}(t) \cdot W_g(i,j)
\label{Grouping}
\end{equation}

\subsection{LGMD Cell}
After all signals arriving at G-Layer, they are linearly integrated to form the feed forward excitation ($FFE$) which also corresponds to the membrane potential ($MP$):
\begin{equation}
MP(t) = \sum_{x=1}^{row}\sum_{y=1}^{col}G_{x,y}(t)
\label{MP}
\end{equation}
where $row$ and $col$ are the rows and columns of G-Layer. The membrane potential in the LGMD cell is exponentially mapped and regularized via a sigmoid transformation, which mimics the activation function of artificial neurons \cite{LGMD1-Glayer}:
\begin{equation}
U(t) = (1 + e^{-|MP(t)| \cdot (n \cdot k)^{-1}})^{-1}
\label{Sigmoid}
\end{equation}
where $U$ indicates the sigmoid membrane potential (SMP). The coefficient $k$ regularizes the output within $[0.5,1)$.

\subsection{FFI Pathway}
With regard to LGMD1 models, e.g. \cite{LGMD1-Glayer,LGMD1-Yue2006,LGMD1-Yue2009}, and biological research in LGMD2 \cite{LGMD2-1997}, although the ventrally located dendritic trees in LGMD1 (subfields B and C in Fig. \ref{circuitry}) are absent from LGMD2, we apply similar FFI pathway for constructing LGMD2 neuron model to fulfill its characteristics revealed in \cite{LGMD2-1997}. Without such a directly inhibitory mechanism, either LGMD1 or LGMD2 represents high firing rates during rapid luminance changes over large area of receptive field, the situations which are inhibited in both LGMDs of locusts.

In such a separate pathway as shown in Fig. \ref{model}, FFI cell $F_t$ at time step $t$ is taken the average value from absolute luminance changes captured by all photoreceptors (Eq. \ref{ffi}). Then it is fed into a first-order low-pass filter (Eq. \ref{ffi_lowpass}) representing a few milliseconds time delay which is in accordance with the biological research \cite{LGMD1-1996}:
\begin{equation}
F_t = \sum_{x=1}^{row}\sum_{y=1}^{col}|P_{x,y}(t)| \cdot n^{-1}
\label{ffi}
\end{equation}
\begin{equation}
\frac{d \overset{-}{F_t}}{d t} = \frac{1}{\tau_2}(F_t - \overset{-}{F_t})
\label{ffi_lowpass}
\end{equation}
where $\overset{-}{F_t}$ denotes the postponed FFI to be conveyed directly to the LGMD cell as illustrated in Fig. \ref{model}. Once the FFI output exceeds its threshold level $T_{ffi}$, LGMD2 neuron will be directly inhibited.

\subsection{Spike Frequency Adaptation Mechanism}
As proposed in Section \ref{ReSec}, we also apply a SFA mechanism in the spiking initiation zone sieving the sigmoid membrane potential \cite{SFA-2009Role}, which is computationally depicted as a conditional first-order high-pass filter in the following formulation:
\begin{equation}
U^{'}(t) = \left\{
\begin{aligned}
&\sigma_{hp} \cdot (U^{'}(t-1) + U(t) - U(t-1)), \\
&if \ (U(t) - U(t-1)) \leq T_{sfa} \\
&\sigma_{hp} \cdot U(t), \quad else \\
\end{aligned}
\right.
\label{sfa_highpass}
\end{equation}
where $U{'}(t)$, $U{'}(t-1)$ indicates the filtered SMPs at two successive frames. $T_{sfa}$ is a very small positive real number. $\sigma_{hp}$ denotes a coefficient calculated by $\sigma_{hp} = \tau_{3}/(\tau_3+\tau_i)$, where $\tau_3$ indicates a time constant in the high-pass filter and $\tau_i$ is the time interval between successive frames, both of which are also in milliseconds.

\subsection{Spiking Mechanism}
Compared to the state-of-the-art LGMD1 models, e.g. \cite{LGMD1-Yue2006,LGMD1-robot2005,LGMD1-DSN-competing,LGMD1-Yue2009}, the LGMD2 neural network could produce more than one spikes at each time step:
\begin{equation}
S^{spike}_t = \left\{
\begin{aligned}
0, \quad &if \quad U^{'}(t) < T_{sp}\\
1, \quad &if \quad T_{sp} \le U^{'}(t) \le T_{sp}+\sigma_{sp}\\
2, \quad &else
\end{aligned}
\right.
\label{spiking mechanism}
\end{equation}
where $T_{sp}$ indicates the spiking threshold, $\sigma_{sp}$ implies steps which partition SMP over the threshold into sections. Finally, a potential collision detection is given by:
\begin{equation}
Collision = \left\{
\begin{aligned}
True, \quad &if \quad \sum_{i=t-N_{ts}}^{t}S^{spike}_i \ge N_{sp}\\
False, \quad &otherwise
\end{aligned}
\right.
\label{collision}
\end{equation}
where $N_{sp}$ denotes the number of continuous elicited spikes and $N_{ts}$ indicates the number of successive frames.

\subsection{Network Parameters Setting}
All the parameters of the proposed visual neural network are decided empirically with consideration of functionalities. Table \ref{lgmdsParams} illustrates the parameters setting of LGMD2 case. No network training and learning methods are currently involved in the framework. More concretely, the adaptable parameters including the $col$, $row$ and $\tau_i$ are decided by the physical properties (the resolution and frames per second) of input visual streams. The weights of local convolution matrix $W$ in dual-channels are set at $0$ for the center cell, $0.25$ for the four nearest neighbors and $0.125$ for the four diagonal units, pertaining to kernel radius $r$ is $1$. The weights in $W_g$ of G-Layer are set equally at $1/9$. It is also necessary to note that the time constant $\tau_3$ in SFA high-pass filter could vary within a wide range of milliseconds so as to adjust the adaptation rate. For constructing LGMD2 which derive from the general LGMDs neural circuit in Fig. \ref{model}, we put forward the bias in all ON channels by defining very small real numbers for both $\theta_1$ and $\theta_3$ which exactly blocked the ON pathway. Intriguingly, such a supralinear computation between signals in different channels also allows us to build a neuron which has a totally opposite collision selectivity compared to LGMD2, i.e, it only reacts to the brighter objects embedded in dark background moving in depth.

\begin{table}[!t]
	\caption{The Parameters of LGMD2 Visual Neural Network}
	\label{lgmdsParams}
	\centering
	\begin{tabular}{llllll}
		\toprule
		\multicolumn{6}{c}{Parameter: Name, Value}                                                  \\
		\cmidrule{1-6}
		Name          & Value     	     & Name       & Value        & Name       & Value           \\
		\midrule
		$col$	      &adaptable         &$T_{sfa}$   &$0.001$       &$\tau_1$    &$5 \sim 50$ms    \\
		$row$         &adaptable         &$W$         &$0 \sim 0.25$ &$\tau_2$    &$5 \sim 100$ms   \\
		$n$           &$col \cdot row$   &$W_g$       &$1/9$         &$\tau_3$    &$400 \sim 1000$ms\\
		$r$           &$1$               &$k$         &$1$           &$\tau_i$    &adaptable        \\
		$w_{bias}$    &$0.3$             &$N_{ts}$    &$4$           &$\theta_1$  &$0 \sim 0.1$     \\
		$\sigma_{1}$  &$0.1$             &$N_{sp}$    &$4 \sim 8$    &$\theta_2$  &$1 \sim 6$       \\
		$\sigma_{sp}$ &$0.1$             &$T_s$       &$10$          &$\theta_3$  &$0 \sim 0.1$     \\
		$T_{sp}$	  &$0.65 \sim 0.78$  &$T_{ffi}$   &$10$          &            &                 \\
		\bottomrule
	\end{tabular}
\end{table}

\section{Experiments and Analysis}
\label{EASec}
In this section, we carry out systematic experiments to demonstrate the unique characteristic of the LGMD2 neural network. All the experiments can be  categorized into two types of tests: the off-line tests and the on-line tests. In the off-line tests, the input stimuli consist of synthetic and recorded video streams. For comparison, both the performance of LGMD1 \cite{LGMD1-Glayer} and LGMD2 neuron models are presented against the synthetic stimuli. In the on-line tests, the LGMD2 neural network was implemented in a miniature robot for real time experiments.

\subsection{Software and Hardware Setting}
The proposed framework was set up in Visual Studio $2013$ (Microsoft Corporation) and Keil (uVision$4$) for handling off-line and robotic experiments respectively. Data analysis were realized in Matlab $2015b$ (The MathWorks, Inc. Natick, USA). The computer used was a laptop (DELL INSPIRON) with two $2.30$ GHz CPUs and Windows $7$ operating system. The parameters of LGMD2 were adopted from Table \ref{lgmdsParams}; the comparative LGMD1 model in off-line tests with parameters setting-up was suggested in \cite{LGMD1-Glayer}. The input image frames were all converted to the gray-scale with intensity valued within $[0,255]$. The resolutions of simulated and real physical recording stimuli were respectively $320 \cdot 240$ and $432 \cdot 240$. The spiking-threshold $T_{sp}$ was set at $0.78$ for off-line and $0.65$ for robotic tests respectively.

\begin{figure}[!t]
	\centering
	\includegraphics[width=0.3\textwidth]{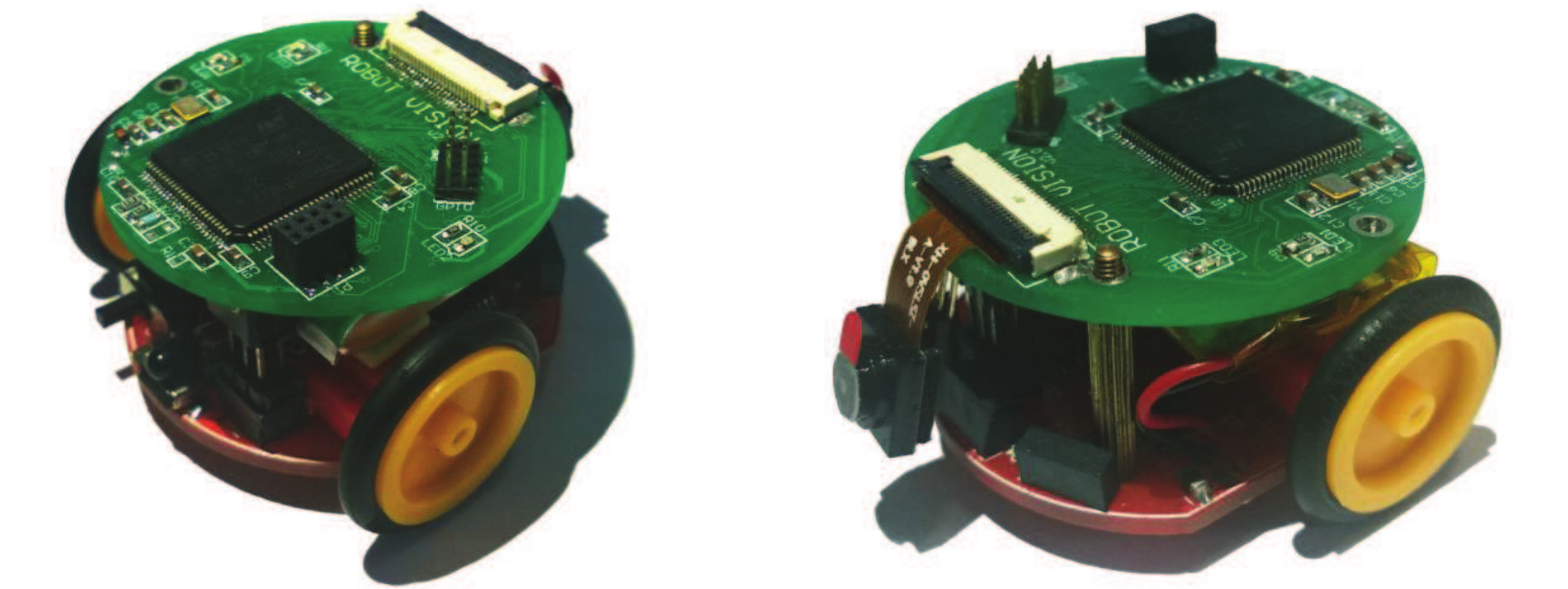}
	\caption{The micro-robot prototype: Upper board (green) executes LGMDs-based vision control. Bottom board (red) is the motion actuator. A mini camera module as the 'eye' of Colias is assembled to the upper board. Two wheels and the battery are assembled to the bottom board.}
	\label{colias_proto}
\end{figure}

As depicted in Fig. \ref{colias_proto}, the mobile robot platform used in real time experiments is named 'Colias'. It is an open-hardware modular micro-robot which is developed to be used in swarm robotic applications \cite{Colias-introduction,Colias-investigation}. On the whole, Colias consists of two main components. One is the motion actuator with diameter of $4$cm, which is deployed on the bottom to provide power and motion controls (the red board in Fig. \ref{colias_proto}). It applies an AVR micro-controller with $8$ MHz clock source. Two micro DC motors and two diameter $2.2$cm wheels are employed to actuate Colias \cite{Colias-motion}. Another one is the extension vision module which is placed on the top of Colias (the green board in Fig. \ref{colias_proto}). Three LEDs are embedded in this module to be the indicators of different real-time status. With the help of a full-duplex serial port used as the debugging interface, Colias can send image samples and model data to the host in real-time, when the debugging mode is allowed. Meanwhile, it can receive varied configuration commands from the host.

To be more specific, a miniature camera is assembled to the upper board representing as an ‘eye’ of the robot, which is essential in the vision-based control of robotic applications. A low voltage CMOS image sensor OV7670 module is utilized. The low-cost camera is capable of operating up to $30$ frames per second (fps) in VGA with output support for RGB422, RGB565, and YUV422. The angle of view could reach approximately $70$ degrees. All these features make the camera suitable for using in such micro robots \cite{Colias-Hu,Colias-Cheng}. We chose a resolution of $72 \cdot 99$ pixels at $30$ fps with the output format of $8$-bit YUV422. Second, the micro-controller is an ARM® Cortex™-M4F core, which is deployed as the main processor for monitoring all the modules and serving the image processing task. The $32$ bit MCU STM32F407 clocked at $168$ MHz provides the necessary computational power to have a real-time image stream processing. Its $192$ Kbyte internal SRAM supports the image buffering and computing. In both LGMD1 \cite{Colias-Hu} and the proposed LGMD2 cases, though very limited SRAM, the time cost for vision model implementation plus motion decision is always less than $30$ milliseconds. There is a digital camera interface (DCMI) which is an embedded one for transmitting of the captured images. DCMI can sustain a data transfer rate up to $54$ MHz. In our case, through such an interface, we can collect the images within different neural layers (e.g. Fig. \ref{robot-appr-set}, \ref{colias-views-dark-appr}, \ref{colias-views-light-appr} and \ref{robot-trans-set}). We can also retrieve varied types of real-time network outputs (e.g. membrane potential, spikes) from the micro-robot.

\subsection{Challenged against Synthetic Stimuli}
In this subsection, we started the experiments from testing the proposed LGMD2 neuron model using computer-simulated stimuli and comparing it with a LGMD1 computational model \cite{LGMD1-Glayer}. All the simulated visual stimuli used in the experiments could be categorized to the following visual motion types: approach-recession, translation, elongation-shortening, whole-field luminance change, and the sinusoidal grating movements. Each stimulus included a dark/light object moving against bright/dark backgrounds. There was no background noise in those synthetic simulated scenarios. In the grating tests, we examined the output of LGMD2 neuron model challenged by grating movements with a wide range of spatial and temporal frequencies.

\begin{figure}[!t]
	\begin{minipage}[t]{0.24\textwidth}
		\centering
		\centerline{\includegraphics[width=1in]{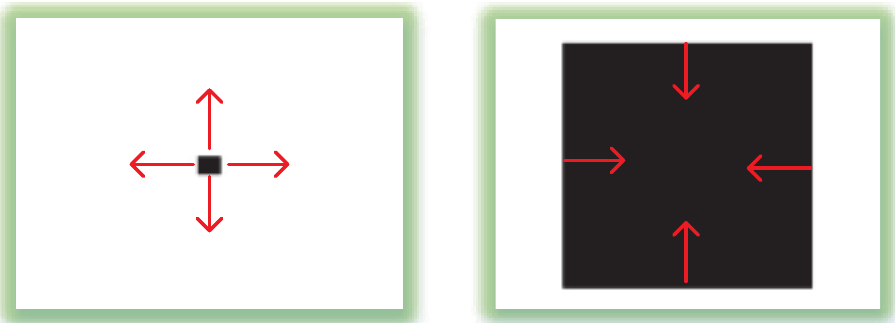}}
	\end{minipage}
	\hfill
	\begin{minipage}[t]{0.24\textwidth}
		\centering
		\centerline{\includegraphics[width=1in]{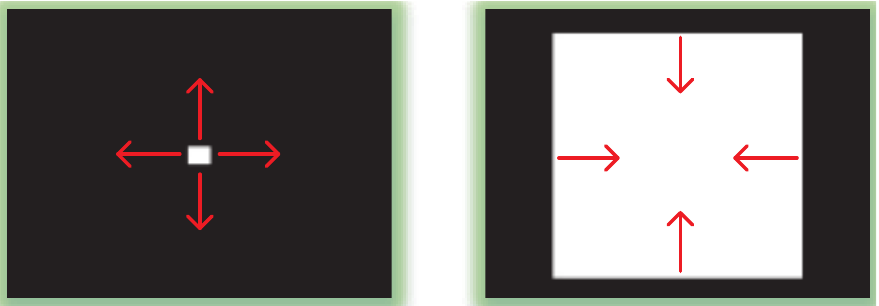}}
	\end{minipage}
	\vfill
	\centering
	\subfloat[]{\includegraphics[width=0.23\textwidth,height=0.11\textheight]{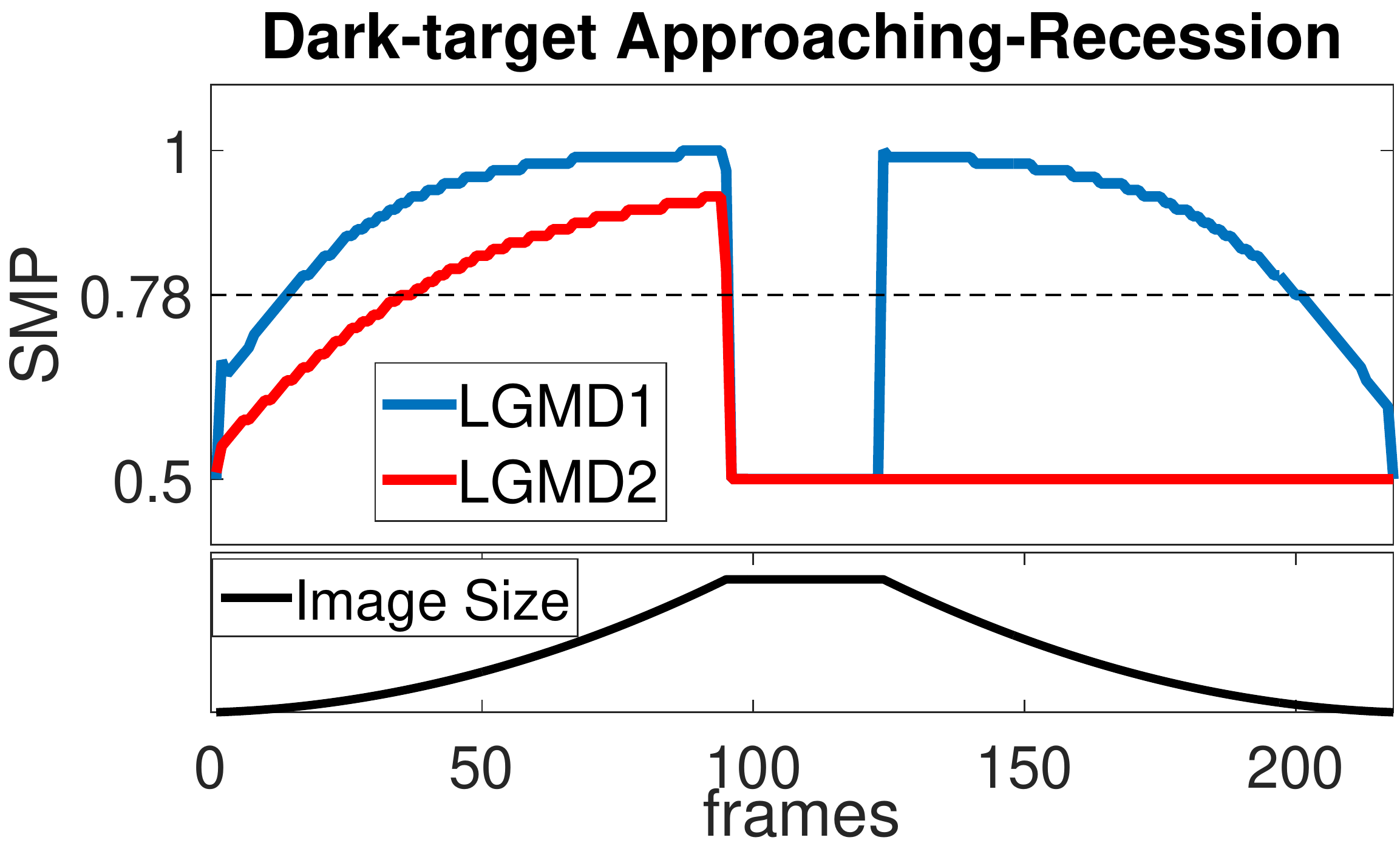}
		\label{das}}
	\hfill
	\subfloat[]{\includegraphics[width=0.23\textwidth,height=0.11\textheight]{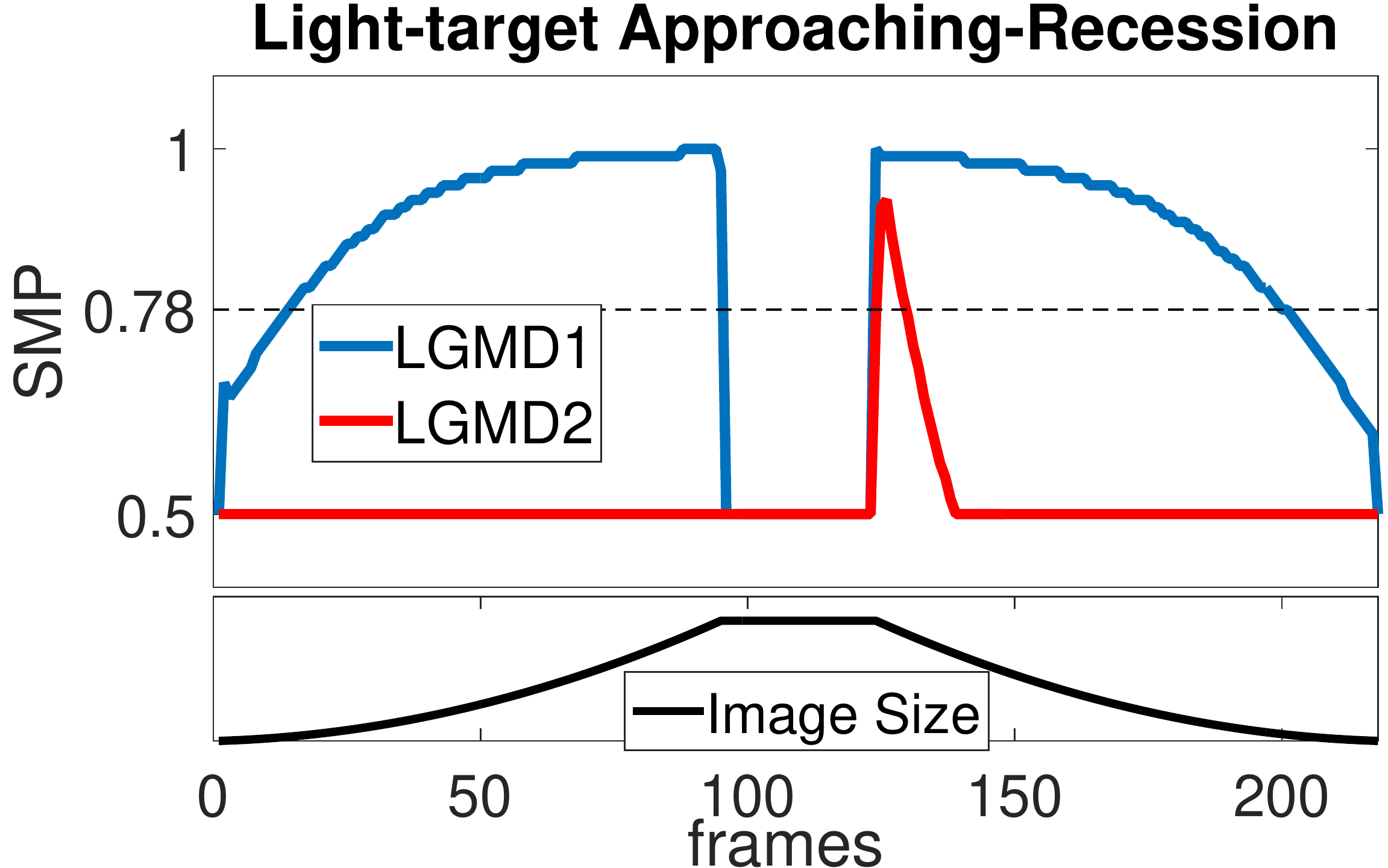}
		\label{las}}
	\caption{The outputs of LGMD2 neural networks challenged by approaching and receding stimuli in comparison with LGMD1: (a) The SMPs of LGMD2 and LGMD1 in responding to dark object approaching and receding against bright background, along with the change of image-size depicted at bottom. The horizontal dashed line indicates the spiking-threshold level. X-axis indicates the time window in frames. (b) Light object approaching and receding against dark background, with other notations the same as (a).}
	\label{simu-looming}
\end{figure}

\begin{figure}[!t]
	\centering
	\includegraphics[width=0.3\textwidth]{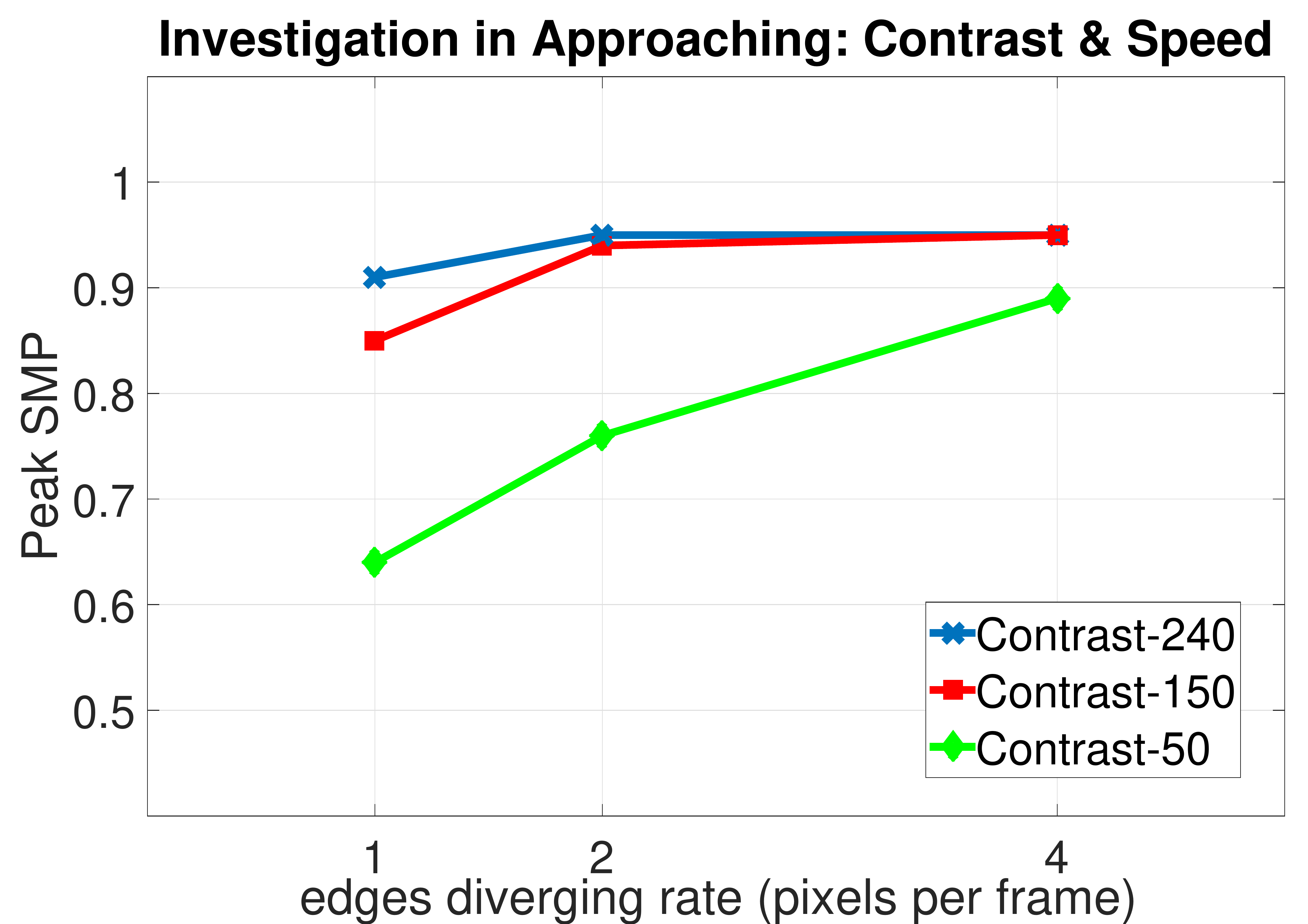}
	\caption{The output of LGMD2 (peak-SMPs) affected by contrast and approaching speed (represented by edge expanding rate).}
	\label{bar-c-s}
\end{figure}

Firstly, we want to check if this LGMD2 model possesses similar unique selectivity as a LGMD2 neuron in locusts does. As shown in Fig. \ref{das}, when challenged against a dark approaching object, both LGMDs elicit rapid increased potential as size of the projected object in the retina grows. However for a dark receding object, LGMD2 is completely inhibited (Fig. \ref{das}), while LGMD1 elicits high-level potential for all depth movements. For a light (or white) approaching-receding object (Fig. \ref{las}), LGMD1 responses to it in a way similar to the dark object moving in depth, which matches the biological research results that LGMD1 neuron is sensitive to all movements in depth regardless background \cite{LGMD1-1996,LGMD2-1997}. As expected, the LGMD2 model demonstrated totally reverse selectivity to light receding objects versus approaching ones revealed its unique preference to the light-to-dark luminance change. This matches the characteristic of a real LGMD2 neuron in locusts perfectly. Moreover, it is also clear that, compare to the LGMD1's response to a receding object, there is a fast adaptation for LGMD2 - the SMP falls down sharply after being briefly excited (Fig. \ref{las}). The two bio-plausible structures in the proposed LGMD2 model - the separated ON/OFF pathways and the SFA mechanism contribute heavily to LGMD2's unique selectivity to dark looming objects.

\begin{figure}[!t]
	\begin{minipage}[t]{0.24\textwidth}
		\centering
		\centerline{\includegraphics[width=1in]{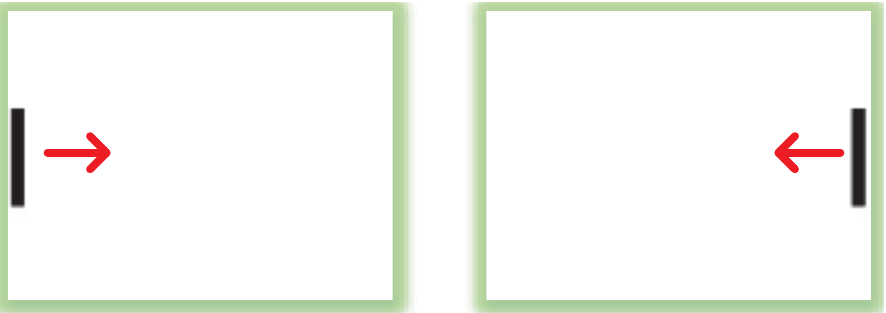}}
	\end{minipage}
	\hfill
	\begin{minipage}[t]{0.24\textwidth}
		\centering
		\centerline{\includegraphics[width=1in]{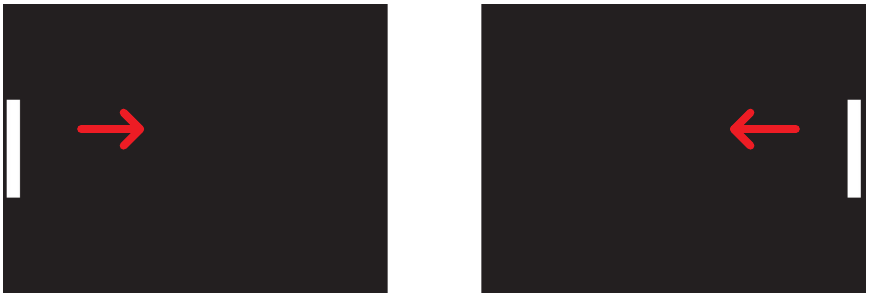}}
	\end{minipage}
	\vfill
	\centering
	\subfloat[]{\includegraphics[width=0.23\textwidth,height=0.11\textheight]{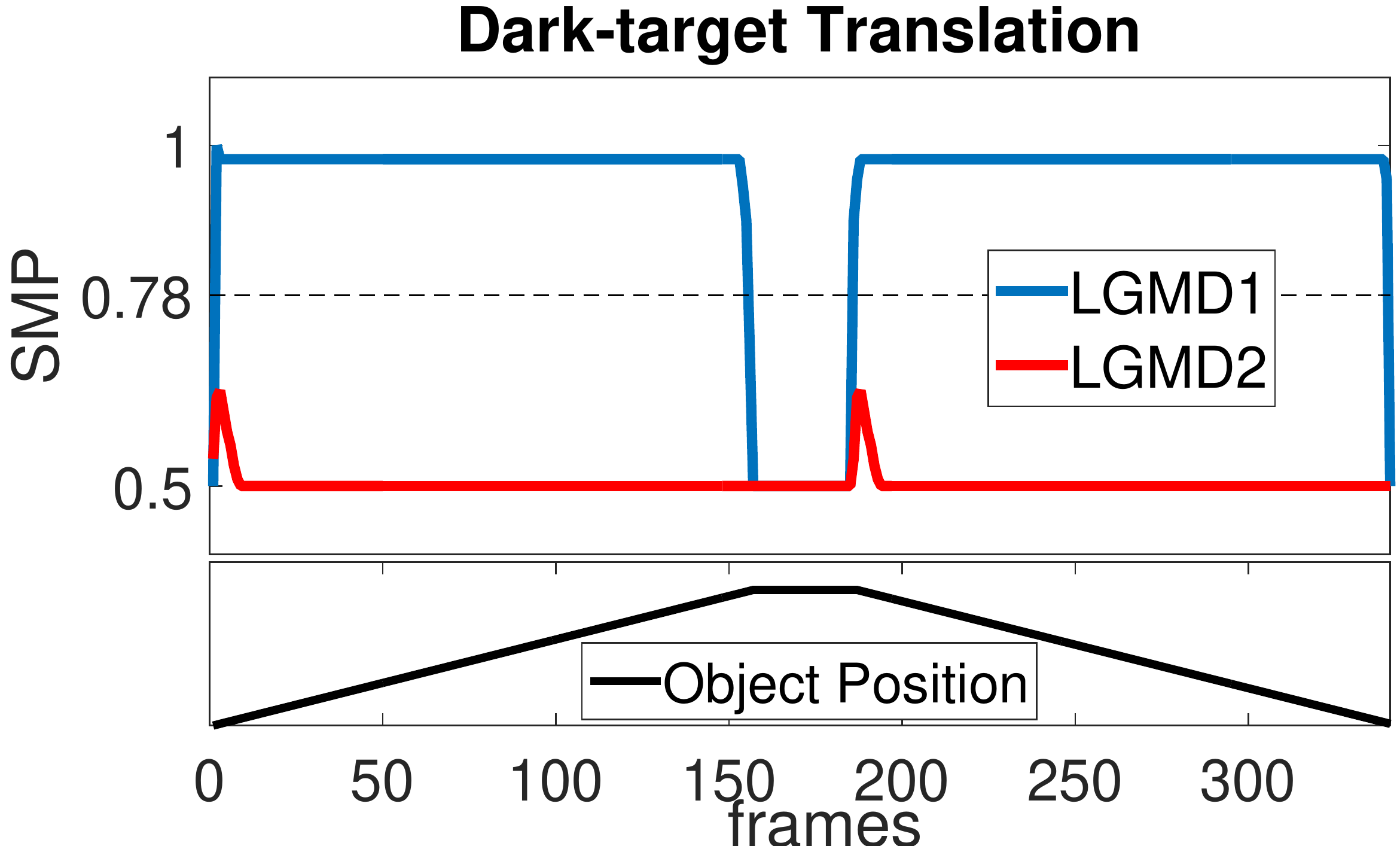}
		\label{dts}}
	\hfill
	\subfloat[]{\includegraphics[width=0.23\textwidth,height=0.11\textheight]{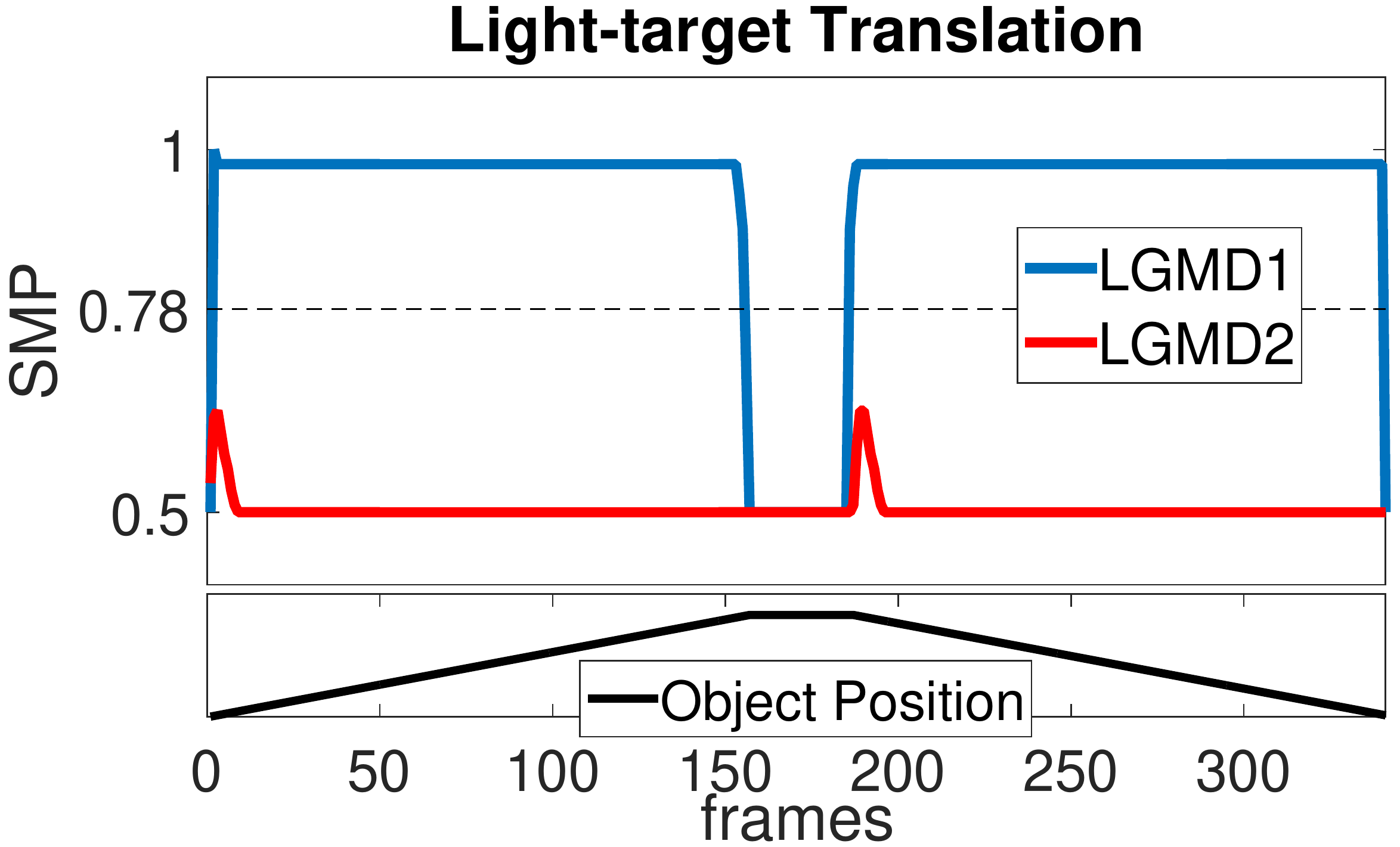}
		\label{lts}}
	\vfill
	\vspace{0.1in}
	\begin{minipage}[t]{0.24\textwidth}
		\centering
		\centerline{\includegraphics[width=1in]{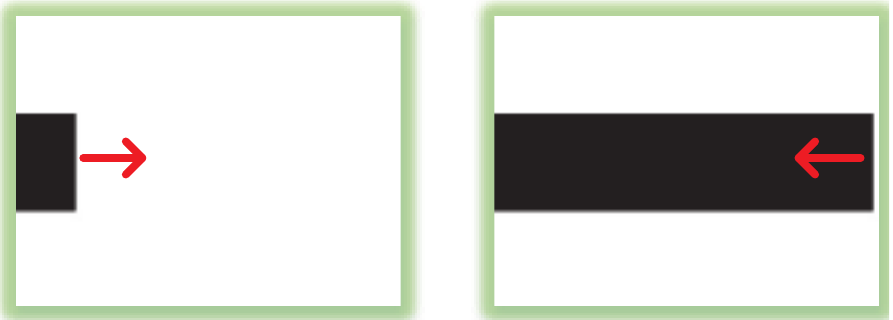}}
	\end{minipage}
	\hfill
	\begin{minipage}[t]{0.24\textwidth}
		\centering
		\centerline{\includegraphics[width=1in]{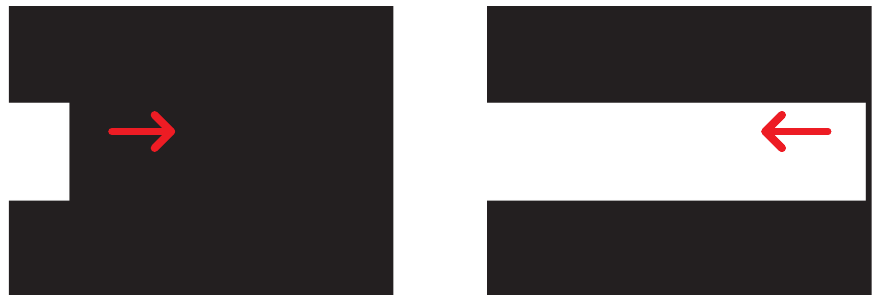}}
	\end{minipage}
	\vfill
	\centering
	\subfloat[]{\includegraphics[width=0.24\textwidth,height=0.11\textheight]{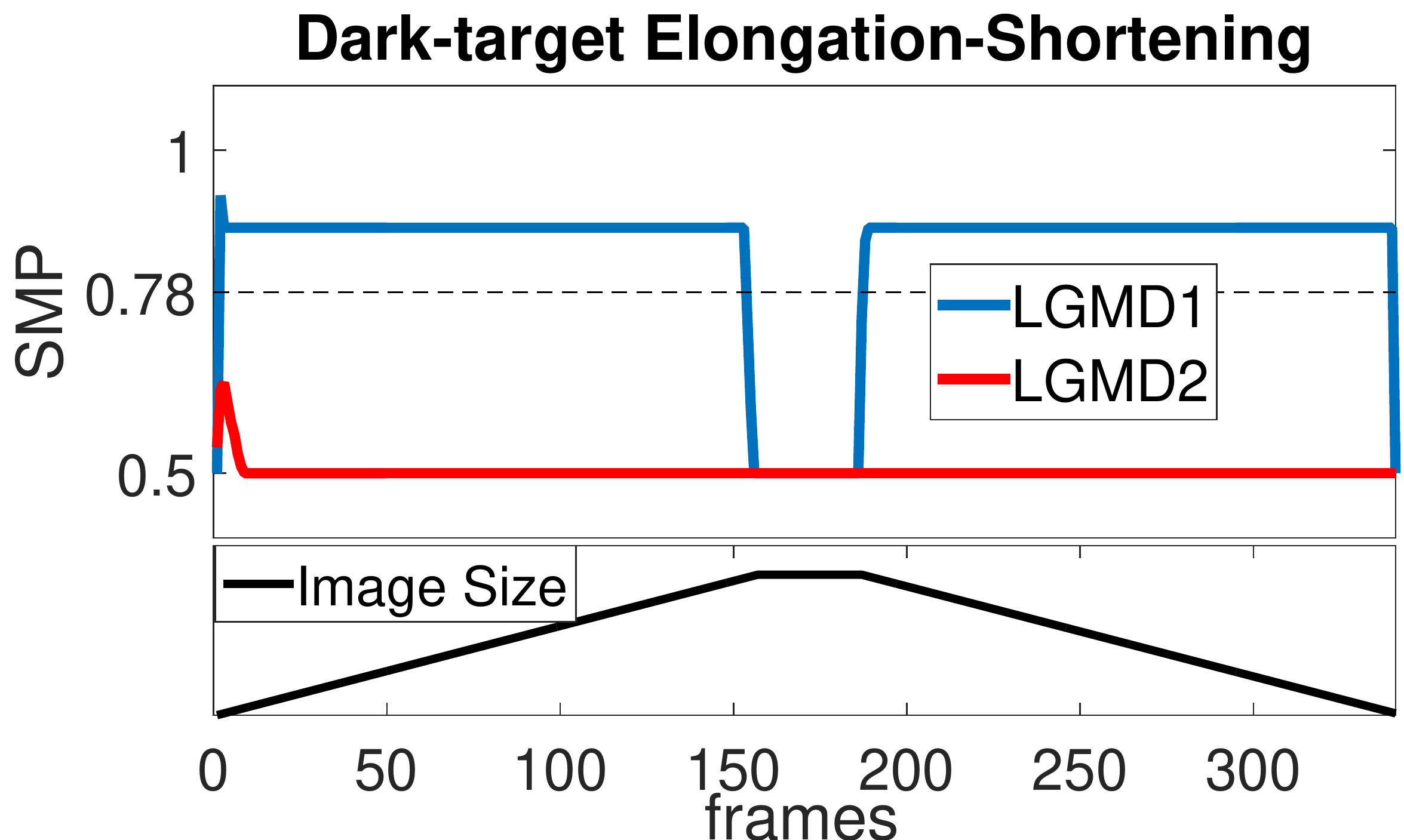}
		\label{des}}
	\hfill
	\subfloat[]{\includegraphics[width=0.23\textwidth,height=0.11\textheight]{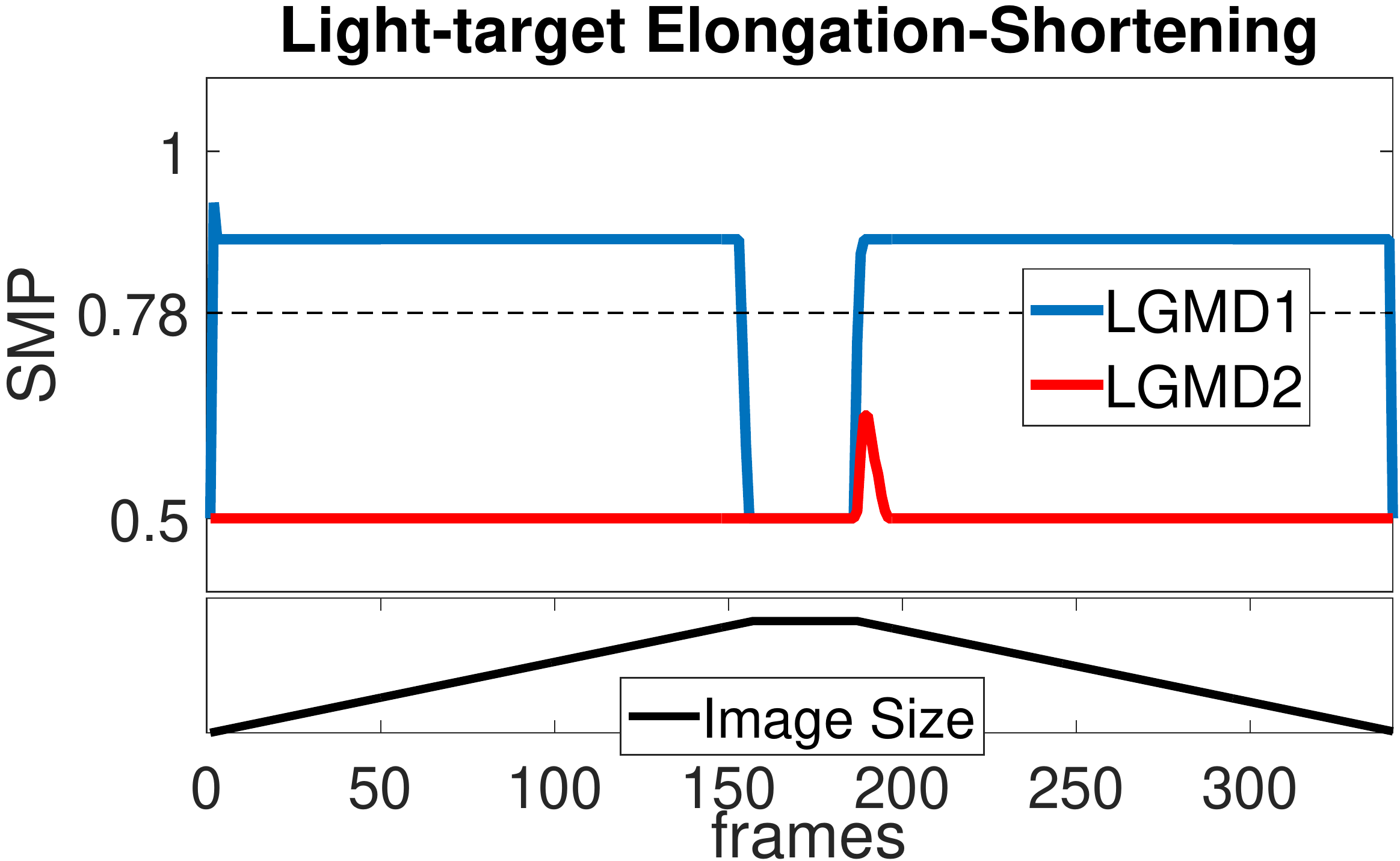}
		\label{les}}
	\caption{The outputs of LGMDs neural networks challenged by movements on X-Y planes at constant speed: (a) The LGMDs SMPs under dark object translation against bright background, along with position indicator depicted at bottom, whereby the rightward movement corresponds the increment of position. (b) Light object translation against dark background. (c) Dark elongating and shortening against bright background, with image-size depicted at bottom. (d) Light elongating and shortening against dark background.}
	\label{simu-trans-elong}
\end{figure}

\begin{figure}[!t]
	\begin{minipage}[t]{0.24\textwidth}
		\centering
		\centerline{\includegraphics[width=1.2in]{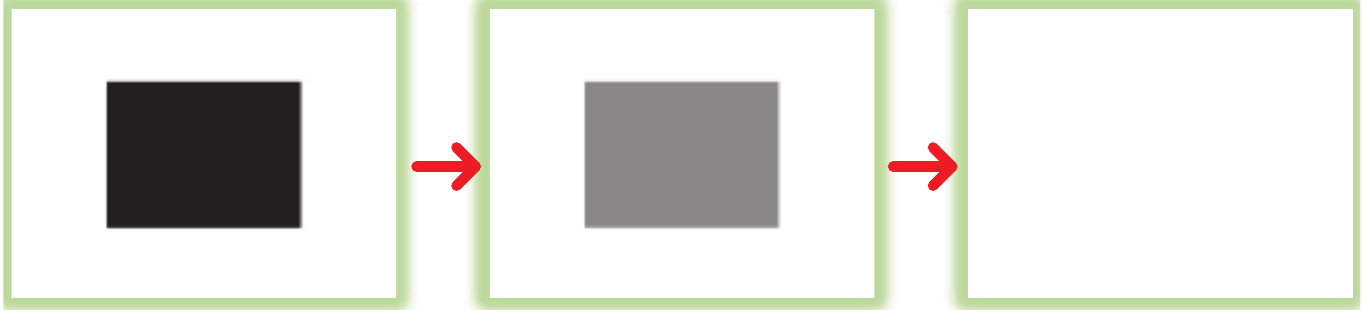}}
	\end{minipage}
	\hfill
	\begin{minipage}[t]{0.24\textwidth}
		\centering
		\centerline{\includegraphics[width=1.2in]{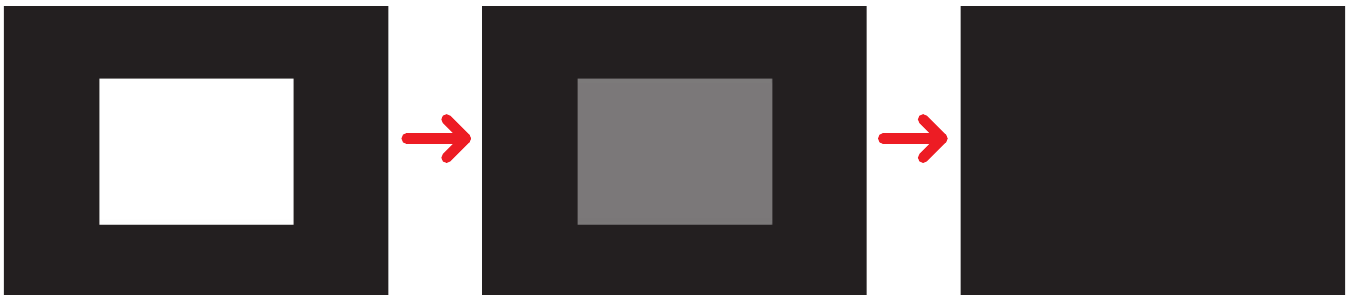}}
	\end{minipage}
	\vfill
	\centering
	\subfloat[]{\includegraphics[width=0.24\textwidth,height=0.11\textheight]{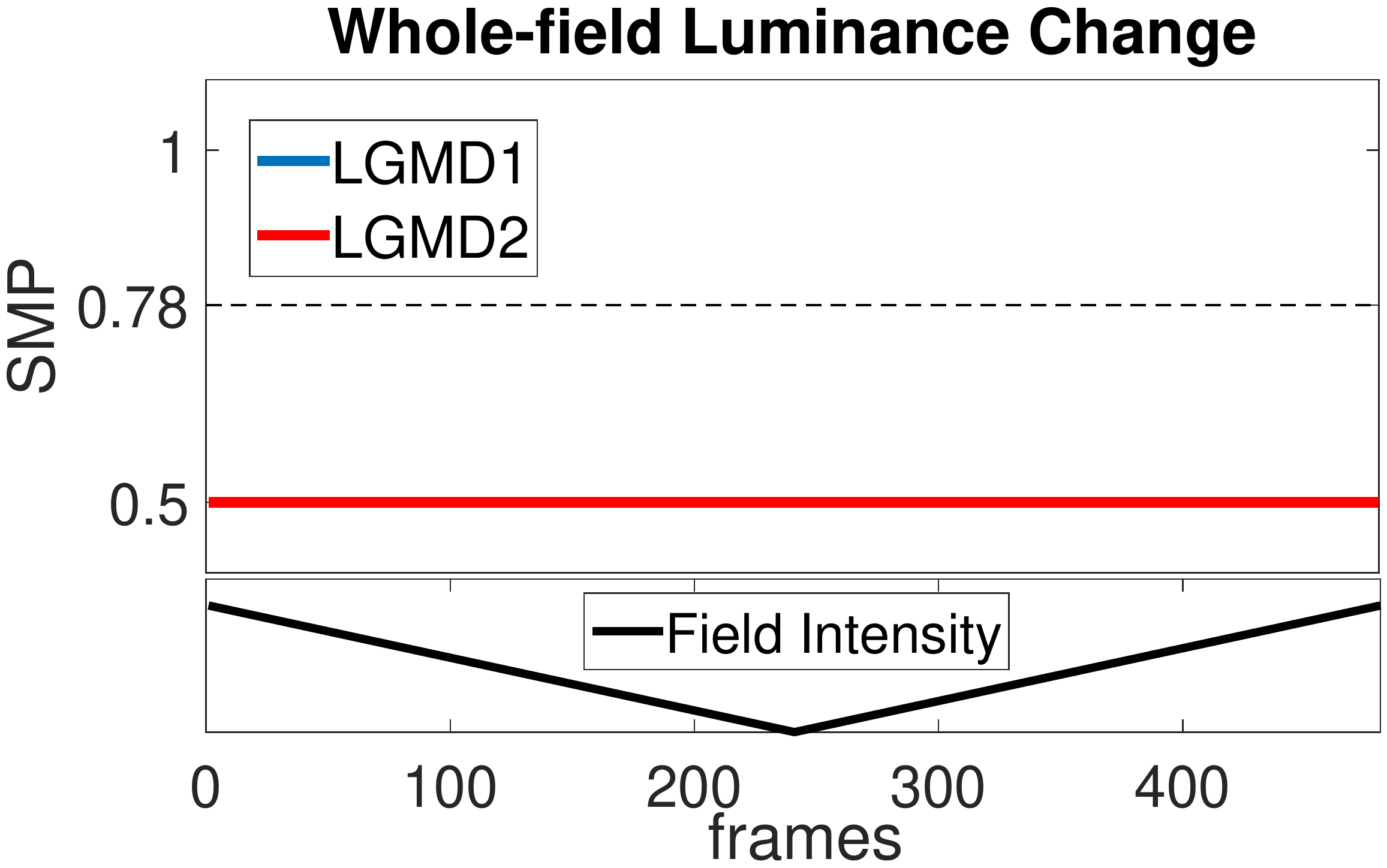}
		\label{wf1}}
	\hfill
	\subfloat[]{\includegraphics[width=0.23\textwidth,height=0.11\textheight]{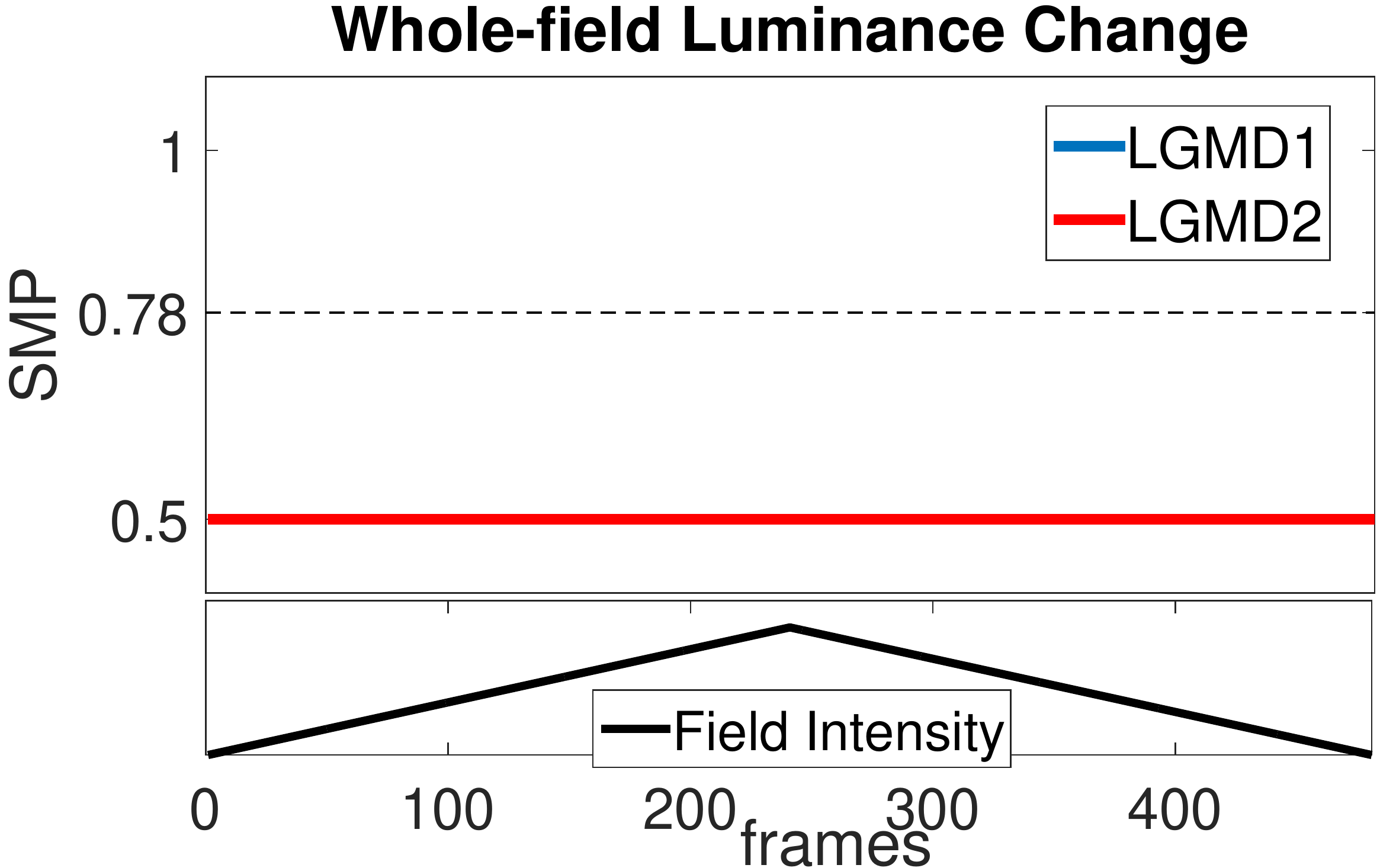}
		\label{wf2}}
	\caption{The results under whole-field luminance change within a fixed area in the receptive field: (a) The LGMDs SMPs during illumination-darkening against bright background, along with field-intensity change depicted at bottom. (b) The LGMDs SMPs during darkening-illumination embedded in dark background.}
	\label{simu-wf}
\end{figure}

\begin{figure}[!t]
	\centering
	\subfloat[]{\includegraphics[width=1.5in]{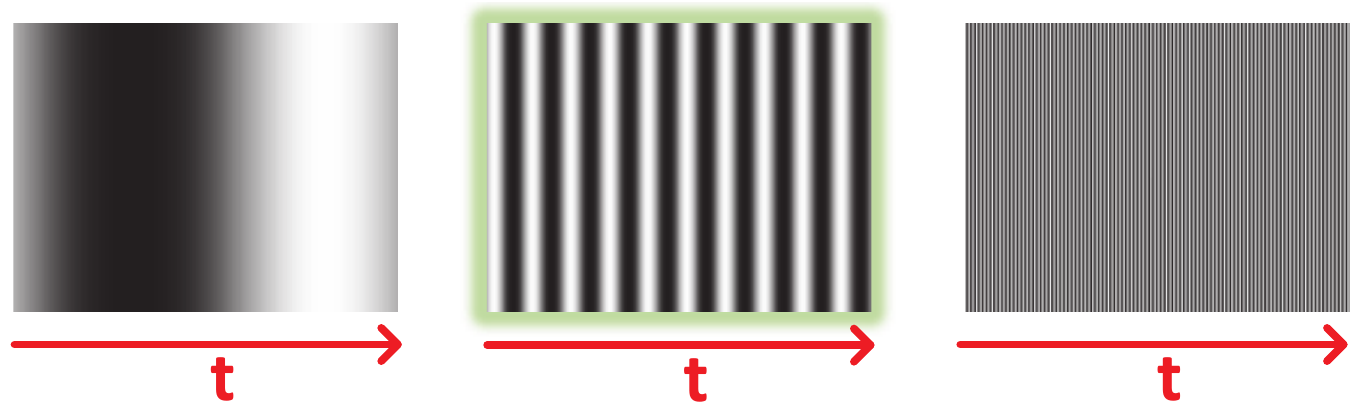}
		\label{simu-grating-stimuli}}
	\vfill
	\vspace{-0.1in}
	\subfloat[]{\includegraphics[width=0.48\textwidth]{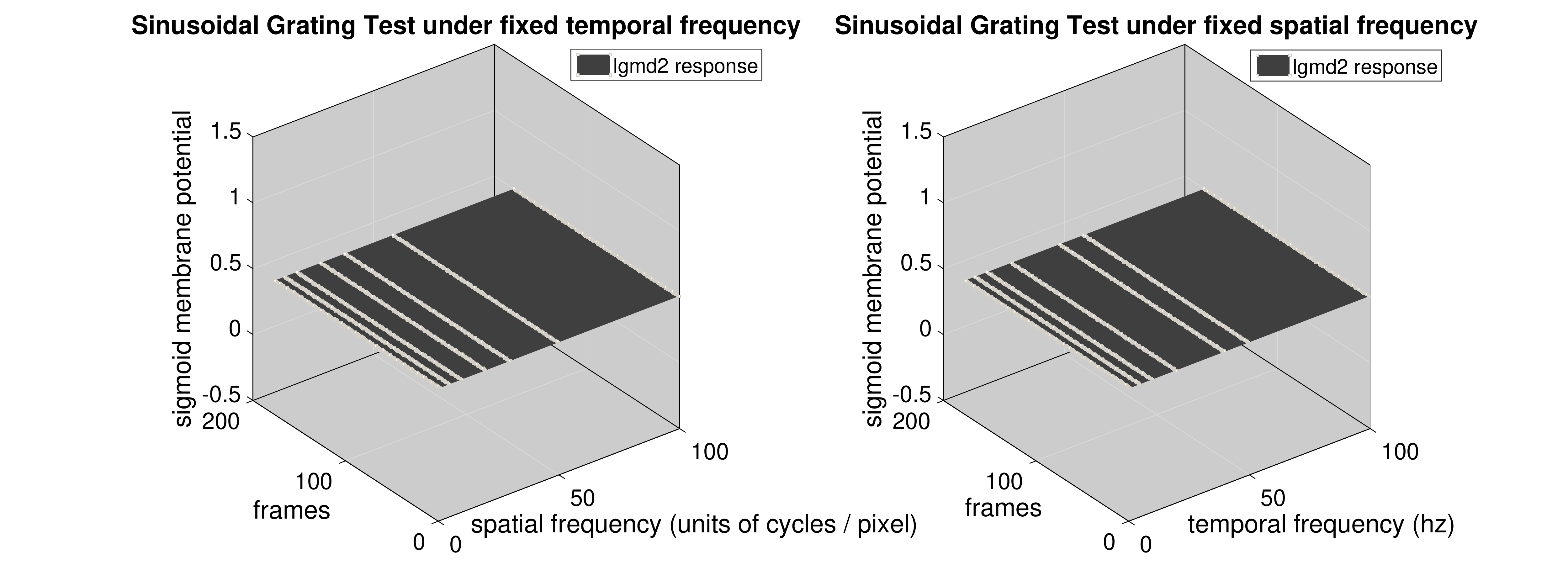}
		\label{lgmd2-gratings}}
	\caption{The outputs of LGMD2 neural network challenged by two sets of sinusoidal gratings, each of which has a fixed time duration of 6-seconds: (a) Three example grating movements along 't' (the fixed time), vary in spatial and temporal frequencies. (b) LGMD2 responses: the left is under fixed temporal-frequency in 20 Hz with varied spatial-frequencies in 1, 5, 10, 20, 30, 50 and 100 units of cycles per pixel respectively; the right is under fixed spatial-frequency in 10 units with varied temporal-frequencies in 1, 5, 10, 20, 40, 50 and 100 Hz respectively. X-axis and Y-axis denote the frames and frequencies. Z-axis indicates SMP levels. The white-lines on 3D-surface imply the specific challenged frequencies.}
	\label{simu-gratings}
\end{figure}

We then look deeper into LGMD2's peak response to different object approaching speeds and different contrasts between the moving object and background. Fig. \ref{bar-c-s} illustrates the outputs of LGMD2 model influenced by contrast and approaching speed. As shown in the figure, contrast can affect peak response, especially when approaching speed is low. It is interesting that the influence of contrast shrink as approaching speed increases. This suggests the LGMD2 neuron may be more effective at a critical moment when predatory becomes very close to the animal.

For the X-Y planes stimuli (i.e. translation movements), as can be seen in Fig. \ref{simu-trans-elong} with either dark or light translation on two directions at constant speed, the LGMD2 neuron model only shows weak and brief response at the beginning of each movement (Fig. \ref{dts}, \ref{lts}), which well conforms to the biological research \cite{LGMD2-1997}. Its responses are rigorously sieved by the SFA mechanism when constant number of photoreceptors are activated by translation movement. Compared to LGMD2, the LGMD1 neuron model exhibits much higher-level potential sustaining to the end of each movement. In both experiments, no distinctive directional motion cues have been extracted by any one of the LGMDs - indicates both the LGMD2 and LGMD1 neuron models are collision selective and not sensitive to translation movements.

The elongation and shortening stimuli represent a situation that an object moving across field of view very close to the retina as shown in Fig. \ref{des} and \ref{les}. Unlike normal translation, the single moving edge elicits the light-to-dark luminance change during dark-elongation and light-shortening; otherwise it gives rise to the dark-to-light luminance change. The LGMD2 model only responds briefly to dark-elongating and light-shortening - conforms to its unique selectivity to the light-to-dark luminance change only (Fig. \ref{des} and \ref{les}), whereas LGMD1 reacts to all situations.

With the similar stimuli in the biological research \cite{LGMD2-1997}, we also simulate the whole subfield luminance change embedded in light/dark background. As illustrated in Fig.\ref{simu-wf}, both LGMDs models are rigorously inhibited when illumination becomes brighter or darker, which appropriately reconcile with the results in biological research \cite{LGMD2-1997}. Similarly, for the systematic grating tests, Fig. \ref{simu-gratings} shows the proposed LGMD2 neuron model remains quiet against grating movements with a broad range of spatial/temporal frequencies. The results demonstrate robustness and potential of LGMD2 neuron model against visual clutters in real world which is critical important for a practical collision detecting system.

\subsection{Challenged against Real Physical Stimuli}
In this subsection, we design experiments to test LGMD2 neuron model further against real physical visual stimuli. We recorded each movement as off-line data for the experiments. Compare to the synthetic scenarios, there are background noise in real physical world recording such as the light flash and shadows, etc. In addition, unlike the simulated movements, the object's moving speeds could not be controlled to, or maintained at, a constant level. Therefore, the visual challenges to present to the proposed collision-detecting system are 'real'.

\begin{figure}[!t]
	\begin{minipage}[t]{0.5\textwidth}
		\centering
		\centerline{\includegraphics[width=0.5\textwidth]{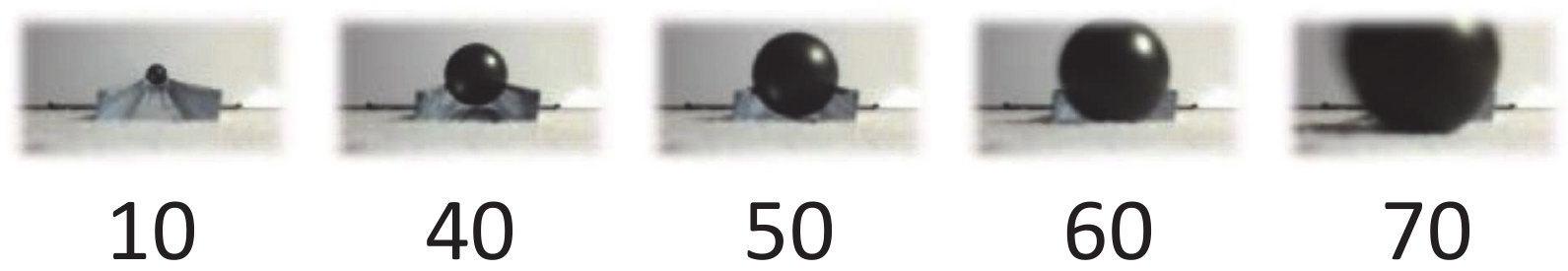}}
	\end{minipage}
	\vfill
	\vspace{-0.1in}
	\centering
	\subfloat[]{\includegraphics[width=0.36\textwidth]{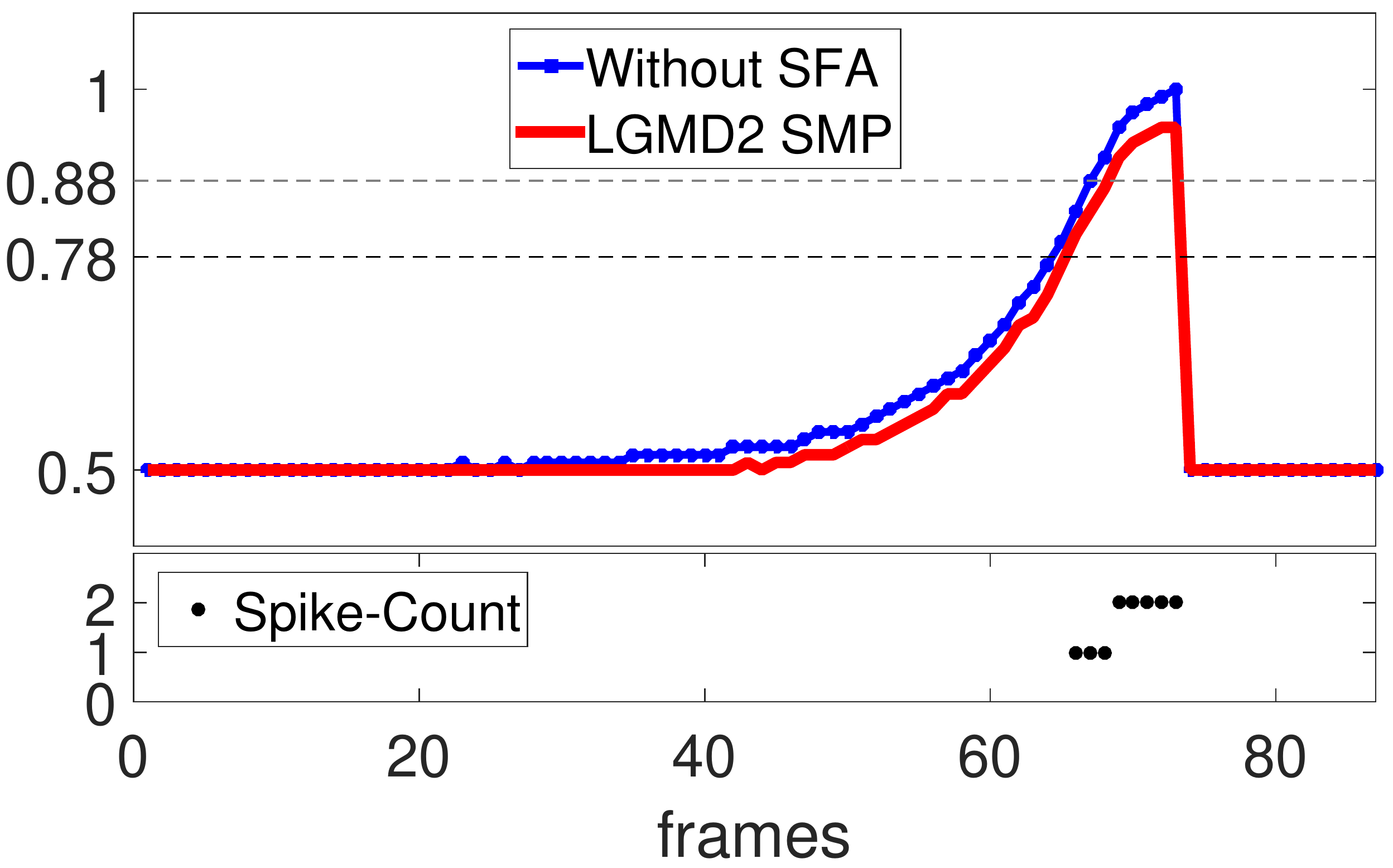}
		\label{real-appr}}
	\vfill
	\vspace{0.1in}
	\begin{minipage}[t]{0.5\textwidth}
		\centering
		\centerline{\includegraphics[width=0.5\textwidth]{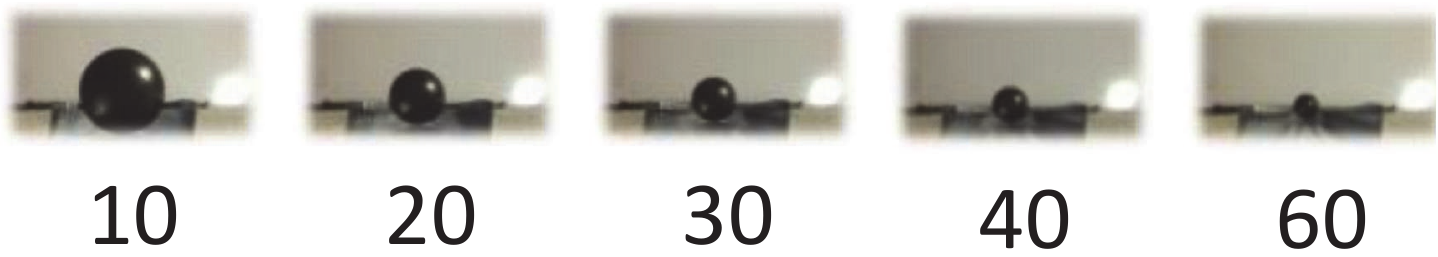}}
	\end{minipage}
	\vfill
	\vspace{-0.1in}
	\centering
	\subfloat[]{\includegraphics[width=0.36\textwidth]{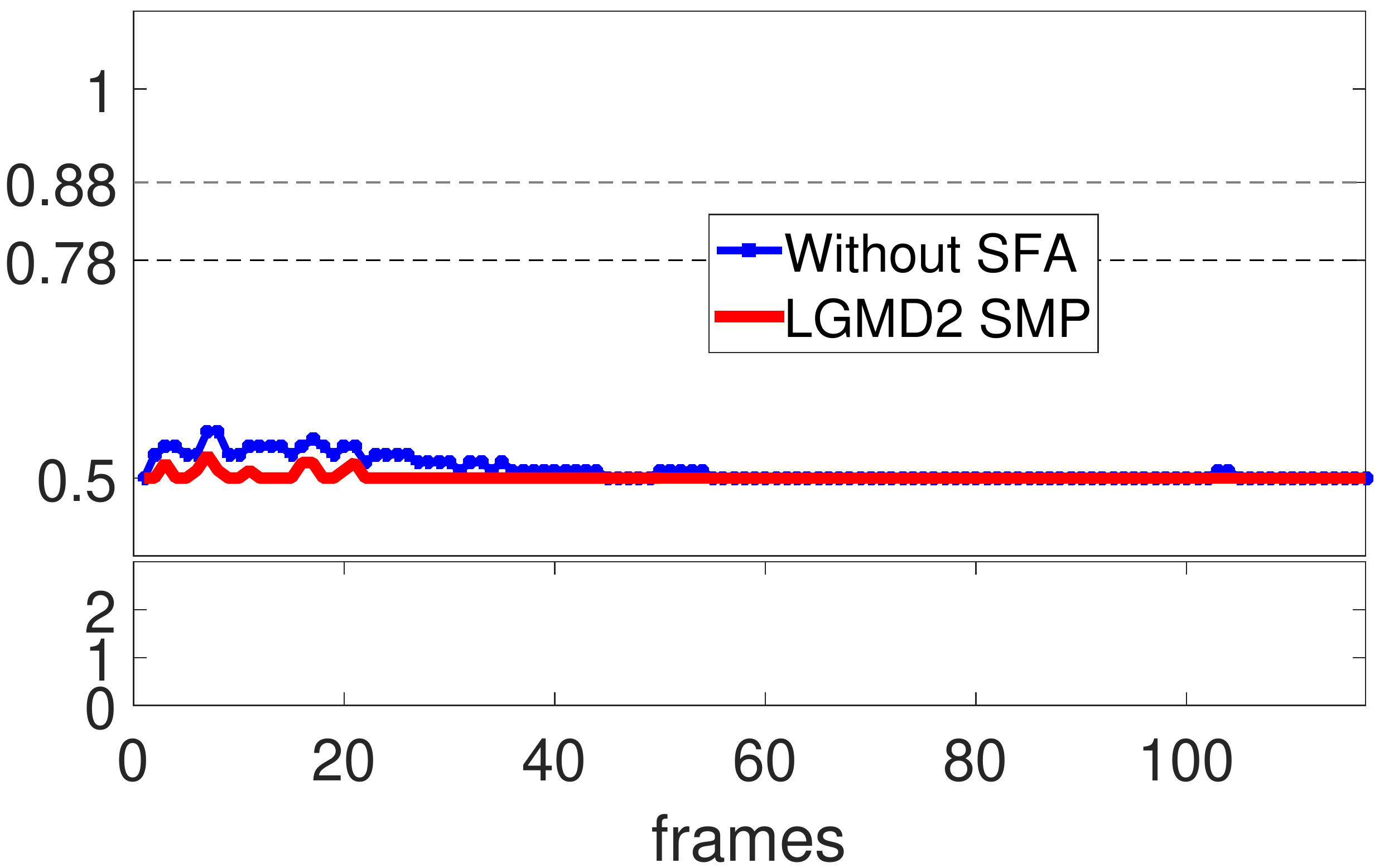}
		\label{real-reced}}
	\caption{The outputs of LGMD2 processing a dark approaching and receding object embedded in bright background. A few samples of snapshots labelled with frame number are shown on top of each plot. (a) approaching case: LGMD2 SMPs before/after the SFA mechanism are depicted in blue/red lines respectively. The two horizontal dashed lines denote two predefined threshold levels. The elicited spikes represented with asterisks are showing at bottom: 1-spike when SMP is between the first and second thresholds, 2-spikes if it exceeds the higher threshold. (b) Receding case with similar notations.}
	\label{real-appr-reced}
\end{figure}

In the first type of real physical stimuli, we examine the LGMD2 model's performance under dark moving object in depth against bright background. As illustrated in Fig.\ref{real-appr-reced}, it is no surprise that the LGMD2 model detects the direct collision to the dark object. It elicits increasing potential as the object closing in, and then it is activated to generate high frequency spikes (after frame-60 in Fig.\ref{real-appr}). At the end of looming, it is directly inhibited by FFI. In the case of approaching, the SMP is only slightly attenuated through SFA mechanism, since it overcomes such adaptation like the real neuron does \cite{SFA-2009,SFA-2009Role,SFA-2014}. On the other hand, when challenged by dark object receding in bright background (Fig.\ref{real-reced}), the LGMD2 model keeps quiet even at the beginning of receding.

\begin{figure}[!t]
	\centering
	\subfloat[]{\includegraphics[width=0.25\textwidth]{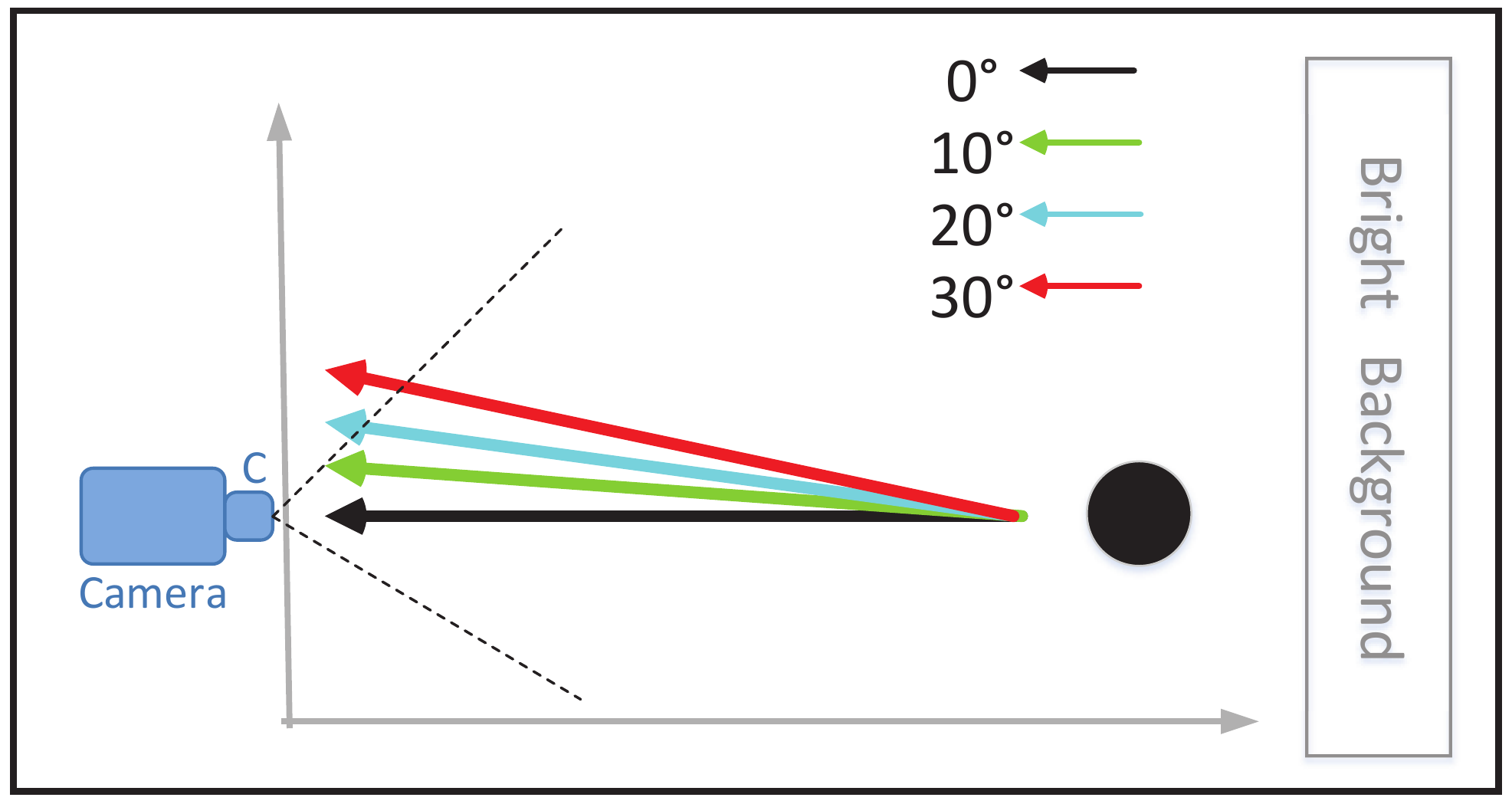}
		\label{aa-illu}}
	\vfill
	\vspace{-0.1in}
	\subfloat[]{\includegraphics[width=0.45\textwidth]{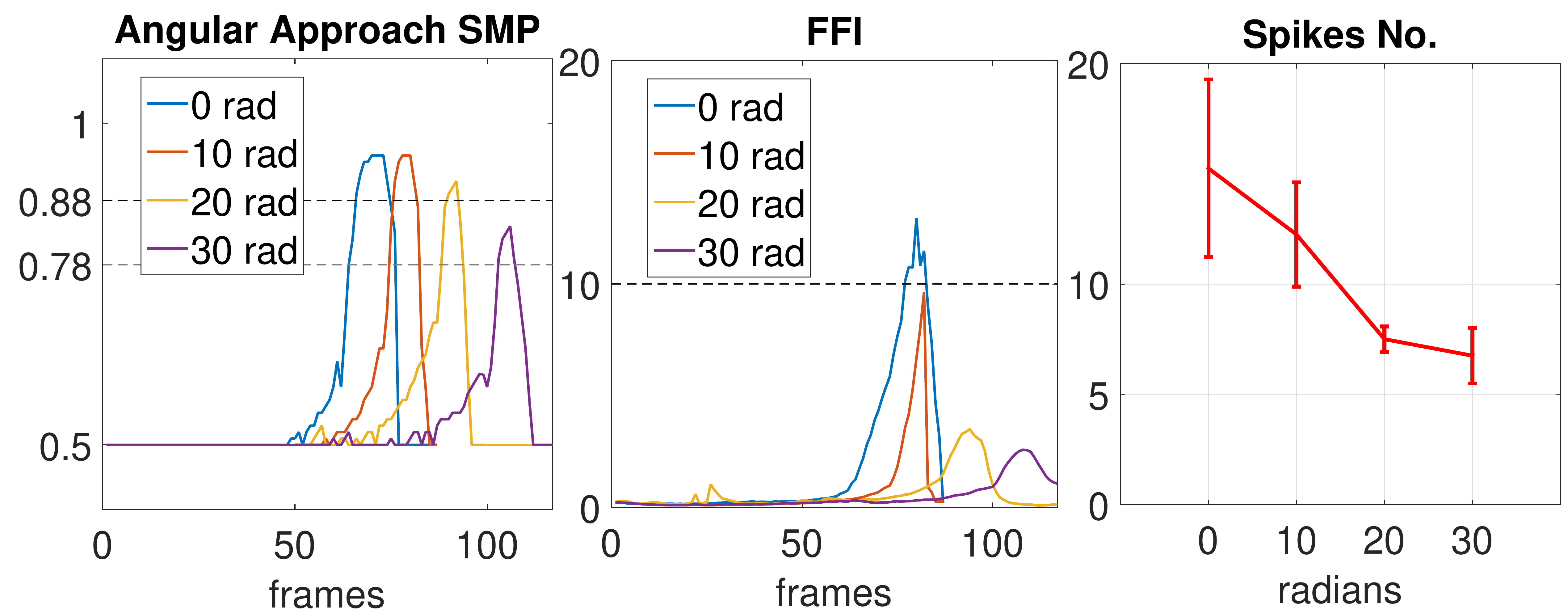}
		\label{real-aa-results}}
	\caption{The systematic 'angular-approach' experiments: (a) The experiment setting-up: the camera is fixed; a dark object against bright background approaches from four distinct angles. (b) The results from left to right plots are LGMD2 SMP (two dashed lines indicate spiking thresholds), FFI (with threshold level), and statistical count of spikes (with mean-variance information each throughout ten repeated tests).}
	\label{real-aa}
\end{figure}

Similarly, we also systematically inspect LGMD2 model's performance challenged against approaches from different angles. Fig.\ref{aa-illu} illustrates the experiment setting of dark object approaching at different angles: a direct collision corresponds $0$-degree angular approach and others represent the near-miss scenes. Fig.\ref{real-aa-results} illustrates that as approaching angle increased, LGMD2 neuron peaked much later, and the peak responses of both SMP and FFI decline. More intuitively, the statistical results from repeated tests demonstrate the LGMD2 neuron model spikes at much lower frequency along with the increasing approaching angle in the near-miss scenes.

\begin{figure}[!t]
	\centering
	\subfloat[]{\includegraphics[width=0.35\textwidth]{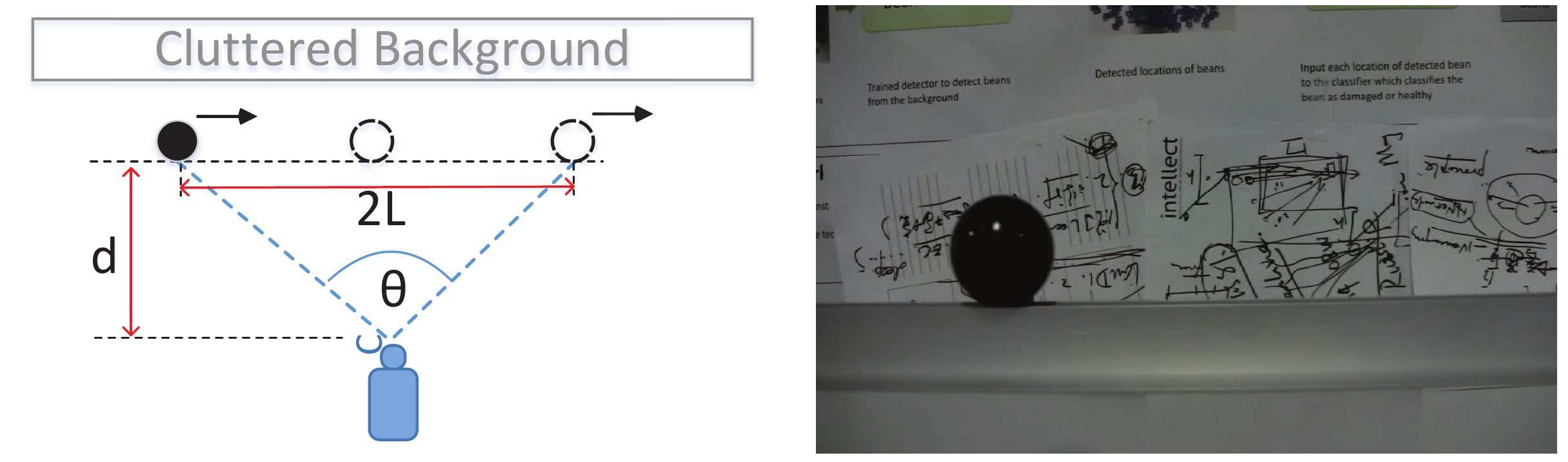}
		\label{trans-illu}}
	\vfill
	\vspace{-0.1in}
	\subfloat[]{\includegraphics[width=0.45\textwidth]{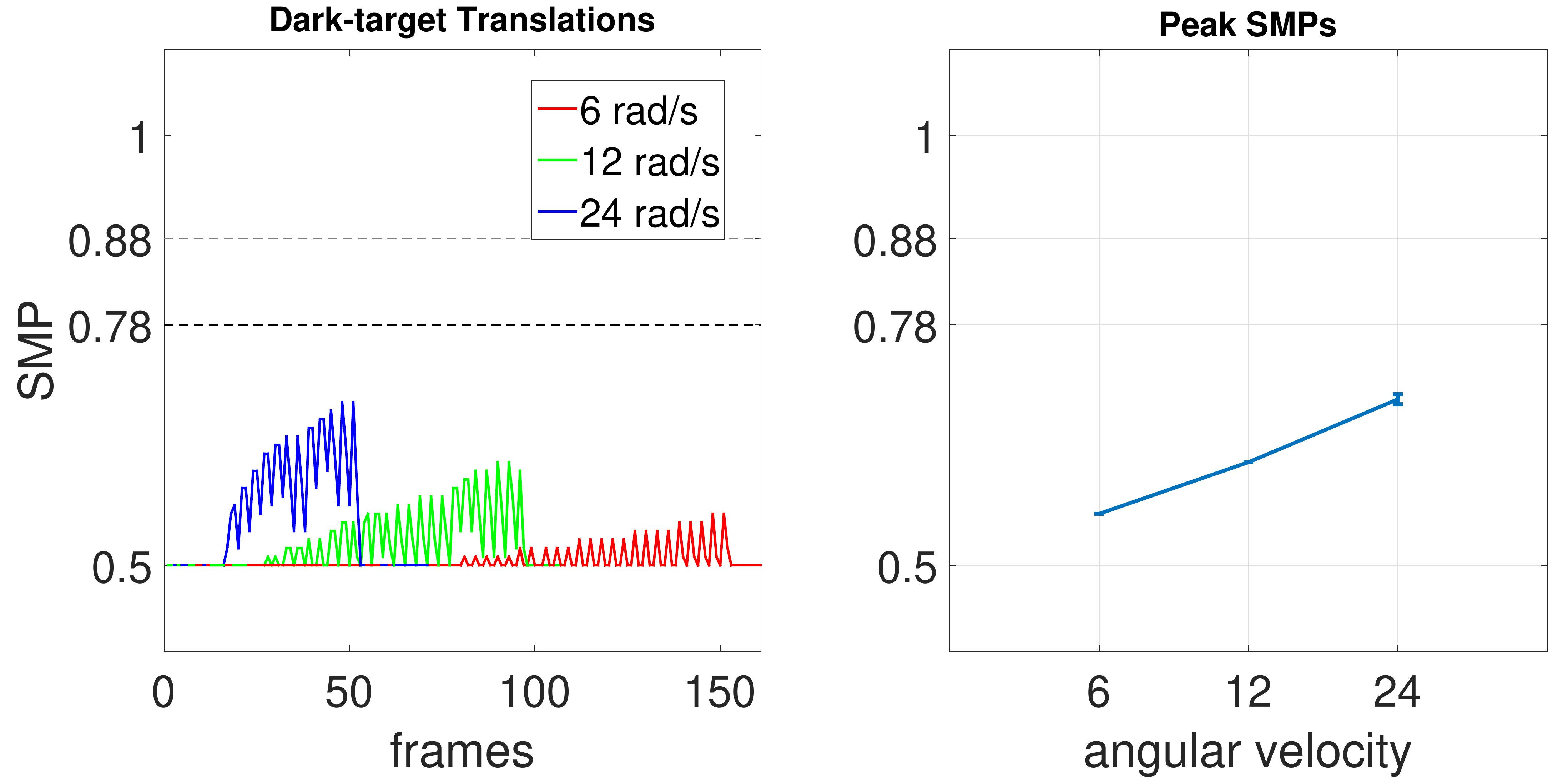}
		\label{real-trans-results}}
	\caption{The systematic translation experiments: (a) The experiment setting-up: a dark object translating against cluttered background - $d$ indicates the perpendicular distance from camera to the route of moving object, $2L$ denotes the length of view-arc along translation route. The angular size $\theta$ can be calculated by $\theta = 2tan^{-1}(L / 2d)$. An example snapshot picked up from the recorded video sequence is also exhibited. (b) The LGMD2 SMPs and statistical results of peak-response under three varied angular velocities, each of which was repeated five times.}
	\label{real-trans}
\end{figure}

\begin{figure}[!t]
	\begin{minipage}[t]{0.5\textwidth}
		\centering
		\centerline{\includegraphics[width=0.5\textwidth]{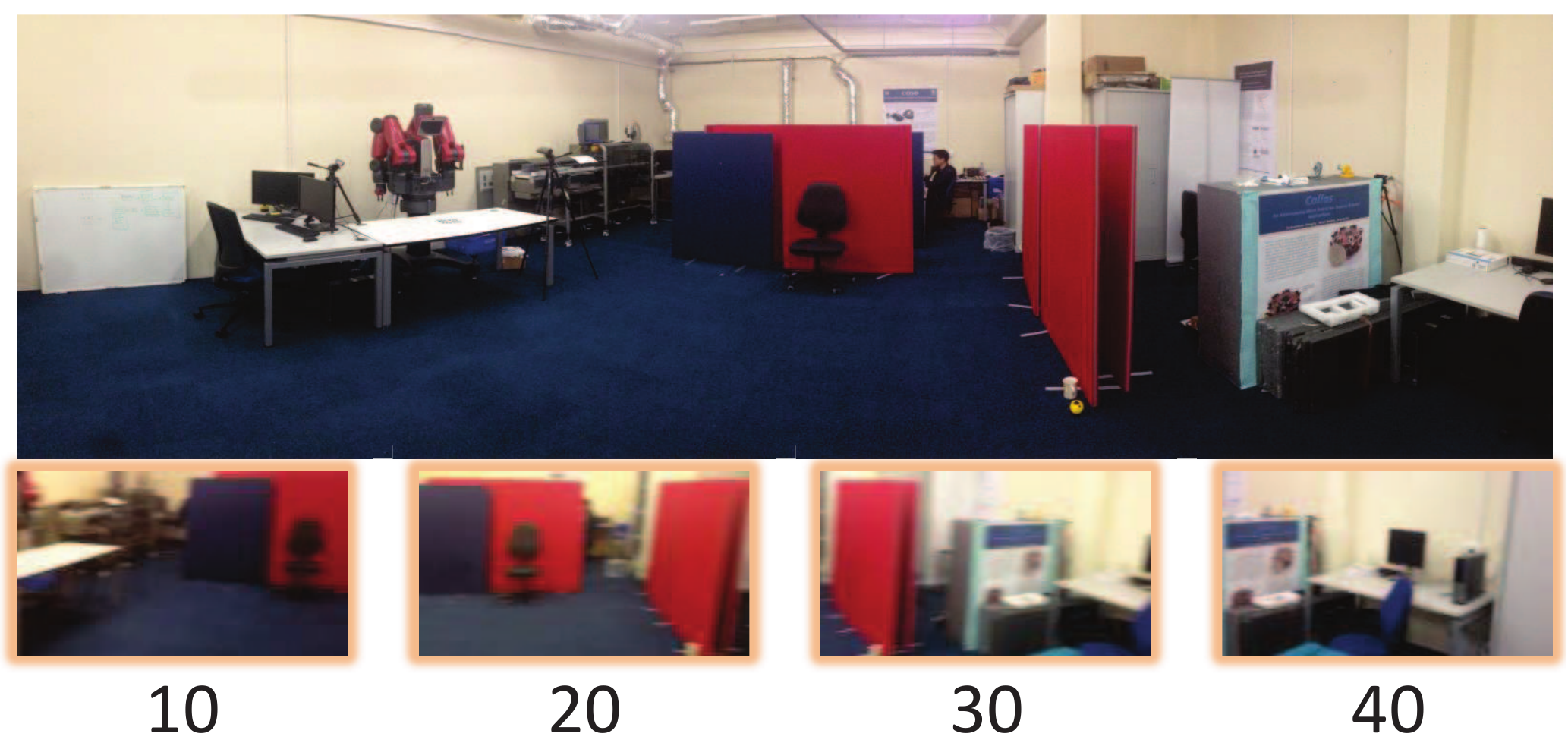}}
	\end{minipage}
	\vfill
	\vspace{-0.1in}
	\centering
	\subfloat{\includegraphics[width=0.36\textwidth]{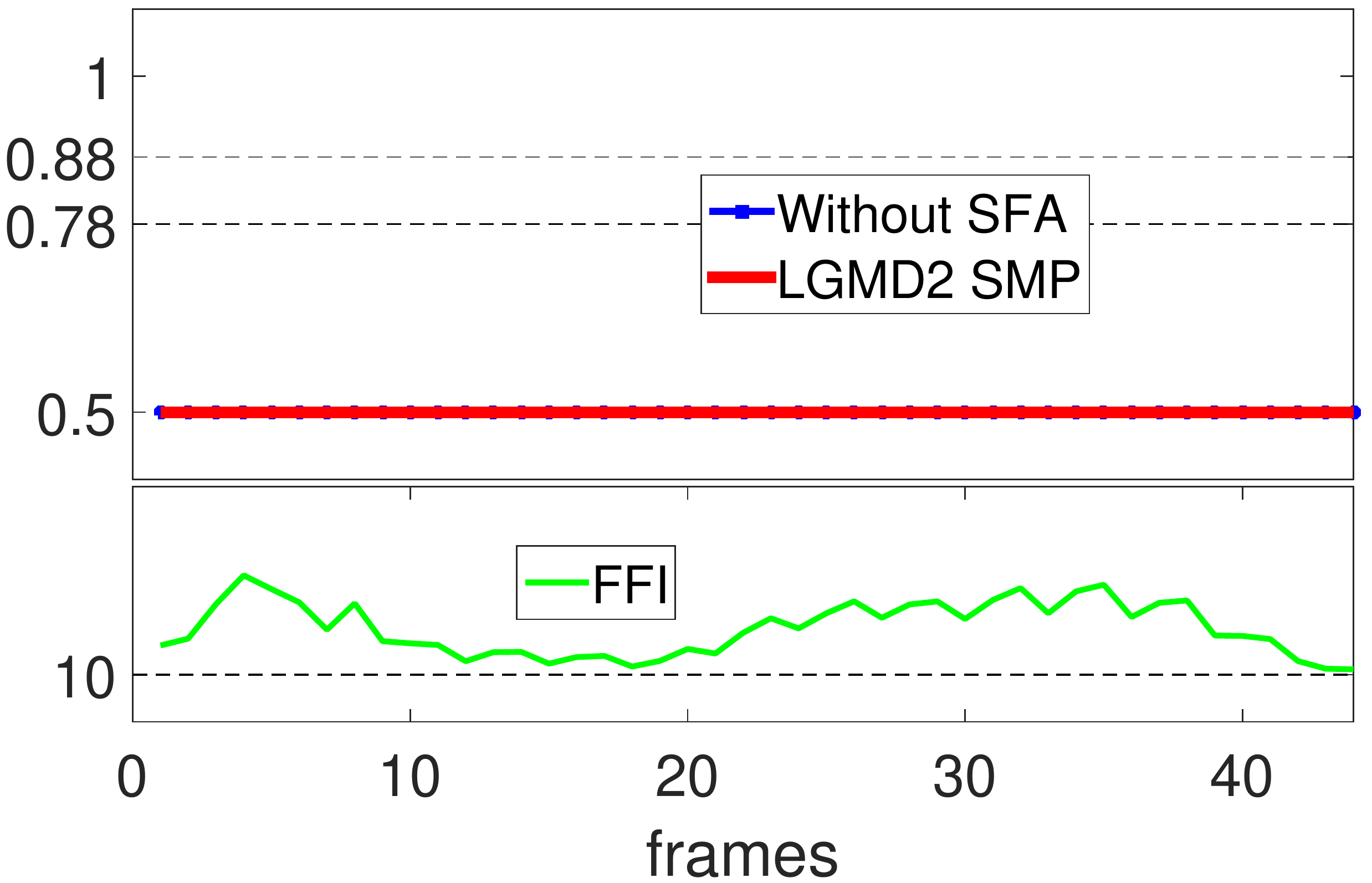}}
	\caption{The results of processing a turning cluttered scene. The panoramic view with specific snapshots are shown at the top. LGMD2 neural responses including the SMP and FFI are depicted with dashed thresholds respectively.}
	\label{real-wf}
\end{figure}

In the second type of real physical stimuli, LGMD2 neuron model will be challenged by a few sets of translation stimuli against cluttered background as illustrated in Fig.\ref{trans-illu}. It is necessary to state here that the angular speeds were not constant yet with a bit acceleration for each translation movement. Therefore, we calculate the average angular velocity for each set of translation movements (Fig. \ref{trans-illu}). The results in Fig.\ref{real-trans-results} illustrate under all translating stimuli, LGMD2 neuron model exhibits low-level responses in comparison with those in approach courses (Fig.\ref{real-appr} and \ref{real-aa-results}), whereby different lengths of time window imply different translation movements at different angular speeds. Interestingly, the statistical results also demonstrate the response of LGMD2 neuron model to speeds - the peak potential steadily climbs up with increasing translation angular speed.

In the last type of real physical stimuli, the LGMD2 model will be challenged against a turning scenario in cluttered scene implying rapid luminance change over a large part of the field of view. As illustrated in Fig.\ref{real-wf}, the LGMD2 neuron is strictly suppressed within the intact turning of view, since the FFI climbs significantly to overstep its threshold and remains at very high level till the end of movement. The result demonstrates that with a similar FFI-pathway like the LGMD1 models, e.g. \cite{LGMD1-Yue2006,LGMD1-Yue2009,LGMD1-Glayer}, the LGMD2 neuron model can also deal with the situation appropriately even a large amount of photoreceptors are highly activated.

To sum up all off-line experiments, the results are satisfactory - the proposed LGMD2 neural network (or LGMD2 model) fulfills the characteristics of the LGMD2 neuron in locusts. The bio-plausible structures - ON/OFF pathway and SFA mechanism, have been proved crucial in realizing its specific collision selectivity to dark-looming objects amongst other kinds of visual stimuli.

\subsection{Robot Experiments}
In the real time (or on-line) experiments, the LGMD2 neural network was implemented in a micro-robot called 'Colias' (Fig. \ref{colias_proto}). To examine its performance in robotic applications and deepen the understanding of LGMD2's unique characteristics, we set up two types of real time experiments: the arena tests and the systematic tests.

\subsubsection{In Arena Tests}
\begin{figure}[!t]
	\centering
	\includegraphics[width=0.49\textwidth]{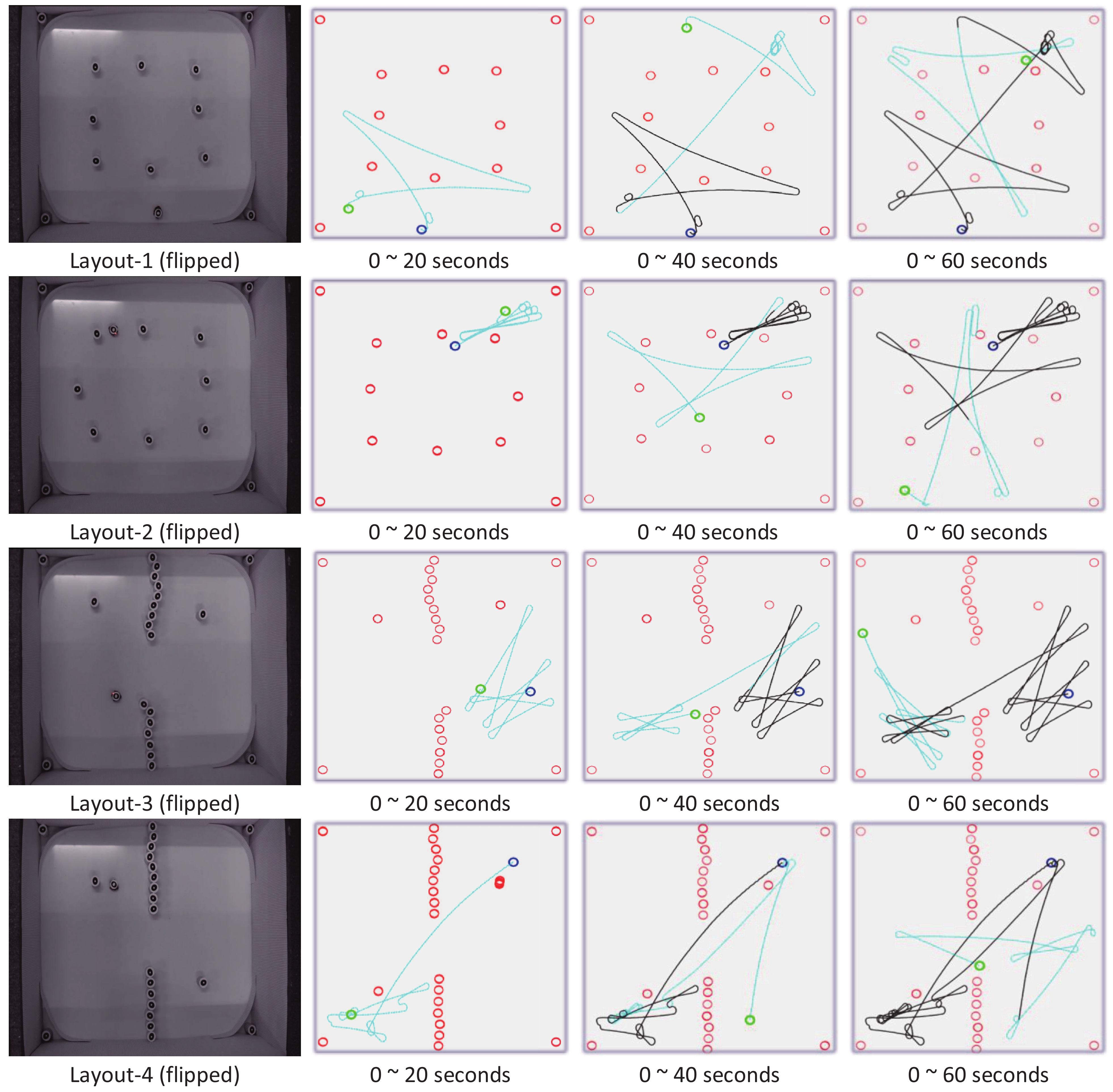}
	\caption{Arena tests: the top-down views of the arena and trajectories of the robot using LGMD2 as collision detector. For each layout, each experiment lasts for one-minute. The trajectories of the robot are in blue for most updated routes or dark lines for routes in previous time window(s). The blue circles indicate the initial positions of the robot, whilst the green ones denote the updated sites at the end of each specific time window. The red circles indicate static obstacles.}
	\label{arena}
\end{figure}

In the first type of on-line experiments, we inspect the effectiveness of LGMD2 neuron model as a quick collision detector for a robot. In the experiments, we put a Colias mini robot, which has implemented LGMD2 as its collision detector, in an arena with a few ($10 \sim 20$) obstacles for different layouts (Fig. \ref{arena}). The arena is approximately $110 \cdot 110$cm. The white internal walls of the arena and the white surface of obstacles are marked with densely distributed dark patterns. A CCD camera is fixed above the arena to form the top-down view to record performances of the Colias robot. There are also specific patterns on top of the micro-robot and the obstacles for a practical multi-robots localization system \cite{Colias-localization}. therefore, we can get the very precise trajectories of all moving objects with the specific pattern in the arena throughout each test.
 
In the arena tests, the Colias with LGMD2 implemented is initialized to go forward autonomously until a potential collision detected. Once a collision detected, it turns right or left randomly with a large angle which is more than $180$ degrees to avoid imminent collisions. After each avoidance behavior, it resumes to go forward, and so on. Fig.\ref{arena} illustrates four experiment layouts with robot's trajectories captured in different time windows. The experiments have demonstrated the robustness of LGMD2 neuron model as a collision detector for autonomous robots in navigation and path exploration. In addition, similar test results with high success rates have been demonstrated partially in \cite{LGMD2-BMVC}.

\subsubsection{Systematic Characteristic Investigations}
\begin{figure}[!t]
	\centering
	\subfloat[]{\includegraphics[width=0.4\textwidth]{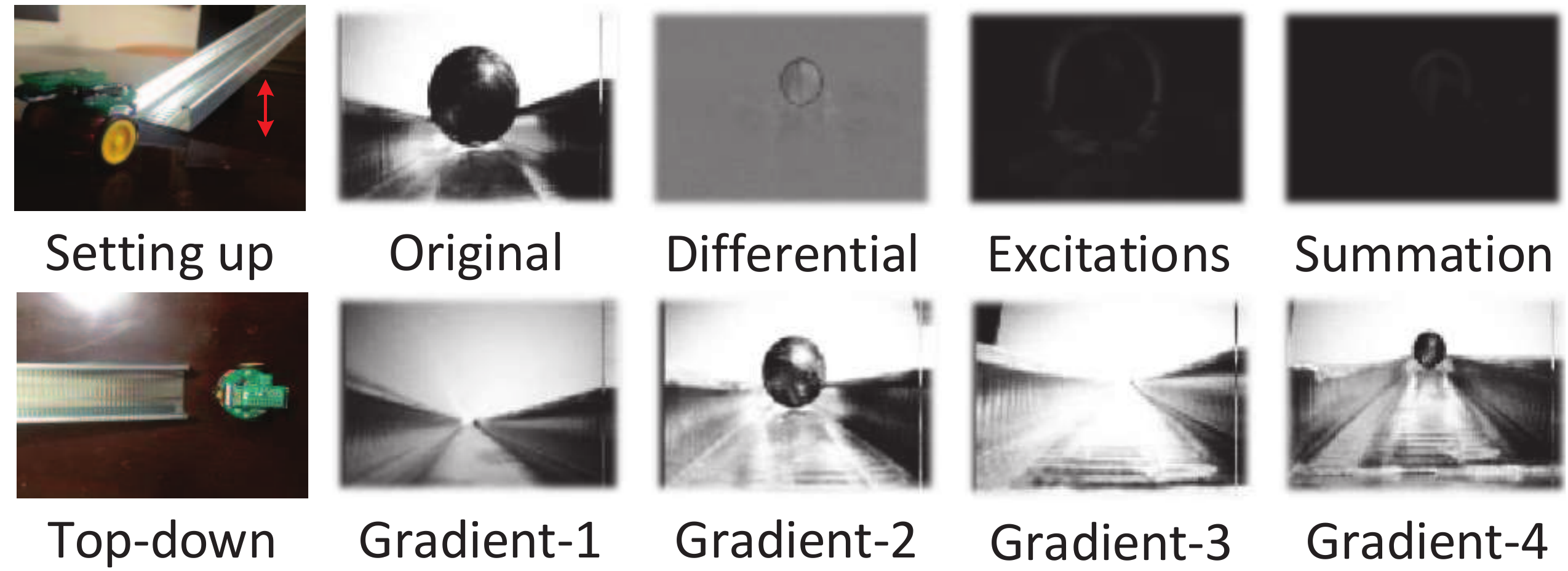}
		\label{robot-appr-set}}
	\vfill
	\vspace{-0.1in}
	\subfloat[]{\includegraphics[width=0.23\textwidth,height=0.11\textheight]{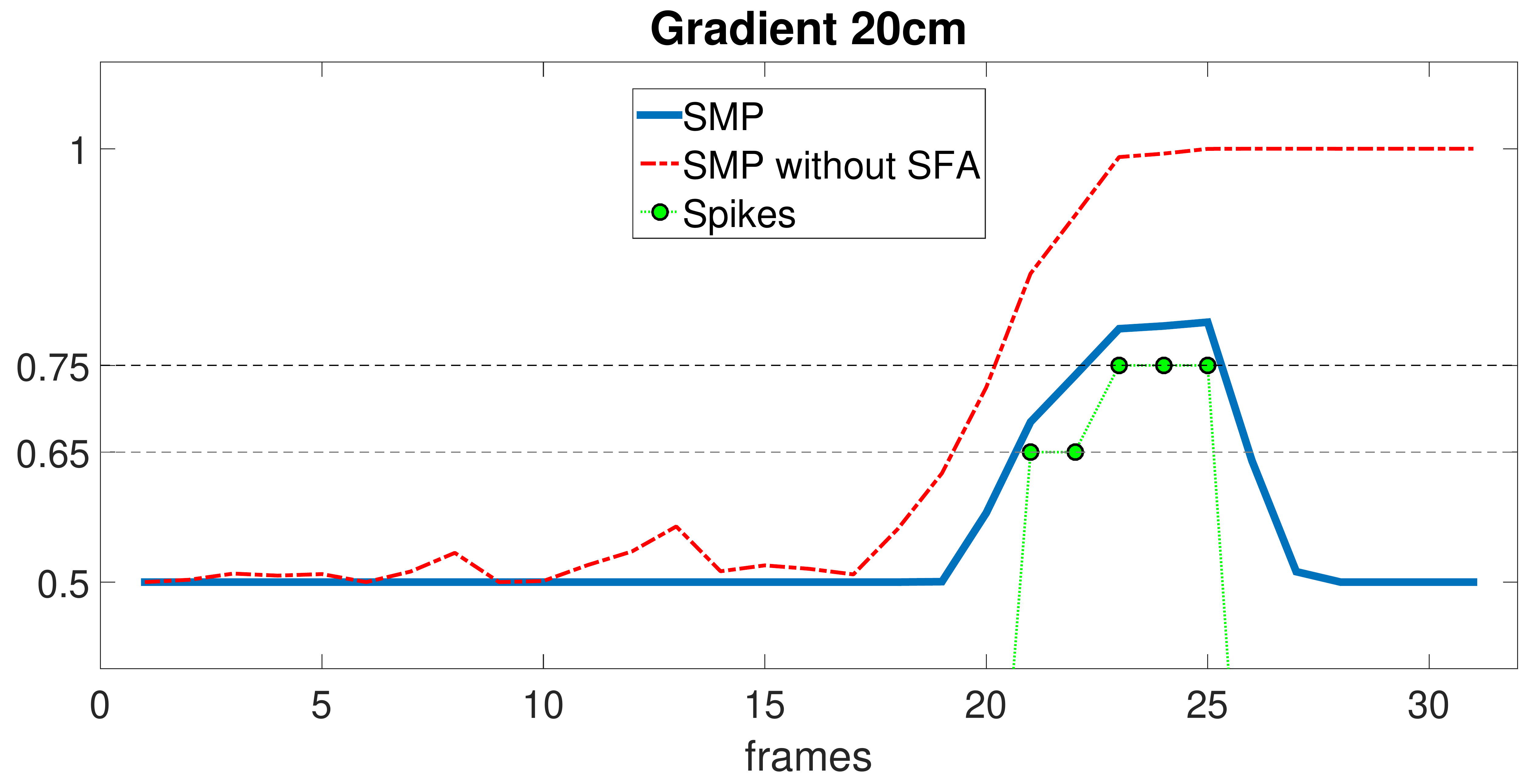}
		\label{g1}}
	\hfill
	\subfloat[]{\includegraphics[width=0.23\textwidth,height=0.11\textheight]{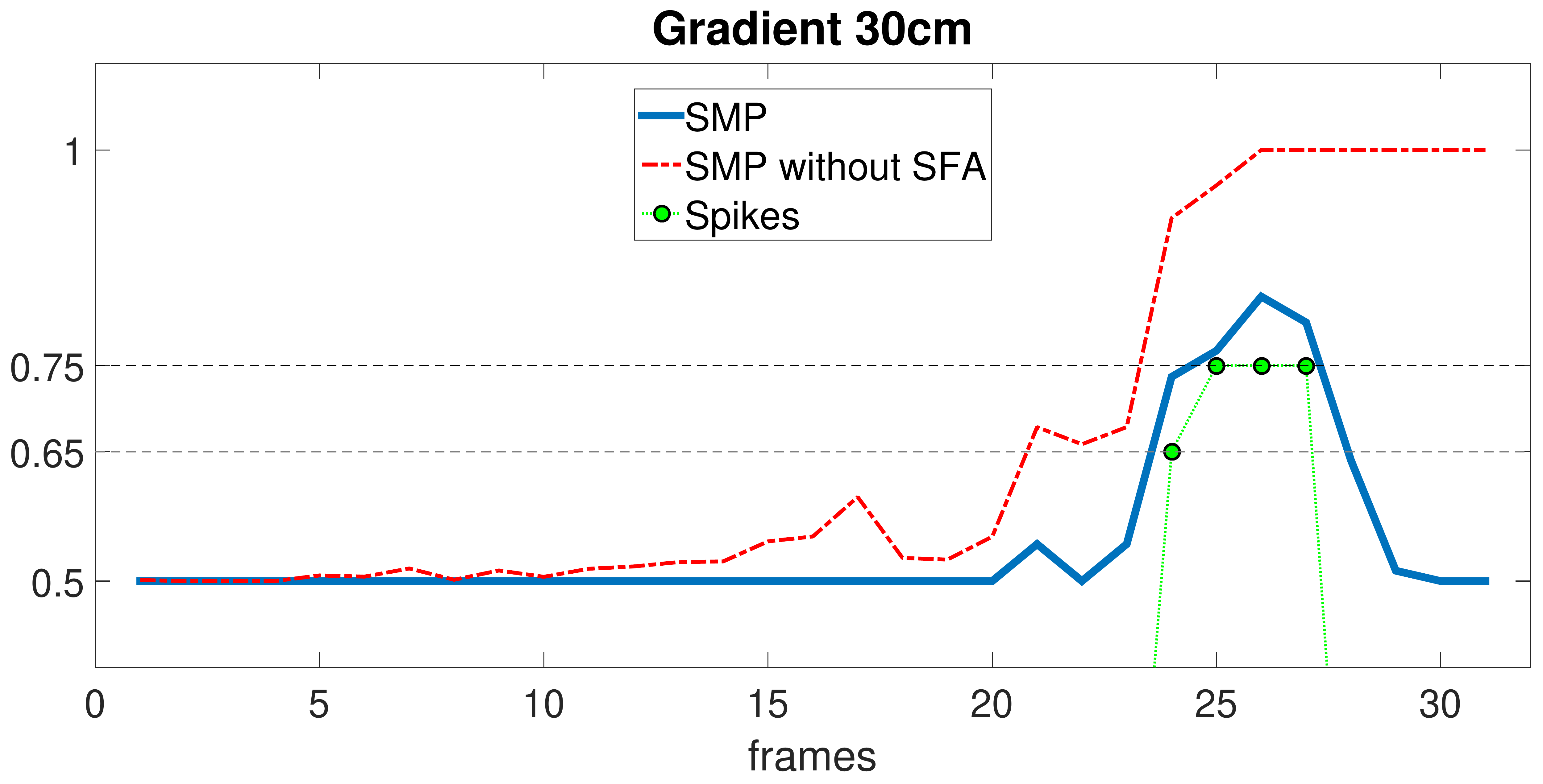}
		\label{g2}}
	\vfill
	\vspace{-0.1in}
	\subfloat[]{\includegraphics[width=0.23\textwidth,height=0.11\textheight]{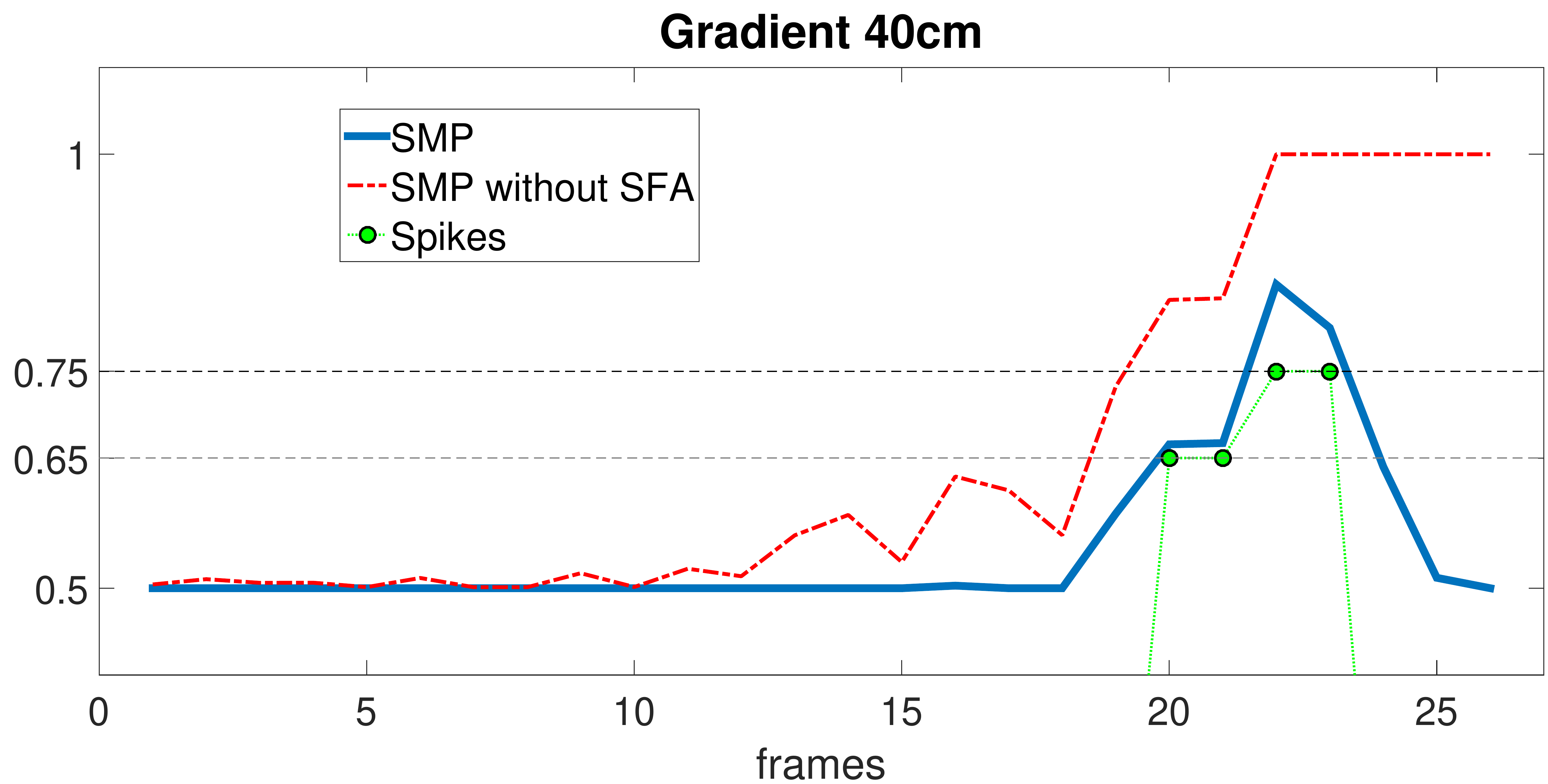}
		\label{g3}}
	\hfill
	\subfloat[]{\includegraphics[width=0.23\textwidth,height=0.11\textheight]{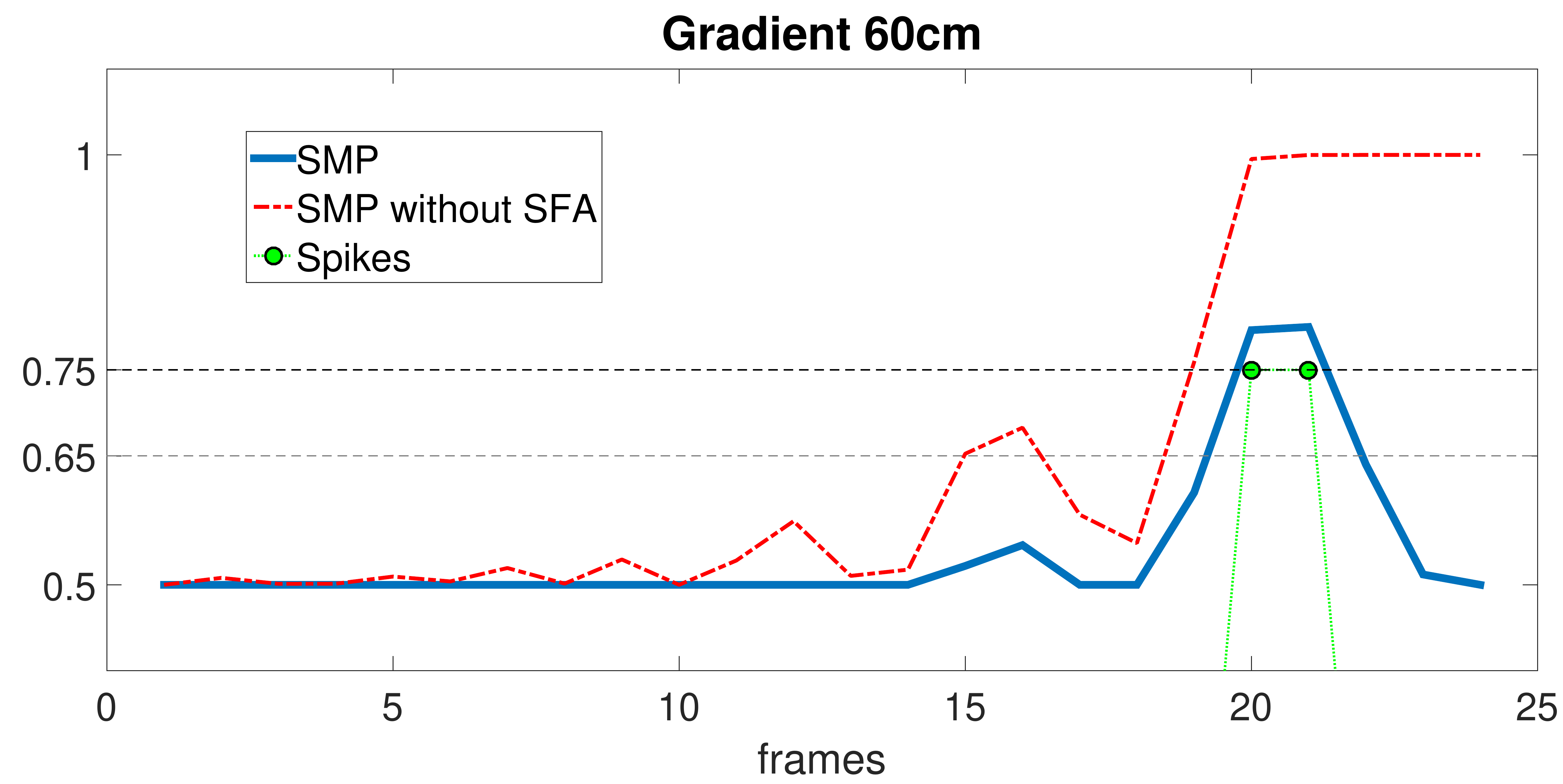}
		\label{g4}}
	\caption{The results of robotic experiments challenged by overhead-approaching stimuli: (a) The experiment setting and sampled maps in different layers of the LGMD2 model in Colias at different approaching gradients. (b), (c), (d), (e) LGMD2's neural responses when challenged at four varied gradients (20, 30, 40, 60cm): the sigmoid membrane potentials (SMPs) of the LGMD2 model with and without SFA filtering are drawn in dashed red and blue lines respectively; two threshold levels are depicted in gray-dashed lines; the spikes are marked at different threshold levels whereby the SMPs (with SFA) surpassing them.}
	\label{colias-overhead}
\end{figure}

\begin{figure}[!t]
	\centering
	\subfloat[]{\includegraphics[width=0.3\textwidth]{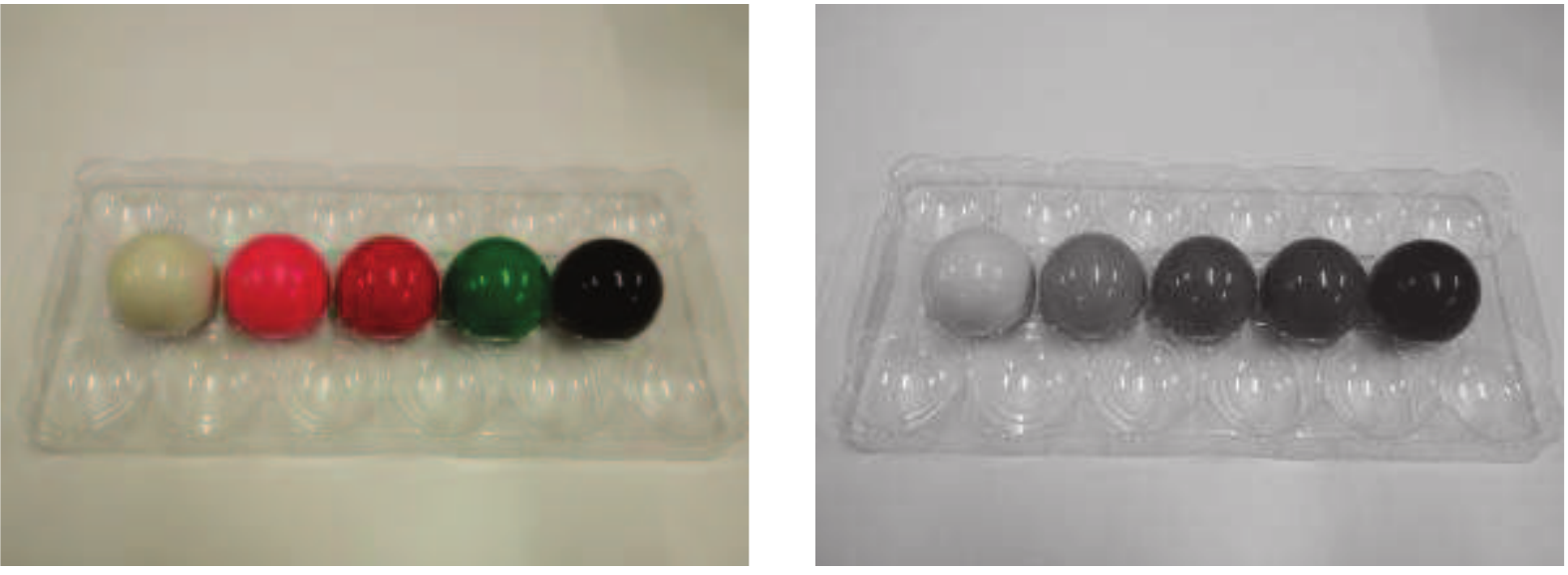}
		\label{balls}}
	\vfill
	\vspace{-0.1in}
	\subfloat[]{\includegraphics[width=0.3\textwidth]{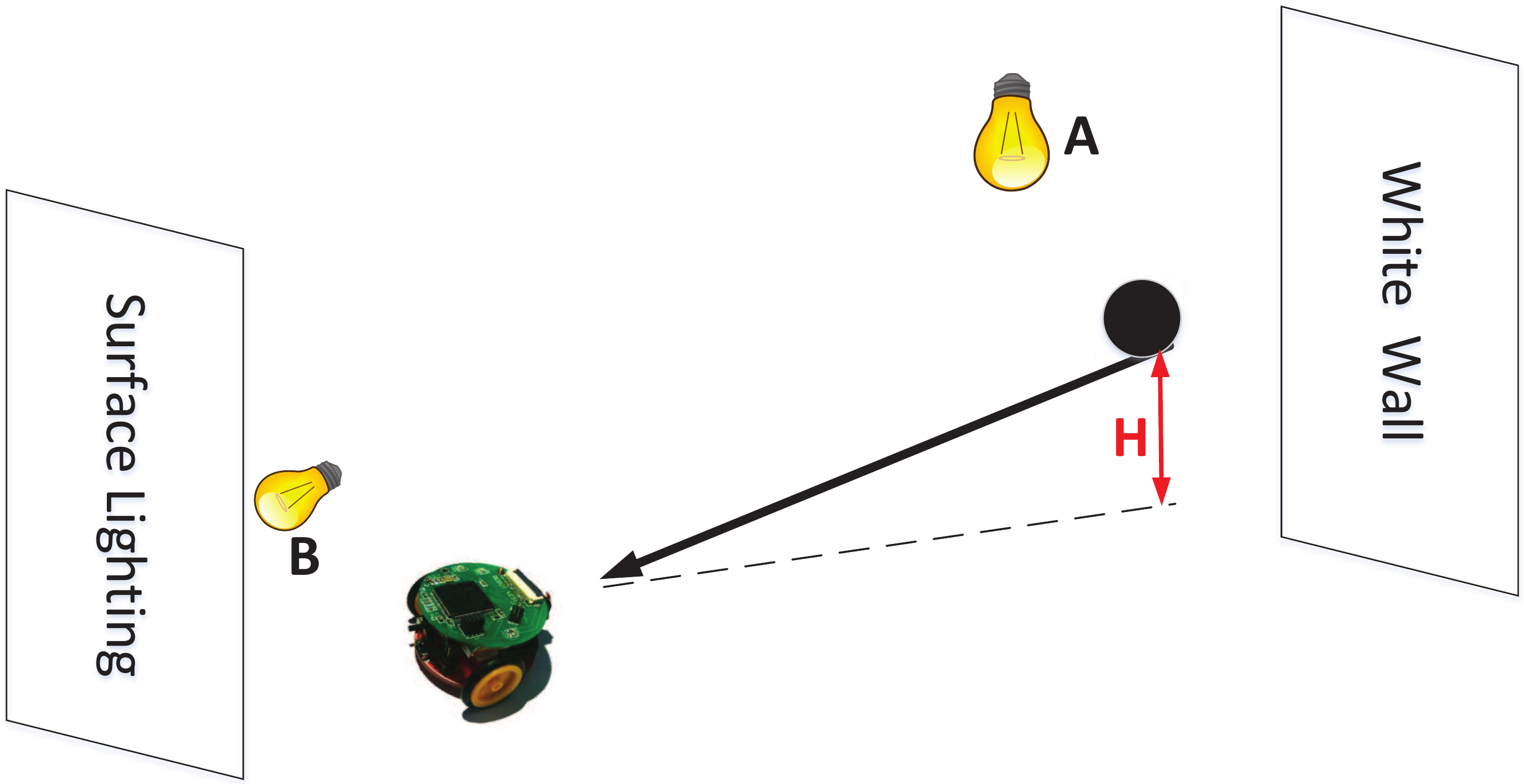}
		\label{balls-appr-set}}
	\caption{Experiment setting-up with different gray-scale approaching objects in collision courses: (a) Five balls in RGB and gray-scale, are used to approach the micro-robot in bright and dark environments respectively. (b) The environment setting-up: all the objects approach with a fixed gradient $H$ indicating approximately the same speed; there are two light sources - A is for top-down global illumination and B is for the local surface lighting behind Colias, which are use to create bright/dark scenarios separately.}
	\label{colias-gray-balls-appr-set}
\end{figure}

\begin{figure}[!t]
	\centering
	\subfloat[]{\includegraphics[width=0.4\textwidth]{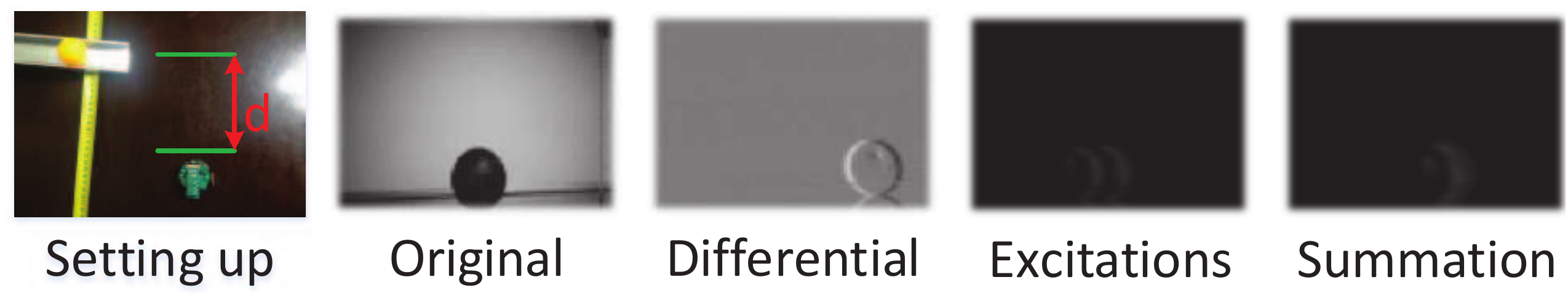}
		\label{robot-trans-set}}
	\vfill
	\vspace{-0.1in}
	\subfloat[]{\includegraphics[width=0.23\textwidth,height=0.11\textheight]{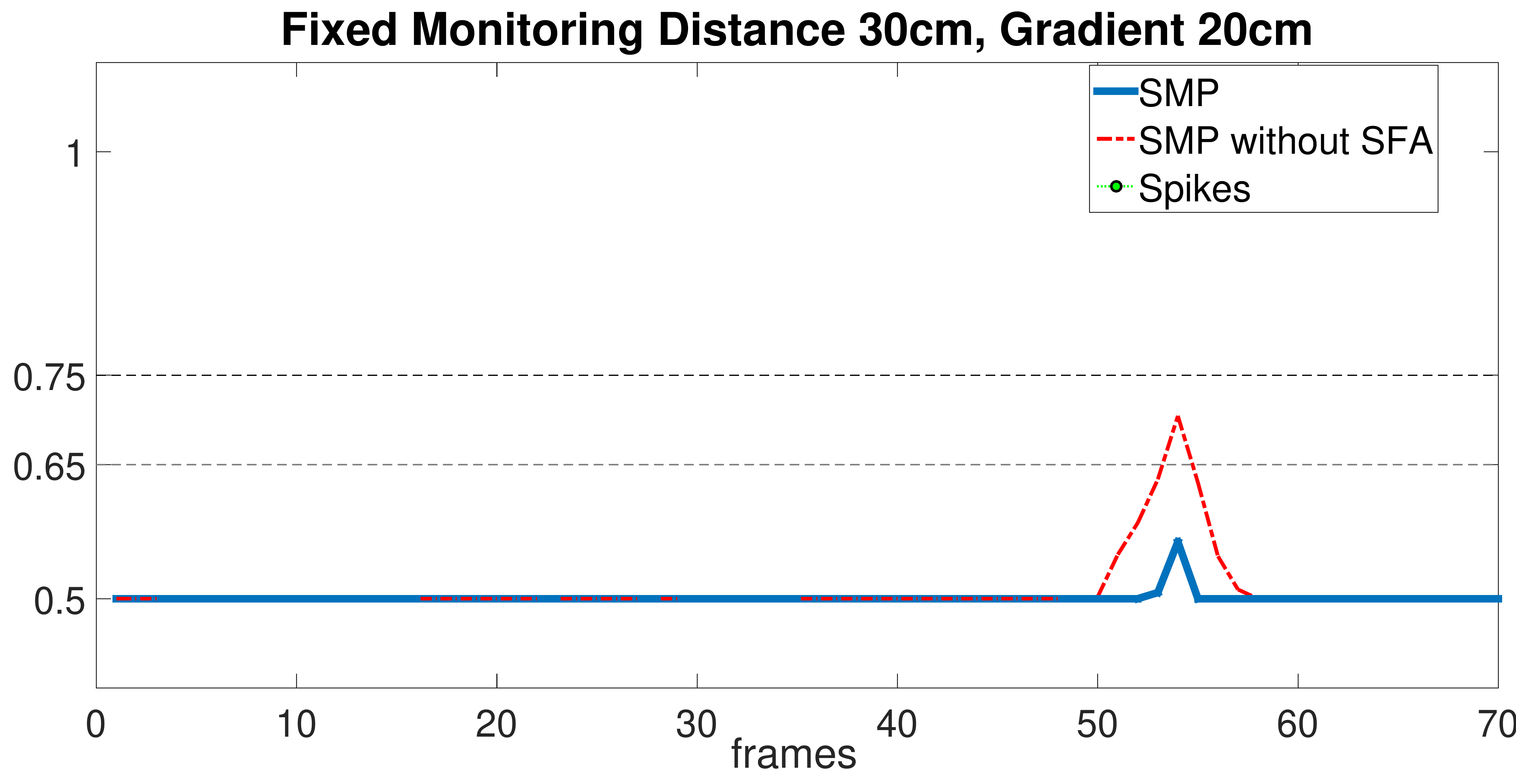}
		\label{dht1}}
	\hfill
	\subfloat[]{\includegraphics[width=0.23\textwidth,height=0.11\textheight]{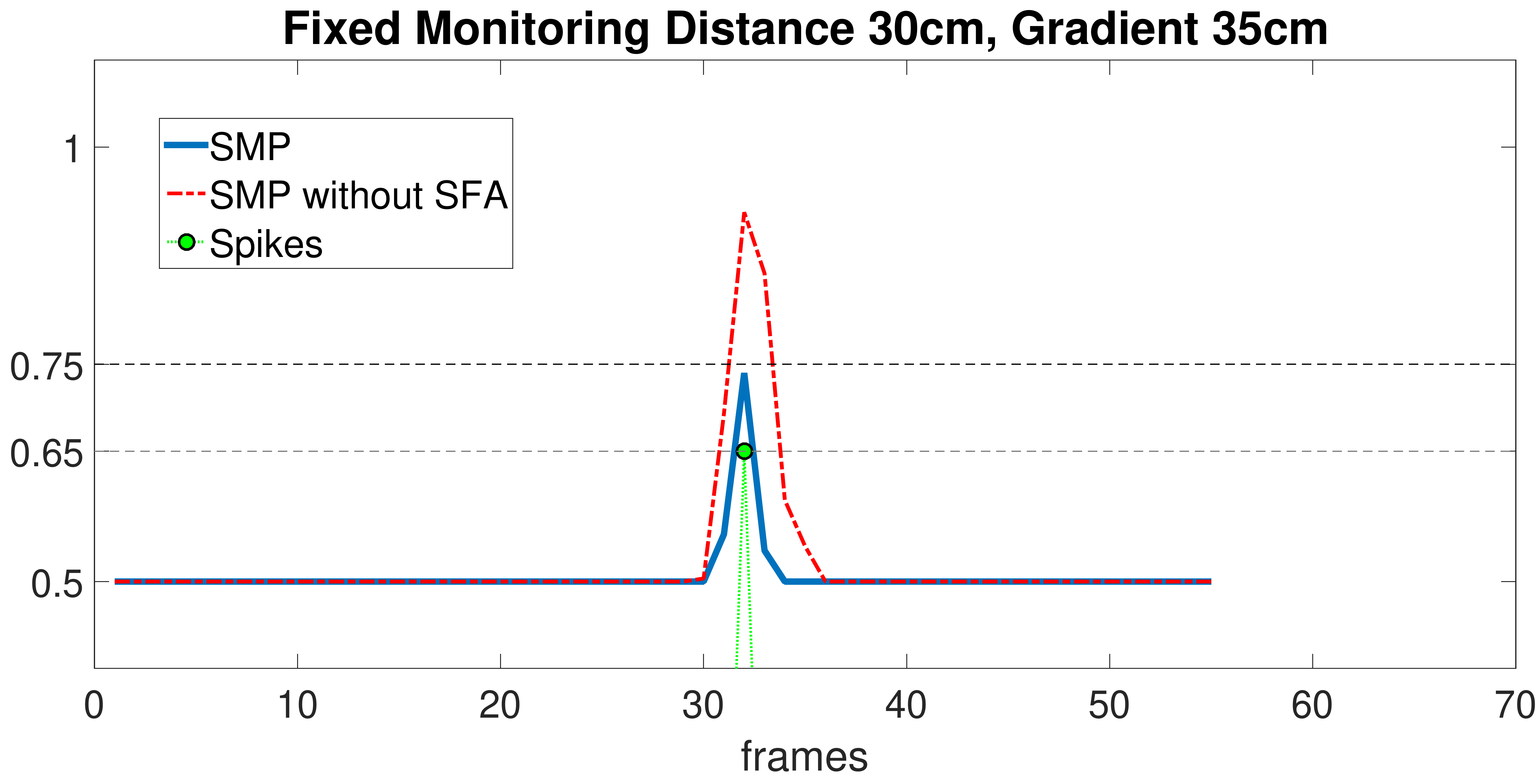}
		\label{dht2}}
	\vfill
	\vspace{-0.1in}
	\subfloat[]{\includegraphics[width=0.23\textwidth,height=0.11\textheight]{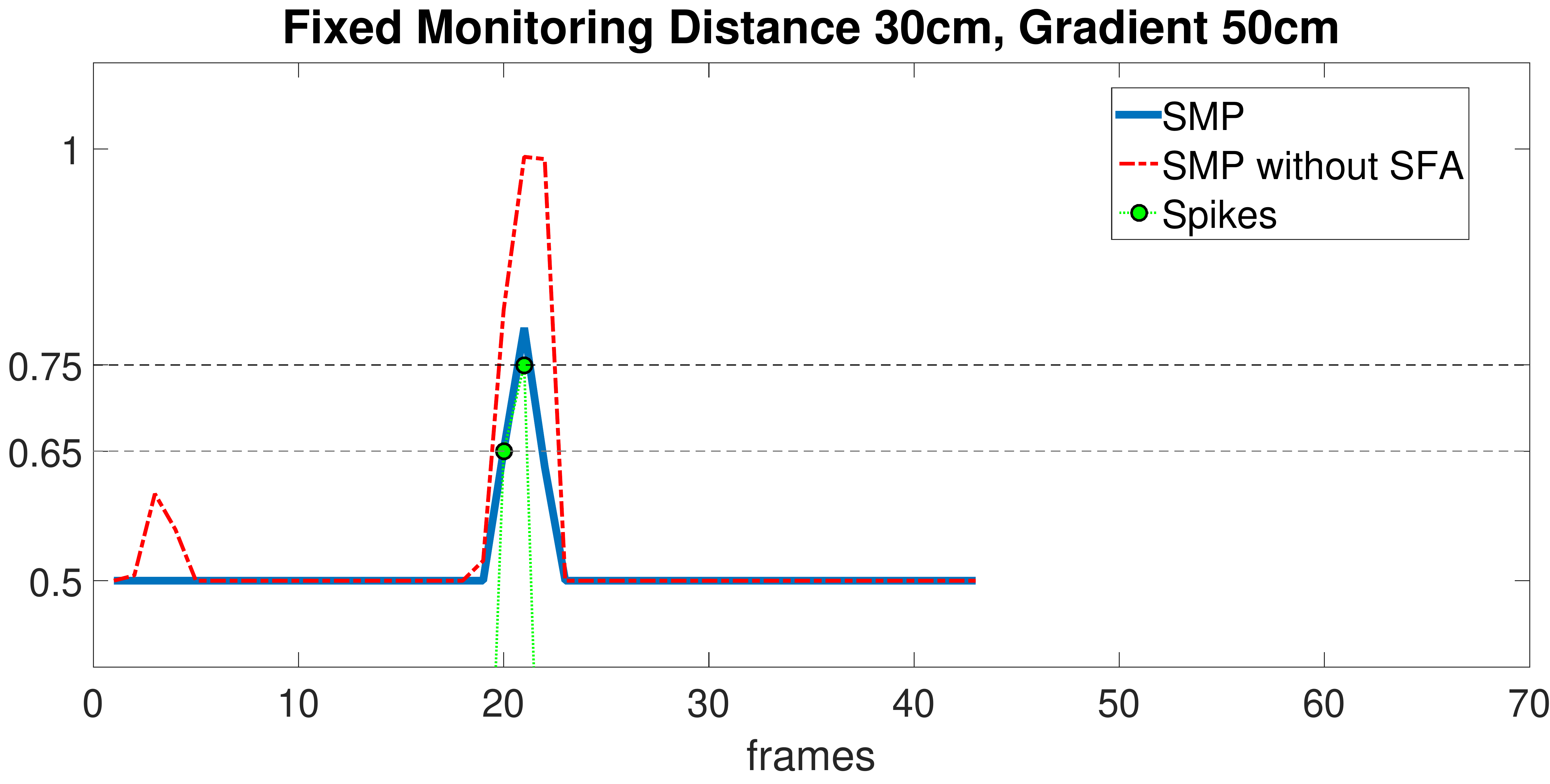}
		\label{dht3}}
	\hfill
	\subfloat[]{\includegraphics[width=0.23\textwidth,height=0.11\textheight]{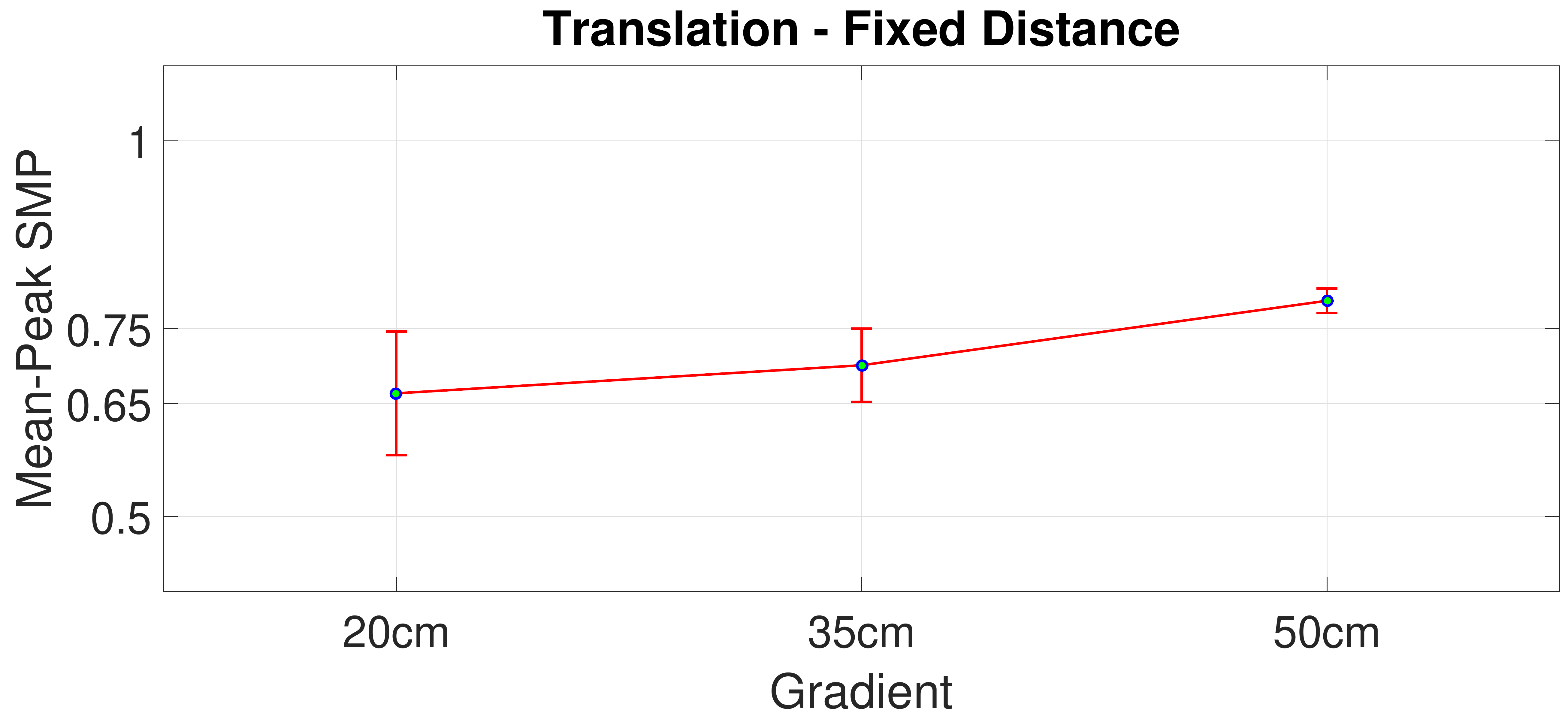}
		\label{bar_dht}}
	\caption{The results of robotic experiments in which LGMD2 is challenged by the first set of systematic translations. (a) The experiment setting-up and sampled views from Colias. In this case, the monitoring distance $d$ was fixed at 30cm and the gradients of slot varied in 20, 35 and 50cm respectively. (b), (c), (d) LGMD2 neural responses: the notations are in accordance with those in Fig.\ref{colias-overhead}. (e) Statistical results of peak-SMPs under translations from varied gradients, each of which repeated ten times.}
	\label{colias-trans-dht}
\end{figure}

For systematic investigation of LGMD2's unique characteristics when implemented in the mini robot Colias, we have designed a few types of real time experiments. The first type of experiments are to challenge the robot (LGMD2) with overhead looming objects in a direct collision course mimicking swooping predators from the bright sky (Fig.\ref{robot-appr-set}). The second type of experiments is to challenge it with translating movements with different velocities, which occur frequently in a visual environment. The translation stimuli involve a dark object crosses the Colias' field of view horizontally at different velocities, or at different distances respectively (Fig. \ref{robot-trans-set}). Further experiments are carried out with five different gray-scaled objects approach  the micro-robot in turn in dark and bright environments in a collision course respectively (Fig.\ref{colias-gray-balls-appr-set}). In all the experiments, we shut down the motion controls unit of the Colias to make it as a motionless observer, and collect its readouts including the sigmoid membrane potential and spikes afterwards.

To form the overhead-approach stimuli, as shown in Fig.\ref{robot-appr-set}, we let a dark ball automatically rolls down towards the front of Colias along a slot set at four different gradients respectively. The results (Fig.\ref{colias-overhead}) demonstrate that the proposed LGMD2 neuron model can recognize all imminent collisions robustly, i.e., the sigmoid membrane potential of the LGMD2 model increases for each approaching ball, and then invokes high frequency spikes. Although the SMPs are attenuated via the SFA mechanism especially at the end of each approaching stimulus after peaking, they all overcome such adaptations during dark objects looming which strongly agrees with the relevant biological research results \cite{SFA-2009Role,SFA-2009}.

For further verifying the unique features of the proposed LGMD2 neuron model against darker/lighter approaching objects, we design experiments to test the LGMD2 model's selectivity in dealing with different gray-level approaching objects. As illustrated in Fig.\ref{balls}, we use five balls (left side image) each with a different RGB color which corresponds to a certain gray-level from bright to dark (right side image). The experiment setting is shown in Fig.\ref{balls-appr-set} - we have prepared two illuminating light sources, one for global illumination (light source A in Fig.\ref{balls-appr-set}), another for local surface illumination behind Colias (light source B in Fig.\ref{balls-appr-set}).

\begin{figure}[!t]
	\centering
	\subfloat[]{\includegraphics[width=0.4\textwidth]{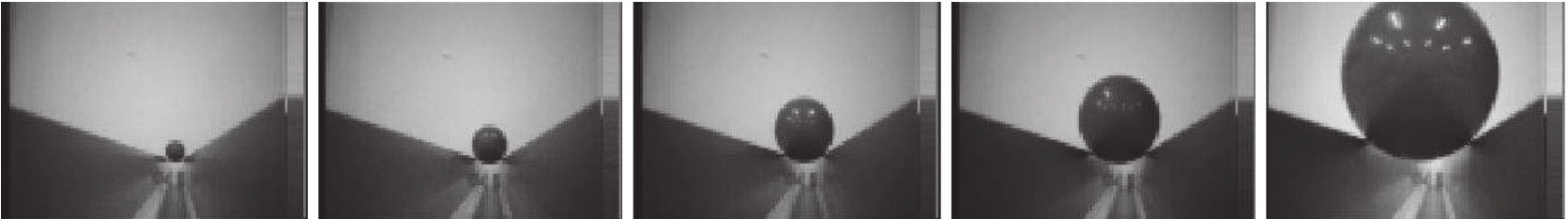}
		\label{colias-views-dark-appr}}
	\vfill
	\vspace{-0.1in}
	\subfloat[]{\includegraphics[width=0.45\textwidth]{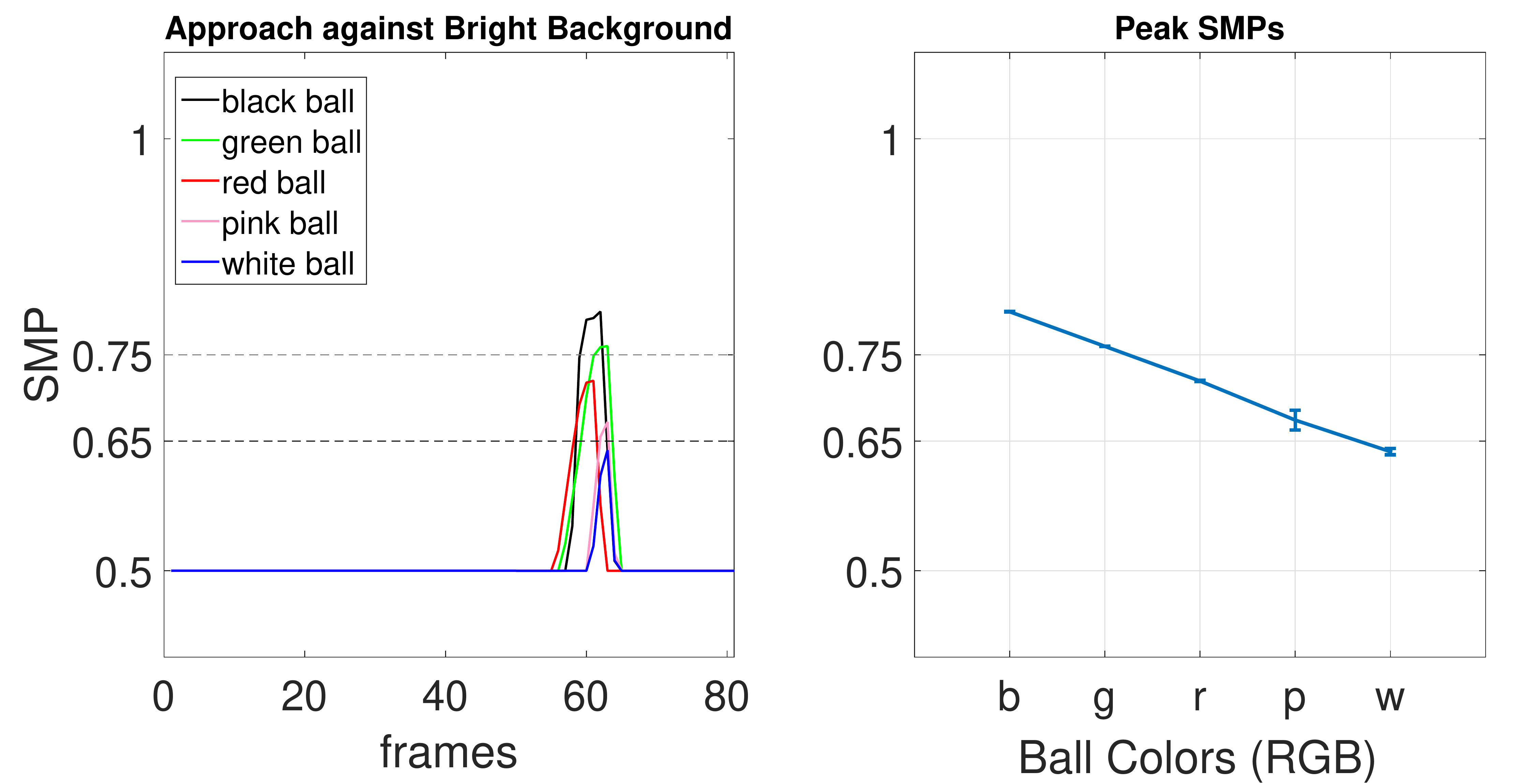}
		\label{dark-balls-apprs}}
	\vfill
	\vspace{-0.1in}
	\subfloat[]{\includegraphics[width=0.4\textwidth]{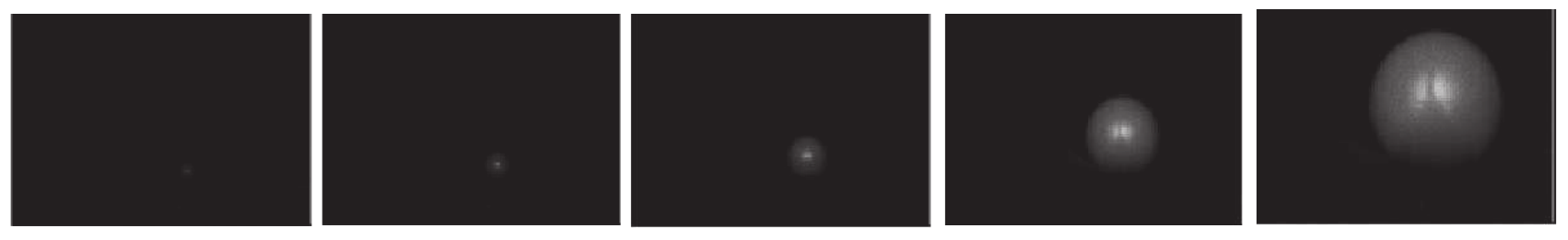}
		\label{colias-views-light-appr}}
	\vfill
	\vspace{-0.1in}
	\subfloat[]{\includegraphics[width=0.45\textwidth]{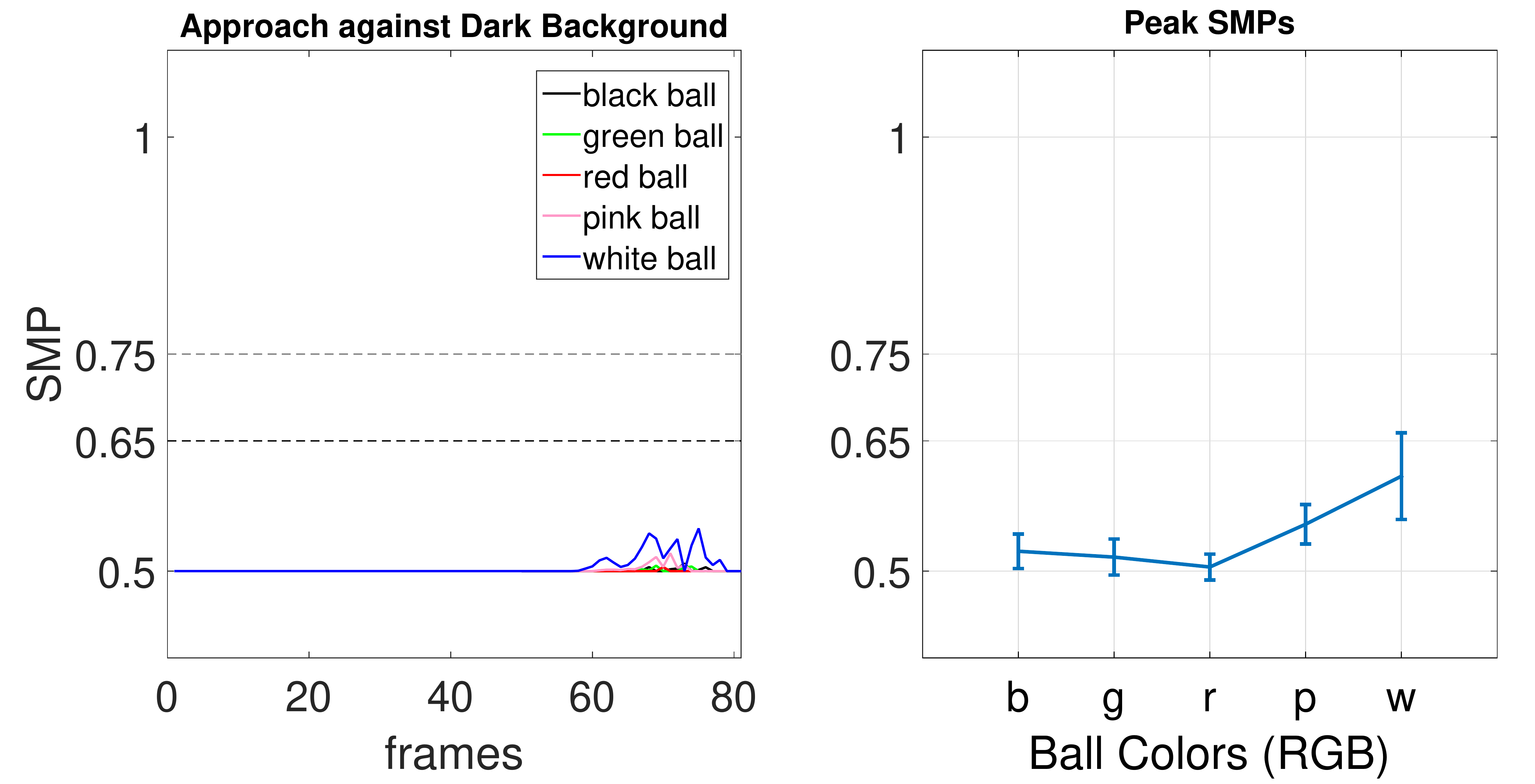}
		\label{light-balls-apprs}}
	\caption{The results of robotic experiments in which the LGMD2 challenged with looming stimuli at five different gray-scales. (a), (c) The sampled views collected during the red ball approaching the micro-robot in bright (a) and dark (c) background respectively. (b) the sigmoid membrane potential of the LGMD2 stimulated by dark objects looming against bright background created by global illumination, with the statistical peak SMPs for different colored (gray-level) balls. For each ball, the looming experiments have repeated ten times. (d) the sigmoid membrane potential of LGMD2 under brighter looming objects against dark background created by local surface illumination, with all the notations are the same as (b).}
	\label{colias-gray-balls-appr}
\end{figure}

In the first round of experiments, only the global illumination (source A in Fig.\ref{balls-appr-set}) is applied to make up a purely bright background (as samples of the views from Colias shown in Fig.\ref{colias-views-dark-appr}). The micro-robot is stimulated by those gray-scaled approaching objects which are all obviously darker than the background. The results in Fig.\ref{dark-balls-apprs} have illustrated LGMD2 neuron model's outputs showing steeply increased potentials for each dark looming object. The vision system could successfully recognize four gray-scaled approaching objects as collisions, yet the white object looming against bright background gave rise to a relatively weaker response which could not activate the LGMD2 model for high-frequency spikes. Intuitively, the statistical results from repeated tests in Fig.\ref{dark-balls-apprs} reveal that LGMD2 neuron model performs stably with small variances. More importantly, it is noticed that a darker object leads to a stronger peak response, which means the model is sensitive to the contrast between looming object and background that matches the results shown in Fig.\ref{bar-c-s}.

In the second round of experiments, we changed the illumination by replacing the global source with the local surface lighting (B in Fig. \ref{balls-appr-set}). As shown in the sampled views collected by Colias depicted in Fig.\ref{colias-views-light-appr}, all targets including the black ball are lighter than the background in this case. As a result, each looming stimuli bring about the dark-to-light luminance change. The results illustrated in Fig. \ref{light-balls-apprs} demonstrate the proposed LGMD2 neural network is not sensitive to this type of collisions with lighter approaching objects. This is exactly consistent with the revealed neural properties of LGMD2 neuron in juvenile locust which is only sensitive to the light-to-dark luminance change \cite{LGMD2-1997,LGMD2-cockpit}. Amongst other gray-levels, although the white looming object  leads to strongest responses, the peaks are all far below the defined threshold as depicted in the results in Fig.\ref{light-balls-apprs}.

When challenged against translation movements, we investigate the effects of different velocities of and varied distances to the stimuli. In the experiments, we let a dark ball automatically roll down along the slot which then horizontally crossed the Colias' field of view as illustrated in Fig.\ref{robot-trans-set}. In the first case, the distance is fixed at $30$cm whilst gradient of the slot varies at $20$cm, $35$cm and $50$cm respectively. Indeed, the translating velocity of a ball rise up along with the increasing gradient of the slot. Fig.\ref{colias-trans-dht} illustrates that the LGMD2 neuron model only responds with brief excitations against translation movements. It is also clear that the neural responses are weakened dramatically by the SFA mechanism. Without SFA, the higher velocity stimuli could activate the LGMD2 neuron model. In addition, the statistical results in Fig.\ref{bar_dht} also reveal the response of LGMD2 neural network to the speed of translation movements - the SMP peaks at higher level with higher speed of translation movements.

\begin{figure}[!t]
	\centering
	\subfloat[]{\includegraphics[width=0.23\textwidth,height=0.11\textheight]{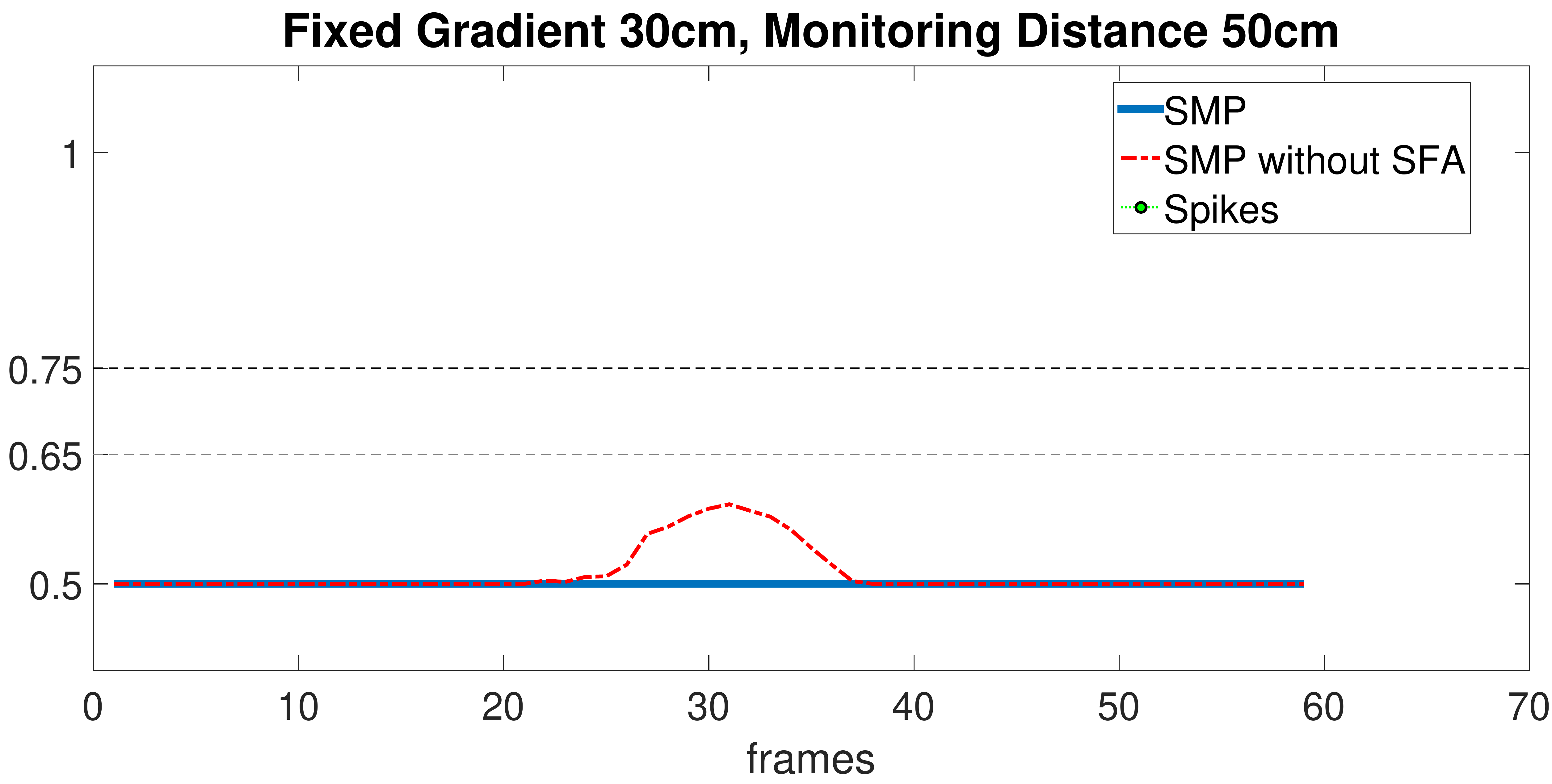}
		\label{hdt1}}
	\hfill
	\subfloat[]{\includegraphics[width=0.23\textwidth,height=0.11\textheight]{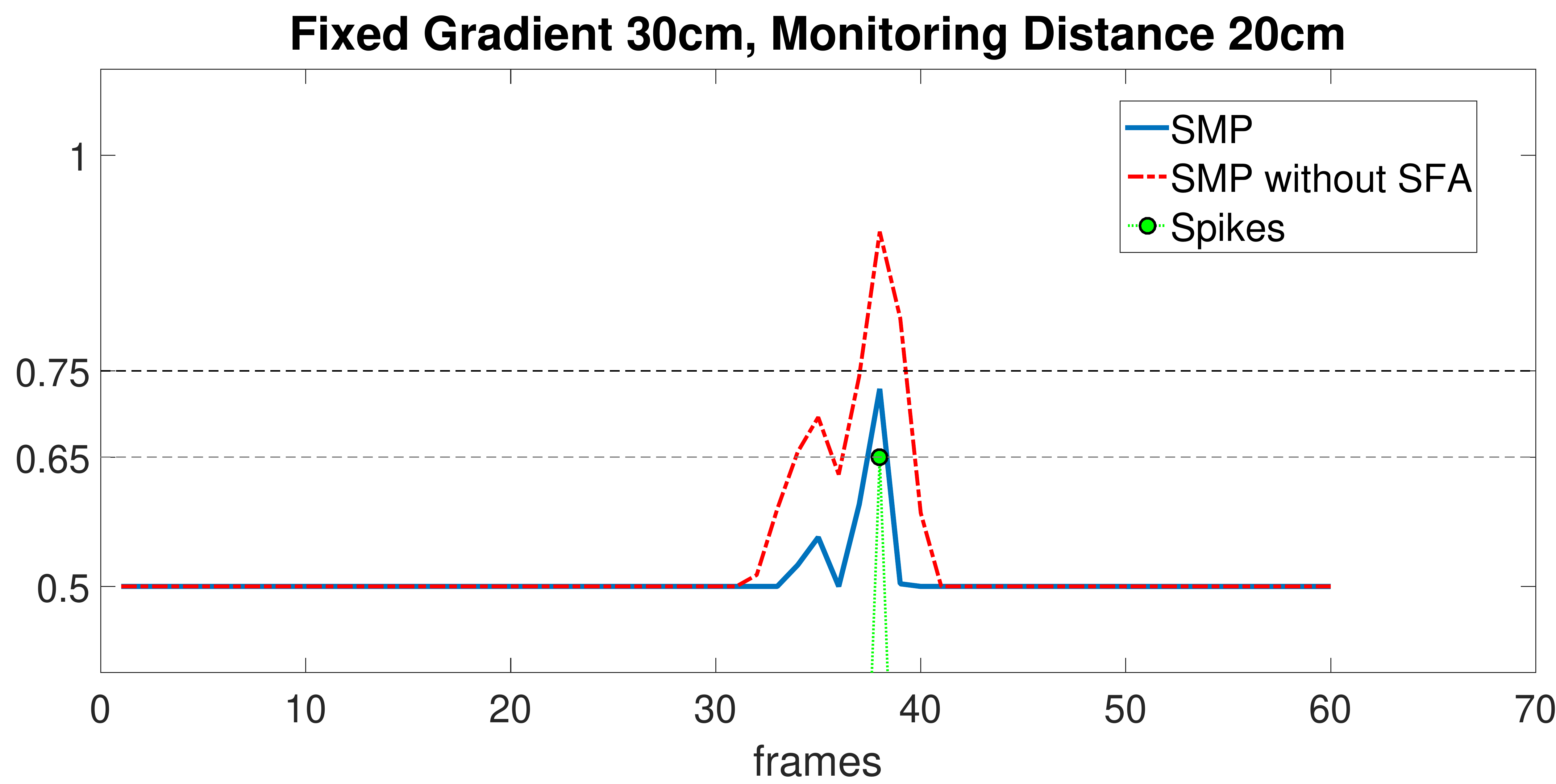}
		\label{hdt2}}
	\vfill
	\vspace{-0.1in}
	\subfloat[]{\includegraphics[width=0.23\textwidth,height=0.11\textheight]{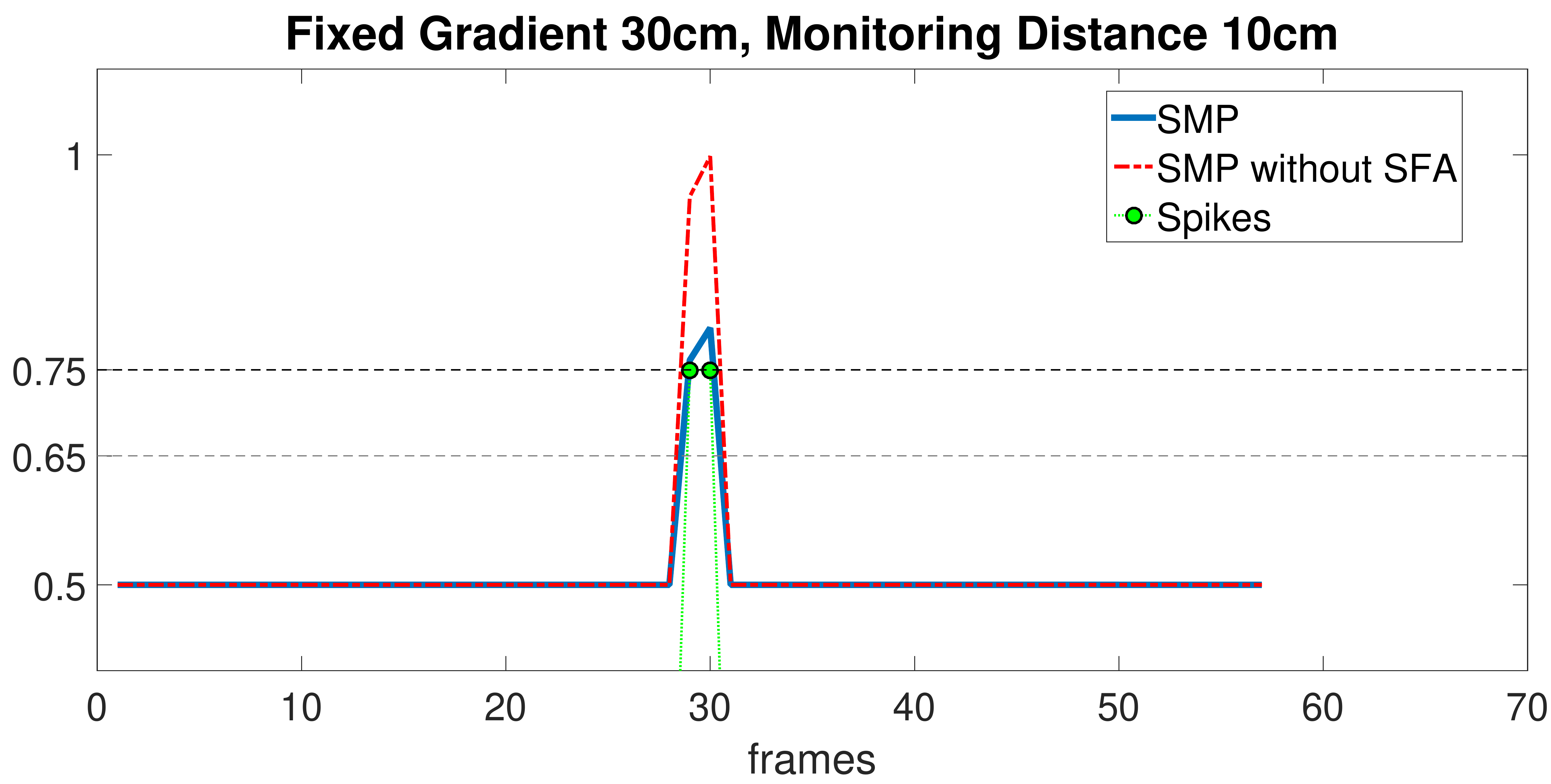}
		\label{hdt3}}
	\hfill
	\subfloat[]{\includegraphics[width=0.23\textwidth,height=0.11\textheight]{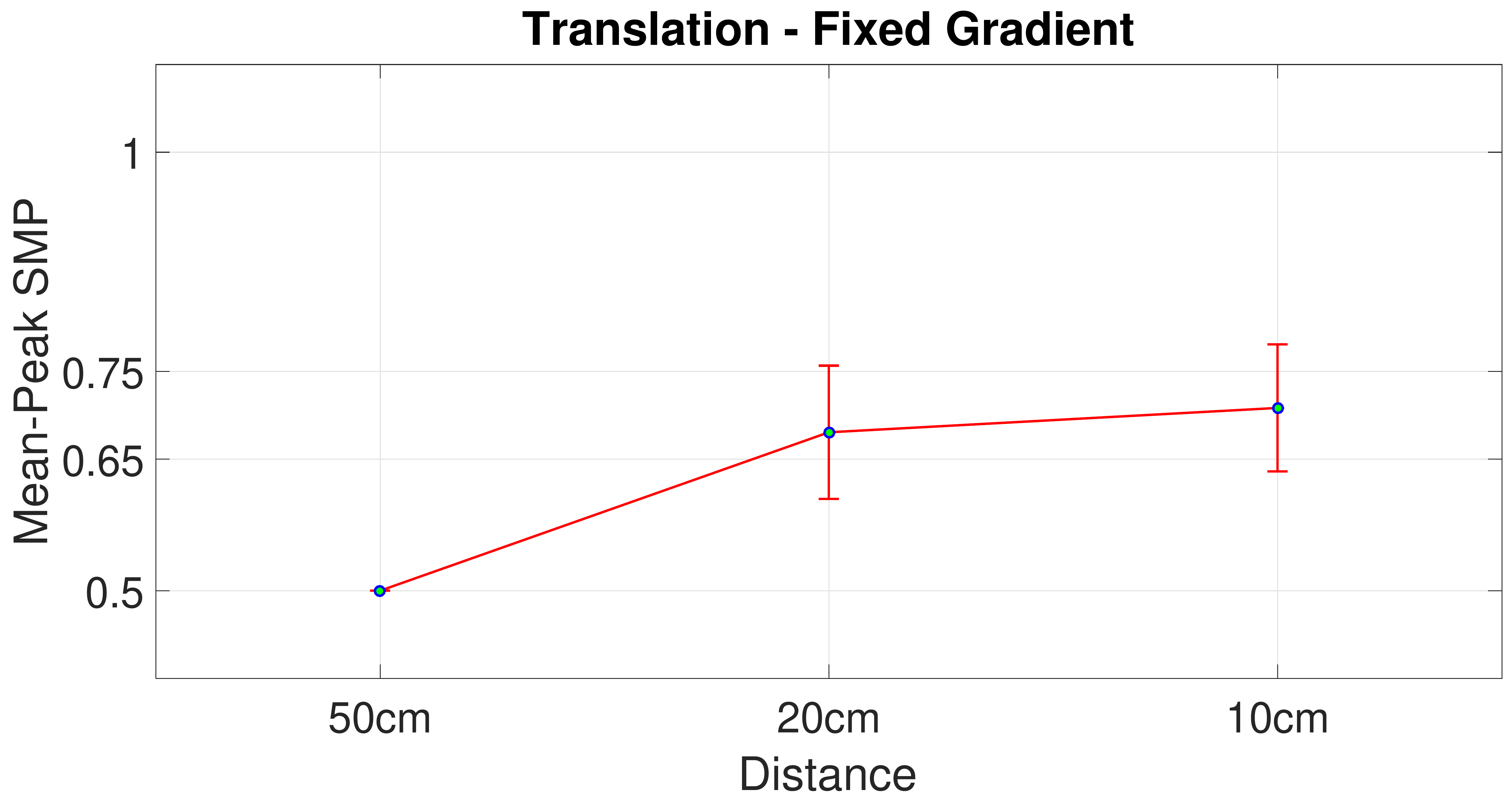}
		\label{bar_hdt}}
	\caption{The results under the second set of translations. In this case, the slot-gradient is fixed at 30cm and the monitoring distances $d$ varied at 50, 20, 10cm respectively. (a), (b), (c) LGMD2 neural responses. All notations are similar. (d) Statistical results of peak-SMPs at different monitoring distances: ten repeated tests for each specific distance.}
	\label{colias-trans-hdt}
\end{figure}

In the second set of translation movement experiments, the slot gradient is fixed at $30$cm implying approximately the same translating speed, while the monitoring distances varied at $50$cm, $20$cm and $10$cm respectively. The results (Fig. \ref{colias-trans-hdt}) is as expected - further distance leads to weaker response of LGMD2 neuron model. When the distance between visual stimuli and the micro-robot is far enough ($50$cm in our case), the LGMD2 neuron model remains almost quiet (Fig. \ref{hdt1}). On the other hand, if the distance between the stimuli and the robot is very close ($10$cm in our case), the LGMD2 neuron model could be highly activated (Fig.\ref{hdt3}), i.e. it is conceivable that a translation movement is too close to the receptive field, it could also be treated as a potential threat or collision.

To give a brief summary, the robotic experiments verify that the proposed LGMD2 based vision system could perform robustly and timely for collision detection with very limited hardware. It also fulfill the unique characteristics of LGMD2 neurons in juvenile locusts' visual pathway - responds to dark approaching object against bright background only. Its selectivity may bring benefits to build robust collision detectors for future robots.

\subsection{Discussion}
Through the above systematic experiments, we have shown that the proposed visual neural network, with parallel ON and OFF pathways (Fig. \ref{model}), demonstrates the similar characteristics of a biological LGMD2 neuron in juvenile locust's visual brain \cite{LGMD2-1997,LGMD2-cockpit}. In locusts, both LGMD1 and LGMD2 respond to rapid expanding object representing an imminent collision or a strike from predator. Nevertheless, the biological functions of LGMD2 differ from the LGMD1 in a number of ways \cite{LGMD2-1997}. First, LGMD2 is not sensitive to a light or bright approaching object whereas LGMD1 is. Second, LGMD2 does not respond to dark receding objects at all while LGMD1 is often excited though very briefly. Our proposed LGMD2 neural network has fully exhibited the above two critical features, as shown in the above results (e.g. Fig.\ref{simu-looming}, \ref{real-appr-reced} and \ref{colias-gray-balls-appr}). A shortcoming of LGMD2 neuron is that it can not recognize light object looming which might be rare for a juvenile locust, whereas LGMD1 can. If we could build multiple visual pathways by combining the characteristics of LGMD1, LGMD2 and other relevant neurons, the collision selectivity to approaching versus receding could be further enhanced.

In both flies and locusts, ON and OFF cells process motion information about the same place in the field of view, however this comes about by distinct mechanisms. The characteristic array of directional sensitive tangential cells found in the lobula plate of many flies are not found in locusts \cite{LGMD1-synaptic2015}. Therefore unlike those directional selective neurons \cite{DSN-2007,DSN-2013,LGMD1-DSN-competing,RotationalNN,Circuit-motion,LGMD1-synaptic2015,Joesch_2010}, we model LGMD2 to only react selectively to looming stimuli with changes in extent, rather than the four cardinal directional movements \cite{LGMD2-1997,LGMD1-1996}. The above experiments with X-Y planes stimuli on different directions (Fig.\ref{simu-trans-elong}) demonstrate that we have achieved such specific characteristic of LGMD2 neurons.

A biophysical mechanism, spike frequency adaptation \cite{SFA-2009,SFA-2003,SFA-2009Role,SFA-Gabbiani,SFA-2014}, has been modeled in this work -  it contributes in shaping the collision selectivity, especially for inhibiting response to translation at constant speed (e.g. Fig.\ref{simu-trans-elong}). However, it has little effect on 'acceleration' of stimulus, for example, the approaching (e.g. Fig.\ref{real-aa},\ref{colias-overhead}) and also the accelerated translation (Fig.\ref{real-trans}), which are likely to overcome adaptation. To achieve the strong inhibition following rapid brightness change over large areas in the retina \cite{LGMD2-1997}, a similar FFI pathway is also built in the LGMD2 neuron model, though its morphological structure has not been explored \cite{LGMD2-1997} in LGMD2 but LGMD1.

\section{Conclusion}
\label{ConSec}
In this article, we propose a collision selective visual neural network based on an unique neuron LGMD2 in the juvenile locusts' visual pathway. The LGMD2 neuron is sensitive to looming objects but only responds selectively to dark objects looming against bright background underlie a preference to the light-to-dark luminance change. With parallel and biased ON and OFF channels encoding onset and offset responses separately in a computational structure, the unique selectivity of LGMD2 neuron has been fully demonstrated in this modeling study. In addition, a biophysical mechanism, the spike frequency adaptation mechanism is employed to enhance the LGMD2's specific selectivity to approaching versus translating and receding. The proposed LGMD2 neuron model has been verified with systematic experiments challenged by stimuli ranging from synthetic to real time. The robotic experiments further approve the LGMD2 model's robust performance in collision selectivity implemented in a ground miniature robot. 

Since this bio-plausible collision sensitive model perceives collision cues via spatio-temporal computation, it can cope with cluttered environments without applying complex object segmentation and recognition methodologies. Similar to other neuromorphic computation structures, the proposed LGMD2 model can also be easily realized in VLSI chip for volume production. With similar separated ON/OFF pathways, we will investigate directional motion selectivity in the future.

\section*{Acknowledgment}
This work was supported in part by EU FP7 project LIVCODE(295151), HAZCEPT(318907) and EU Horizon 2020 project STEP2DYNA(691154). We thank Dr. F Claire Rind and Dr. Peter Simmons for the suggestions on gratings tests, Prof. Jigen Peng for the suggestions on model illustration, Dr. Tom{\' a}{\v s} Krajn{\' i}k for running the localization system in the arena tests.

\bibliographystyle{IEEEtran}

\bibliography{IEEEqinbing}




\end{document}